\documentclass[a4paper,onecolumn,11pt]{quantumarticle}
\pdfoutput=1
\usepackage{pgfplots}
\pgfplotsset{compat=1.18}
\usetikzlibrary{arrows.meta}
\usepackage{cite}
\usepackage{amsmath,amssymb,amsfonts,amsthm}
\usepackage{graphicx}
\usepackage{textcomp}
\usepackage{bbm}
\usepackage{mathrsfs}

\usepackage[dvipsnames]{xcolor}
\usepackage{tikz}
\usepackage{hyperref}
\usepackage{enumerate}
\usepackage{enumitem}

\usepackage{xparse}
\usepackage{etoolbox}

\usepackage{algorithm}
\usepackage{algpseudocode}
\newtheorem{lemma}{Lemma}
\newtheorem{theorem}{Theorem}
\newtheorem{definition}{Definition}

\newtheorem{proposition}{Proposition}
\newtheorem{remark}{Remark}
\newtheorem{assumption}{Assumption}

\newtheorem{example}{Example}

\newcommand{\BR}{\bigoplus_{j\neq i}{\mathfrak{B}}(\Hc_j)}

\newcommand{\WCS}{\Wc_{c/s}}
\newcommand{\WC}{\Wc_{c}}
\newcommand{\WS}{\Wc_{s}}
\newcommand{\Hcp}{\Hc=\bigoplus_{i=1}^q \Hc_i}
\newcommand{\ti}{\Gamma}

\newcommand{\Dc}{\mathcal{D}}
\newcommand{\Ec}{\mathcal{E}}

\newcommand{\Nc}{\mathcal{N}}

\newcommand{\Hc}{\mathcal{H}}
\newcommand{\Jc}{\mathcal{J}}

\newcommand{\Sc}{\mathcal{S}}
\newcommand{\Pc}{\mathcal{P}}
\newcommand{\Zc}{\mathcal{Z}}
\newcommand{\Qc}{\mathcal{Q}}
\newcommand{\Ic}{\mathcal{I}}
\newcommand{\Kc}{\mathcal{K}}

\newcommand{\Wc}{\mathcal{W}}
\newcommand{\Uc}{\mathcal{U}}
\newcommand{\Vc}{\mathcal{V}}

\DeclareMathAlphabet{\mathpzc}{OT1}{pzc}{m}{it}

\newcommand{\Rb}{\mathbb{R}}
\newcommand{\Cb}{\mathbb{C}}
\newcommand{\Pb}{\mathbb{P}}

\newcommand{\Nb}{\mathbb{N}}

\newcommand{\Bf}{\mathfrak{B}}

\newcommand{\Hf}{\mathfrak{H}}

\newcommand{\one}{I}
\newcommand{\ones}{\mathbbm{1}}

\newcommand{\tr}{{\rm tr}}
\newcommand{\diag}{{\rm diag}}

\newcommand{\supp}{{\rm supp}}

\newcommand{\ket}[1]{\left| #1 \right>}
\newcommand{\bra}[1]{\left< #1 \right|}
\newcommand{\norm}[1]{\left|\left| #1 \right|\right|}

\newcommand{\braket}[2]{\left< #1 \middle\vert #2 \right>}
\newcommand{\ketbra}[2]{\left\vert #1 \middle>\!\middle< #2 \right\vert}

\definecolor{darkviolet}{rgb}{0.58, 0.0, 0.83}
\definecolor{lavender}{rgb}{0.45, 0.31, 0.59}

\renewcommand{\Pr}{\Pb}

\begin{document}
	\author{Marco Peruzzo}
	\affiliation{Department of Information Engineering, University of Padova, Italy}
	\email{marco.peruzzo@dei.unipd.it}
	
	\author{Mădălin Guţă}
	\affiliation{School of Mathematical Sciences, University of Nottingham, Nottingham, NG7 2RD, United Kingdom}
	
	\affiliation{Centre for the Mathematics and Theoretical Physics of Quantum Non-equilibrium Systems, University of Nottingham, Nottingham, NG7 2RD, United Kingdom}
	
	\email{madalin.guta@nottingham.ac.uk }
	
	\author{Francesco Ticozzi}
	
	\affiliation{Department of Information Engineering, University of Padova, Italy}
	
	\affiliation{Padua Quantum Technologies Research Center, University of Padova, Padua, Italy}
	
	\email{ticozzi@dei.unipd.it}
	
	\title{Stochastic Complementation of Quantum Channels: Stationary States and Metastability}
	
	\maketitle
	
	\begin{abstract}
		We develop a theory of {stochastic complementation and metastability} for discrete-time Markovian quantum dynamics, described by iterations of a quantum channel, as an extension of the corresponding theory for classical Markov chains.
		Given a decomposition of the underlying Hilbert space into orthogonal subspaces we restrict the original  channel to the corresponding block-diagonal algebra, obtaining an hybrid quantum-classical channel. For this reduced dynamics, the {hybrid stochastic complements} decouple the original channel into reduced ones, which can be used to compute the global stationary state one subspace at a time when such state is compatible with the chosen block decomposition.
		Hybrid stochastic complementation also provides a framework for characterizing convergence to stationarity and describe metastable dynamics for channels where the coupling between subspaces is weak, which we name {\em nearly completely reducible}, obtaining a quantum version of the classical Simon-Ando theory. Potential applications and simple yet paradigmatic examples are discussed.
	\end{abstract}
	
	\section{Introduction}
	
	Open quantum systems frequently relax on several widely separated time
	scales. A system may rapidly lose information about its microscopic initial
	condition while retaining, for much longer times, relevant information or geometric structures. Such intermediate regimes are of interest both for
	understanding relaxation and phase transitions in dissipative quantum
	systems and for quantum-information applications, where the lifetime of an
	encoded degree of freedom or a measurement record can be much longer than
	the timescale of the strongest dissipative effects. So far, \emph{metastability} has been
	investigated in open quantum dynamics using spectral methods, which identify
	slowly decaying modes and the associated manifold of long-lived states
	\cite{macieszczak2016towards,macieszczak2021classical,macieszczak2021operational}.
	Metastable dynamics also has a natural interpretation in terms of quantum
	trajectories and effective switching between long-lived phases
	\cite{hscIntroBrown2024}. In the discrete-time setting, spectral and
	trajectory-based aspects of metastability for iterated quantum channels have
	also been studied in \cite{hscIntroJin2024}.
	
	A complementary question is whether a useful 
	description of metastable behavior can be derived directly from a geometric picture, and in particular a {\em physically specified
		partition} of the Hilbert space, rather than by spectral arguments. Such partitions naturally arise, for example,
	from state conditioning with nondemolition measurements, Zeno subspaces, approximately
	conserved energy sectors, and symmetry eigenspaces \cite{hscIntroBraginsky1980,hscIntroFacchiPascazio2008,hscIntroAlbertJiang2014}. When transitions between
	these sectors are weak, one often finds fast equilibration within each sector,
	followed by slower redistribution of their populations (these scenarios are discussed in detail below, in a dedicated part of this introduction). The challenge is to
	formalize this intuition and provide sufficient conditions for the emergence of these separate timescales, in a manner that respects complete positivity and
	retains an effective description of the degrees of freedom \emph{within} each sector.
	
	In a classical commutative framework, this is attained in Meyer's theory of \emph{stochastic
		complementation} for Markov chains \cite{mayer-sc}. Given a partition of a
	finite state space, the stochastic complement associated with a selected
	class incorporates both direct transitions inside the class and all
	excursions outside it that eventually return. For an irreducible chain, the
	complements are themselves irreducible stochastic matrices, and their local
	stationary distributions, together with a smaller coupling matrix,
	reconstruct the global stationary distribution. The same construction is
	closely connected with the Simon--Ando theory of nearly completely
	reducible dynamics \cite{HS-AA:61,PC-RA:77}: rapid relaxation within weakly
	coupled classes can precede the slow exchange of probability between them.
	Extending this picture to quantum systems is not immediate. A quantum state
	may have both populations in several orthogonal subspaces and coherences
	between them: a partition in mutually exclusive alternatives cannot simply
	be imposed on the full dynamics while neglecting those
	coherences.\\[2mm]

	To this aim, we consider a finite-dimensional Hilbert-space decomposition
	$\mathcal H=\bigoplus_{i=1}^q\mathcal H_i$ and a discrete-time quantum
	channel $\Phi$. 
	We then introduce the {\em associated hybrid channel}
	$\Phi_{\oplus},$ which acts on the
	block-diagonal algebra $\bigoplus_i\Bf(\mathcal H_i)$. Its blocks
	$\Phi_{ij}:\Bf(\mathcal H_j)\to\Bf(\mathcal H_i)$ are
	completely positive maps that replace classical transition probabilities
	with quantum operations. In particular, the label $i$ can be treated as a
	classical variable, while the unnormalized state on $\mathcal H_i$ encodes
	its probability together with its conditional quantum state, effectively defining an hybrid classical-quantum system. For each
	selected sector, we consider a hybrid stochastic complement (HSC) ${\cal S}_i$ , which 
	includes both the {\em internal dynamics within the sector} and all possible \emph{excursions through
		other sectors} that ultimately return to the selected one.  Crucially, under irreducibility assumptions the HSC is a
	completely positive and trace-preserving channel, not merely a formal
	Schur complement (or Kron reduction, see \cite{kron, bullo}). 
    {The HSC definition and its properties are directly connected to the theory of recurrences and first-return maps for quantum open random walk \cite{LardizabalSouza2015,GrunbaumLardizabalVelazquez2020,GrunbaumVelazquez2018,BardetBernardPautrat2017}. A more detail comparison is provided below. }
    
	Building on the HSC definition, we provide results for hybrid
	classical--quantum scenarios in three main directions. First, we show that the
	HSCs of an irreducible associated hybrid channel are irreducible quantum
	channels and establish how to reconstruct 
	block-diagonal stationary states from the stationary states of the complements. 
    
    Second, we
	quantify weak inter-sector coupling and distinguish the behavior of
	block-populations from that of the full quantum state, thereby exposing the
	role of surviving coherences. Third, under suitable primitivity and
	spectral assumptions, we obtain explicit short-time stabilization bounds
	and describe the intermediate metastable regime in terms of the HSC
	stationary states. In particular, these estimates separate a growing error term and the inter-sector leakage from a term
	that decays with the intra-sector mixing rate. This gives a quantitative
	criterion for determining a window in which local equilibration has occurred but
	global equilibration has not. Our contribution is thus complementary to
	spectral metastability theory: it builds on a physically interpretable
	partition, provides completely positive return dynamics and explicit
	metastability bounds.

	\paragraph*{Motivating scenarios.}
	Several physically relevant settings motivate treating the block structure
	as given.

	{\em Approximate symmetries.}
	Symmetries provide a natural source of the required subspace decomposition.
	A strong unitary symmetry of an open-system dynamics commutes with the
	individual Kraus operators, or with the Hamiltonian and noise operators of a
	continuous-semigroup generator in Lindblad form \cite{hscIntroBucaProsen2012,hscIntroAlbertJiang2014}.
	The introduction of symmetry-breaking perturbations can induce transitions between sectors
	that were exactly decoupled, allowing for separation of local relaxation from
	slow inter-sector exchange. Our results identify the
	stationary states within these sectors and provide quantitative
	description of their metastable behavior. A strong symmetry need
	not eliminate coherences between sectors: this motivates the distinction
	we make between nearly completely reducible channels and nearly strongly reducible
	channels, for which the full state can approach a block-diagonal
	metastable state. The two-qubit examples in this work demonstrate this
	difference explicitly.
	
	{\em Perturbed quantum nondemolition measurements:}
	A quantum nondemolition  
	measurement (QND) is designed to extract information
	about an observable without inducing transitions between its eigenspaces;
	this principle underlies repeated measurements and the stabilization of
	quantum observables \cite{hscIntroBraginsky1980,hscIntroBauerBernard2011,
		hscIntroBauerBenoistBernard2013}. If the measured observable has spectral
	projections $\{\Pi_i\}$, its mutually exclusive outcomes provide a natural
	partition $\mathcal H=\bigoplus_i\operatorname{ran}\Pi_i$. In a nominal
	QND dynamics, these sectors are invariant, even if nontrivial evolution is
	allowed within degenerate eigenspaces. Unwanted couplings or detector imperfections can weakly break the nondemolition property: a population initially associated
	with one outcome may then leak into another sector. Repeated monitoring may
	quickly suppress coherences between different outcomes while the recorded
	sector changes only rarely, creating a natural
	mechanism for the emergence of two-stage relaxation. HSCs
	provide quantum return maps that incorporate arbitrary sequences of such
	transitions for the \emph{nonselective}
	(or averaged), and can be used to study the associated long-lived
	block-populations. 
	
	{\em Generalized quantum Zeno dynamics:}
	A closely related situation is generated by alternating a projective
	measurement with a free-evolution step. A nonselective measurement with projectors
	$\{\Pi_i\}$, induces a perfectly block-dephasing map
	$\mathcal W_{\mathrm c}$; one cycle of a stroboscopic protocol can therefore
	be written as
	\(
	\Phi_\tau
	= \mathcal E_\tau
	 \mathcal W_{\mathrm c},
	\)
	where $\mathcal E_\tau$ represents the effective free evolution. In the standard
	Hamiltonian case $\mathcal E_\tau(X)=e^{-i\tau H}Xe^{i\tau H}$,
	transitions between different measured sectors have probabilities of order
	$\tau^2$ per step; in the frequent-measurement limit this leads to Zeno
	confinement and effective evolution within the measured subspaces
	\cite{hscIntroFacchiPascazio2008}. More generally, $\mathcal E_\tau$
	may be a quantum dynamical semigroup or another open-system channel, as
	considered in generalized Zeno dynamics
	\cite{hscIntroLiHerreraKwek2013,hscIntroBurgarth2020}. For a finite
	inter-measurement time, the channels $\Phi_\tau$ induces a
	hybrid structure, as considered in this work; its off-block transitions, induced  $\mathcal E_\tau$  quantify
	departures from ideal sector confinement of ${\cal W}_c$. The HSC then describes the
	resulting return dynamics, including repeated excursions between Zeno
	sectors. There is an important distinction between coherent and dissipative
	leakage: Lindblad jump processes may induce transitions with probability of
	order $\tau$ per step, so fast projective monitoring alone does not in general
	suppress open-system transition as effectively. The metastability estimates apply
	when the inter-sector leakage is actually small and fast
	relaxation is present within each sector.
	
	{\em Thermalization in approximately separated energy sectors.}
	Energy eigenspaces, or suitably coarse-grained energy shells, provide
	another physically-meaningful Hilbert-space partition. Suppose a dynamics
	rapidly mixes states within each sector, while the interactions responsible
	for energy exchange between sectors act on a slower time scale. Under
	suitable mixing assumptions, the resulting intermediate states may have a
	microcanonical distribution, or a more general  equilibrium, within each energy shell, with slowly varying relative
	weights. At longer times, weak coupling to a thermal reservoir can produce
	transitions between energies and, when detailed balance and ergodicity
	hold, relaxation towards a canonical Gibbs state. This scenario is
	consistent with the weak-coupling theory of thermal master equations
	\cite{hscIntroDavies1974,hscIntroSpohnLebowitz1978} and with the
	statistical-mechanical distinction between microcanonical descriptions and
	canonical equilibrium \cite{hscIntroGoldstein2006}. 
	The present framework can quantify the separate processes through intra-sector HSC stationary
	states and a coupling matrix for their equilibrium weights. While the
	microcanonical-to-canonical timescale separation is not automatic, it can be ensured by suitable structural and dynamical assumptions.
    
	\paragraph*{Relation to first-return maps and existing coupling methods.} 

    {The analysis of quantum return dynamics has been considered in the {\em first-return-function} framework of Gr{\"u}nbaum and Vel{\'a}zquez
\cite{GrunbaumVelazquez2018} and its application to quantum Markov
chains by Gr{\"u}nbaum, Lardizabal, and Vel{\'a}zquez
\cite{GrunbaumLardizabalVelazquez2020}. These works express monitored
first returns through resolvent formulas that include both one-step
transitions within the target subspace and excursions outside it,
 establishing renewal identities and splitting rules.
Related work by Lardizabal
and Souza \cite{LardizabalSouza2015} constructs stationary operators
from excursion occupation operators associated with invariant
states of the return map. Bardet, Bernard, and Pautrat
\cite{BardetBernardPautrat2017} develop a systematic theory of passage
times, visit counts, and recurrence for open quantum walks,
including irreducibility-based recurrence classifications and  relations
between mean return times and stationary weights.
These results provide
direct precedents for the structural and stationary-state
properties of hybrid stochastic complements.

The HSC considered here coincides with the integrated first-return maps of the associated hybrid channel.  
    In particular,  Proposition \ref{prop:spectral_radius} in this paper corresponds to the finite-dimensional transience argument used in \cite  [Theorem 2.8]{GrunbaumLardizabalVelazquez2020}; in Theorem \ref{thm:hsc_prop}  the CPTP property can be derived by the recurrence results of \cite{GrunbaumLardizabalVelazquez2020} applied to the hybrid map $\Phi_\oplus,$ while irreducibility and structure of the stationary state is found also in \cite{BardetBernardPautrat2017}. We have maintained the above mentioned results and their proofs in  our framework in order to provide a consistent presentation.         A related approach has recently been proposed by Argoud and Joye in \cite{ArgoudJoye2026}, who consider a special class of quantum channels
on direct-sum Hilbert spaces, combining unitary evolution within
subspaces with measure-and-prepare transfers between them. They reconstruct equilibria through resolvents
of the internal dynamics and finite classical coupling problems, providing a concrete
instance of stationary-state reconstruction through classical
coupling data.

    With respect to these works, in addition to introducing these maps as part of a stochastic-complementation and hybrid system framework, we  accommodate for general completely
positive inter-sector maps and provide stationary-state and intermediate scale dynamics through a finite classical coupling matrix, including the effects of inter-sector coherences
in the original quantum dynamics.  In addition, we develop the scope of these results to a full {\em metastability analysis of the original channel}, including quantitative estimates relating inter-sector leakage and coherence effects to local equilibration.}
	
	\paragraph*{Relation to existing theories of quantum metastability.}
	The typical spectral approaches to metastability identify a metastable manifold from
	slow modes of a Lindblad generator and provides a general description that
	may include approximately disjoint phases, decoherence-free subspaces, and
	noiseless subsystems \cite{macieszczak2016towards}. In the case
	of the so-called \emph{classical metastability}, that manifold is approximately the convex
	hull of a finite family of quantum states, with effective stochastic motion
	among the metastable phases \cite{macieszczak2021classical}. An operational
	approach based on approximate stationarity over a time window provides
	another complementary characterization
	\cite{macieszczak2021operational}. These works are not restricted to
	metastability arising from a predetermined Hilbert-space partition, and in this respect
	are  more general. However, for generic dynamics, the slow manifold may hold no meaningful structure, unless the time-scale separation is between the center manifold - which is always an algebra with respect to a modified product - and the rest \cite{adiabaticalgebraic}. Conversely, stochastic
	complementation starts with a concrete sector decomposition and constructs
	\emph{exact CPTP return channels} whose stationary states supply explicit
	candidate metastable phases.  
	Our approach does not purport to describe the most general quantum
	(spectrally) metastable manifold: in particular, persistent quantum coherences or noiseless subsystem degrees of freedom may require a different approach.
	The distinction is especially relevant to discrete-time quantum channels,
	where metastability has also been formulated directly in spectral and
	trajectory terms \cite{hscIntroJin2024,hscIntroBrown2024}.
	
	\paragraph*{Paper organization.}

	The paper is organized as follows. 
	Section~\ref{sec:hsc} introduces the associated hybrid classical--quantum channel
	and defines its stochastic complements. We establish their complete
	positivity, trace preservation, and irreducibility properties, and show
	how their stationary states, together with a classical coupling matrix,
	reconstruct the stationary state of the hybrid channel and, under an
	additional block-diagonality assumption, of the original quantum
	channel. The equivalent classical results are summarized in Appendix ~\ref{sec:csc}, including the coupling
	theorem for stationary distributions and its connection with nearly
	completely reducible Markov chains and Simon--Ando metastability. Section~\ref{sec:ncr} develops the theory of nearly completely and nearly
	strongly reducible channels. After quantifying deviations from these
	two notions of reducibility, we derive stabilization bounds and describe
	the short-, intermediate-, and long-time regimes, emphasizing the
	distinction between metastability of block-populations and of the full
	quantum state. Section~\ref{sec:approxsymm} illustrates these results with two-qubit
	channels possessing approximately preserved strong-symmetry sectors,
	contrasting the persistence of inter-sector coherences with their rapid
	suppression, while Section \ref{sec:QND} develops the analysis for a joint system (a toy model for a two level atom with extra electron degrees of freedom) undergoing partial quantum-nondemolition measurements and weakly perturbed by an additional Hamitonian coupling. Section~\ref{sec:conclusions} summarizes the conclusions and outlines possible
	extensions. The appendices collect the technical proofs and
	norm estimates used in the main text.

	\paragraph*{Mathematical notation - } In this work, we consider finite-dimensional Hilbert spaces \(\Hc\simeq\Cb^{n}\). Vectors in \(\Hc\) are denoted by \(\ket{\phi}\), while $\bra{\phi}=\ket{\phi}^{\dag},$ where \(\dag\) denotes the complex-conjugate transpose. Given two vectors \(\ket{\phi},\ket{\psi}\in\Hc\), their standard inner product is
$\braket{\phi}{\psi}
    =\sum_{i=0}^{n-1}\bar{\phi}_{i}\psi_{i}\in\Cb,$ where \(\bar{\phi}_{i}\) denotes the complex conjugate of \(\phi_i\).
    We denote the space of linear operators on \(\Hc\) by \(\Bf(\Hc)\)
and equip it with the Hilbert--Schmidt inner product
\(\langle X,Y\rangle_{\mathrm{HS}}=\operatorname{Tr}(X^\dagger Y)\).  Given $\Hc_i, \Hc_j \subseteq \Hc$, ${\mathfrak{B}}(\Hc_i,\Hc_j)$ is the space of bounded linear operators acting between $\Hc_i$ and $\Hc_j$. 
We use capital letters such as \(X,Y\) for operators and \(I\) for the
identity. In the chosen basis, we identify operators with their matrices
in \(\Cb^{n\times n}\). We write
\(\diag(M_1,\ldots,M_k)=\bigoplus_{i=1}^k M_i\)
for the block-diagonal matrix with diagonal blocks \(M_1,\ldots,M_k\).
We define the support of \(X\) by \(\supp(X):=(\ker X)^\perp\).
We use Fraktur letters such as \(\mathfrak F,\mathfrak G\) for operator
subspaces and write \(\mathfrak F\oplus\mathfrak G\) for their direct
sum when they are orthogonal in the Hilbert--Schmidt inner product.
We denote superoperators by calligraphic letters or capital greek letters, for example 
\(\Psi,\Phi, \Ec:\Bf(\Hc)\to\Bf(\Hc)\). A nonzero operator \(X\) is an
eigenoperator of \(\Ec\) with eigenvalue \(\lambda\) if
\(\Ec(X)=\lambda X\). We denote the spectrum of \(\Ec\) by
\(\operatorname{sp}(\Ec)\). We denote with $\|\Ec\|_1,\|\Ec\|_2, \|\Ec\|_\infty$ the super-operator norms induced by the trace, Hilbert-Schmidt and operator norms for operators in ${\mathfrak{B}}(\Hc)$. The formal definitions of these norms and the relevant properties are reported in Appendix \ref{app:norms}.

	\section{The Hybrid Quantum Stochastic Complement}\label{sec:hsc}
	In this section we introduce the theory of hybrid stochastic complementation for quantum channels. We show that, under suitable hypotheses, the emerging  stochastic complement provides a tool to compute stationary states and characterize its convergence, including a class of metastable behaviors. 
	
	\subsection{Preliminaries}\label{sec:prem_not}
	We consider a quantum system with finite-dimensional Hilbert space $\Hc$ of dimension $D<\infty$. 
   We further denote with $ \mathfrak{D}(\mathcal{H}) = \left\{ \rho \in \mathfrak{B}(\mathcal{H}) \;:\; \rho = \rho^\dagger \geq 0,\ \mathrm{Tr}(\rho)=1 \right\} 
	$ the set of density operators on ${\mathfrak{B}}(\Hc)$ and with $\rho \in \mathfrak{D}(\mathcal{H})$ the  state of the system.

	We assume that the quantum system undergoes a \textit{
		discrete-time, time-homogeneous Markovian dynamics}. The evolution is described in full generality by the  equation
	$$\rho_{t+1}=\Phi(\rho_t),$$
	where $\rho_t$ is the state of the system at time $t$ and  $\Phi$ is a quantum channel, i.e., a linear, completely positive (CP) and trace-preserving (TP) map which maps $\mathfrak{D}(\mathcal{H})$ into itself. We recall that every CP map admits an operator-sum (Kraus) representation \cite{Nielsen2010} , i.e. $\Phi(\cdot)=\sum_k M_{k} \cdot M^\dag_{k}$, where $\{ M_k\}\subset {\mathfrak{B}}(\Hc)$, and the map is TP if and only if $\sum_k M_k^\dag M_k=I$.
	
	{A subspace $\Vc\subseteq\Hc$ is invariant under $\Phi$ if the channel maps every operator supported on $\Vc$ to an operator supported on $\Vc$: if $\Pi_{\Vc}$ is the orthogonal projector onto $\Vc$, this means that
		\[
		\Phi\left(\Pi_{\Vc}{\mathfrak{B}}(\Hc)\Pi_{\Vc}\right)
		\subseteq
		\Pi_{\Vc}{\mathfrak{B}}(\Hc)\Pi_{\Vc}.
		\]
		A channel $\Phi$ is {\em irreducible} if it admits only  trivial invariant subspaces, i.e. ${0}$ and $\Hc$.}
	
	An additional useful property of a quantum channel is primitivity: we recall that a quantum channel $\Ec$ is \emph{primitive} if  and only if there exists $n \in \mathbb{N}$ such that for every density matrix $\rho \in\Dc(\Hc)$ we have $\Ec^n(\rho) > 0$.
	
	\subsection{Associated hybrid dynamics}

	In analogy to classical theory, where the Markov chain state space was partitioned into subclasses, here we consider a decomposition of $\Hc$ into $q\geq 2$ subspaces:
	\begin{align}\label{eq:hs}
		\Hc=\bigoplus_{i=1}^q\Hc_i, \quad {\rm with} \quad {\rm dim}(\Hc_i)=d_i, \quad {\rm dim}(\Hc)=\sum_i d_i= D<\infty.    
	\end{align}
	
	A fundamental difference from the classical scenario is the presence of {\em coherence} between components of the state supported in different subspaces. In order to retrieve classical-like results, the quantum channel shall be divided into: (i) a part acting on block-populations (block-diagonal components of the system's state), called the associated hybrid channel, 
	and (ii) a part governing block-coherences (off-diagonal blocks of the system's state). This decomposition will be instrumental  in defining the hybrid stochastic complement and its properties. The following subsections detail these ideas.
	
	\subsubsection*{Operator block decomposition}
	Similarly to classical Markov chains, the decomposition of the Hilbert space \eqref{eq:hs} induces a block decomposition of operators $X\in {\mathfrak{B}}(\Hc)$ as
	\begin{equation}\label{eq:blk_op}
		X=\begin{bmatrix}
			X_{11} &X_{12}& \dots &X_{1q}\\
			X_{21}& X_{22} &\dots& X_{2q}\\
			\vdots& \ddots &\ddots& \vdots\\
			X_{q1}& X_{q2}& \dots &X_{qq}\\
		\end{bmatrix}
	\end{equation}
	where $X_{ij} \in {\mathfrak{B}}(\Hc_i,\Hc_j)$. We let the algebra of block-diagonal operators acting on $\Hc$, be $$
	{\cal A}_\oplus=\{X\in {\mathfrak{B}}(\Hc) \ : X=\diag(X_1,\dots X_q),\  X_i\in {\mathfrak{B}}(\Hc_i)\} \cong \bigoplus_{i=1}^q{\mathfrak{B}}(\Hc_i).$$

	\subsubsection*{Block embedding and selection maps.} 
    To introduce block-decompositions for the state and the dynamics, we introduce the maps
	\begin{align*}
		\Jc_{k}:\,&{\mathfrak{B}}(\Hc_k)\hspace{5px}\to {\mathfrak{B}}(\Hc)\\
		& X \hspace{20px} \mapsto J_k X J_k^\dag= (\bigoplus_{i=1}^{k-1}0_i ) \oplus X\oplus (\bigoplus_{i=k+1}^{q}0_i )
	\end{align*}
	where we let  $J_k\!=[0\dots I_k \dots 0]^\top\in {\mathfrak{B}}(\Hc_k,\Hc)$, $I_k$ and $0_k$ be the identity and zero operator on $\Hc_k$. The adjoint of $\Jc_k$ with $\Jc^\dag_k:{\mathfrak{B}}(\Hc)\to {\mathfrak{B}}(\Hc_k)$ is a {\em block-selection} superoperator that for any  $X_k\in {\mathfrak{B}}(\Hc_k)$,  $\Jc_k^\dag(X)=X_{kk}$, i.e. it returns the $k$-th diagonal block of $X$ which coincides with $X_{kk}$. For a more compact notation, we also define the {\em projection map} $\Pc_k(\cdot)=\Jc_k\Jc_k^\dag(\cdot)= \Pi_k \cdot \Pi_k$, where $\Pi_k$ is the orthogonal projector onto $\Hc_k$.

	\subsubsection*{State decomposition}
	We next introduce a state decomposition that separates the diagonal blocks from the remaining coherences. 
	
	\begin{definition}[\bf State decomposition]\label{def:state_decomposition} Let $\rho$ be a density operator on $\Hcp$. We define:
		\begin{enumerate}[label=S.\arabic*)]
			\item the {\em block populations} $\mu :=\sum_i \Jc_i(\mu_i)$ where $\mu_i:=\Jc^\dag_i(\rho)$.
			\item the {\em block coherences} $\rho_{\rm o} := \rho-\mu$
		\end{enumerate}
		Then 
		\begin{equation}\label{eq:state_decomposition}
			\rho=\mu+\rho_{\rm o}=\sum_i\Jc_i(\mu_i)+\rho_{\rm o}.
		\end{equation}
	\end{definition}
	
	The {block-population} $\mu\in \Dc(\Hc)$ represents the restriction of $\rho$ onto the algebra of block-diagonal operators $\bigoplus_{i=1}^q{\mathfrak{B}}(\Hc)$. Then $\mu_i\in {\mathfrak{B}}(\Hc_i)$ is the $i$-th diagonal block of the state and we refer to it as the block-population of the system in ${\mathfrak{B}}(\Hc_i)$.
	\subsubsection*{Channel decomposition}
	A corresponding decomposition can be introduced for the dynamics:
	\begin{definition}[\bf Quantum channel decomposition]\label{def:original_channel_decomposition}
		Let $\Phi$ be a quantum channel on a Hilbert space $\Hcp$.
		We define the following maps
		\begin{enumerate}
        \item the inter-block maps \[\Phi_{ij} :{\mathfrak{B}}(\Hc_j)\to{\mathfrak{B}}(\Hc_i), \qquad \,\Phi_{ij} := \mathcal{J}^\dag_i \Phi \mathcal{J}_j;\]
			
			\item  the associated \;{\em hybrid channel} $\Phi_{\oplus} := \sum_{i,j=1}^q \Jc_i \Phi_{ij} \Jc_j^\dag$;
            \item the reminder map $\Phi_{\Delta}=\Phi-\Phi_{\oplus}.$
		\end{enumerate}
		Then 
		\begin{equation}\label{eq:channel_decomposition}
			\Phi =\Phi_{\oplus}+\Phi_{\Delta}=\sum_{i,j=1}^q \Jc_i \Phi_{ij}\Jc_j^\dag+\Phi_{\Delta}.
		\end{equation}
		
	\end{definition}

	The map $\Phi_{\oplus}$ leaves $\bigoplus_{i=1}^q{\mathfrak{B}}(\Hc_i)$ invariant and captures the evolution of block-population, in particular $\Phi_{ij}$ describe how the block-population in ${\mathfrak{B}}(\Hc_j)$ transition to ${\mathfrak{B}}(\Hc_i)$.
	Instead, the map $\Phi_{\Delta}$ captures all processes involving block-coherences, including:
	(i) generation of block-coherences from block-populations,
	(ii) transfer of block-coherences into block-populations, and
	(iii) the internal evolution of block-coherences.
	
	\begin{remark}[\bf Associated Hybrid systems]
		The map $\Phi_{\oplus}$ and the state $\mu$ can be interpreted as the quantum channel and state associated with a {\em hybrid classical–quantum system} that includes both a classical variable (the block index) and the quantum degrees of freedom within each block. The state of such a system can be  described by a classical probability distribution together with a family of conditional quantum states.
		For this reason, we refer to $\Phi_{\oplus}$ as the {hybrid channel} associated to $\Phi$.
		
		The simplest situation in which an hybrid description emerge is that of non-selective measurements: consider an observable $O=\sum_i i \Pi_i\in\Hf (\Hc),$ with $\textrm{range} (\Pi_i)=\Hc_i$ for all values of a classical label $i \in \{1,\dots,q\}$. According to L\"uder's postulate, the normalized block population
		$\frac{\mu_i}{\operatorname{tr}[\mu_i]} \in \mathfrak{D}(\mathcal{H}_i)$
		represents the quantum state conditioned on the observation of the value $i$, while the probability that the classical variable takes the value i is given by $p_i = \operatorname{tr}[\mu_i]$. Hence, the system {\rm conditioned on a measurement of $O$ with unknown outcome} (a non-selective measurement), is described by the ensemble $\{(p_i, \mu_i/\operatorname{tr}[\mu_i])\}_{i=1}^q$. Equivalently, this ensemble is compactly encoded in the block-diagonal state $\mu = \bigoplus_{i=1}^q \mu_i$. The hybrid channel $\Phi_{\oplus}$ preserve the hybrid structure of states.
		In the algebraic formulation of quantum mechanics \cite{bratteli2012operator}, hybrid classical–quantum systems are described by operator algebras with non-trivial center.   
	\end{remark}
	
	\begin{proposition} [\bf Irreducibility of associated hybrid channel]\label{prop:hc_props} Let $\Phi$ be an irreducible quantum channel on a Hilbert space $\Hcp$, then the associated hybrid channel $\Phi_{\oplus}$ is an irreducible quantum channel as well. 
	\end{proposition}
	The proof of Proposition \ref{prop:hc_props} can be found in Appendix \ref{app:proofs_hc_hsc}.

	\subsection{Introducing the hybrid stochastic complement}\label{sec:hsc_intro}
	The hybrid stochastic complement describes the evolution of the block-population in ${\mathfrak{B}}(\Hc_i),$ which may also involve excursions through some ${\mathfrak{B}}(\Hc_j),$ with $ j\neq i$ before returning.

	For a given target subspace $\Hc_i$, $i\in\{1,\dots,q\}$, 
	we introduce the bi-partition  
	\begin{align}\label{eq:hs_bip}
		\Hc=\Hc_i\oplus \bar \Hc_i, \qquad {\rm with} \qquad \bar{\Hc}_i=\bigoplus_{j\neq i} \Hc_j, \quad i,j\in\{1,\dots,q\}
	\end{align}
	and we let $\bigoplus_{j\neq i} {\mathfrak{B}}(\Hc_j)\cong\{X\in {\mathfrak{B}}(\Hc) \ : X=\bigoplus_{j\neq i} X_{jj},\  X_{jj}\in {\mathfrak{B}}(\Hc_j)\}$ be the algebra of block diagonal operators on  $\bar{\Hc}_i$. { In Definition \ref{def:original_channel_decomposition}, we introduced the map $\Phi_{ii}=\Jc^\dag_i\Phi \Jc_i$. Note that $\Phi_{ii}$ is completely positive as it is given by the composition of CP maps.} In analogy with the classical case (see Section  \ref{sec:csc}) we introduce the following maps. 
	\begin{enumerate}[label=\roman*)]
		\item The evolution of block-population in $\BR$ is described by the map
		\begin{align*}
			\Bar{\Phi}_i:\BR&\qquad \to \qquad \BR,\\
			X & \qquad \mapsto \qquad \bigoplus_{j\neq i} \sum_{s\neq i}\Phi_{js} (X_{ss}).
		\end{align*}
		
		\item The transition of population from ${\mathfrak{B}}(\Hc_i)$ to $\BR$ is described by  
		\begin{align*}
			\Phi_{*i}:{\mathfrak{B}}(\Hc_i)&\qquad\to\qquad \BR,\\
			X& \qquad \mapsto \qquad \bigoplus_{j\neq i}\Phi_{ji}(X_{ii}).
		\end{align*} 
		\item The transition of  population from $\BR$ to ${\mathfrak{B}}(\Hc_i)$ is described by
		\begin{align*}
			\Phi_{i*}:\BR&\qquad  \to \qquad {\mathfrak{B}}(\Hc_i),\\  X& \qquad \mapsto \qquad \sum_{j\neq i} \Phi_{ij}(X_{jj}).
		\end{align*}
	\end{enumerate}
	{ All the above are CP since they are obtained as (orthogonal) sums of the CP maps $\Phi_{ij}$ introduced in Definition \ref{def:original_channel_decomposition}.} The associated hybrid channel acts on block diagonal operators $X=X_{ii}\oplus \bar{X}_{ii}\in {\mathfrak{B}}(\Hc)$, where $X_{ii}\in {\mathfrak{B}}(\Hc_i)$, $ \bar{X}_{ii}\in \bigoplus_{j\neq i} {\mathfrak{B}}(\Hc_j)$ as:
	\begin{align}\label{eq:blk_hch_i}
		\Phi_{\oplus}(X)=(\Phi_{ii}(X_{ii})+\Phi_{i*}(\bar{X}_{ii}))\oplus (\Phi_{*i}({X}_{ii})+\Bar{\Phi}_{i}(\bar{X}_{ii})).
	\end{align}
	\begin{proposition}[\bf Spectral properties of complementation maps]\label{prop:spectral_radius}
		If $\Phi_{\oplus}$ is an irreducible CPTP map the eigenvalues of the CP maps $\Phi_{ii}$ and $\Bar{\Phi}_i$ lie strictly inside the unit disk.
	\end{proposition}
{Proposition \ref{prop:spectral_radius} can be obtained as a consequence of the recurrence analysis of \cite{GrunbaumLardizabalVelazquez2020}; a direct proof in our framework  is found in Appendix \ref{app:proofs_hc_hsc}.}

	We are now ready to introduce the Hybrid Stochastic complement. Consider the following expression:
	\begin{align}\label{eq:hsc_full}
		\Sc_{ii}=  \Phi_{ii}+ \Phi_{i *}\sum_{k=0}^\infty\Bar{\Phi}^k_{i} \, \Phi_{* i}.
	\end{align}
	A probabilistic-inspired interpretation of the above follows from the classical case (see equation \eqref{eq:csc}), since:
	\begin{enumerate}[label=\roman*)]
		\item $\Phi_{ii}$ describes the evolution of the block-population within ${\mathfrak{B}}(\Hc_i)$.
		\item $\Phi_{i *}\sum_{k=0}^\infty \Bar{\Phi}^k_{i} \, \Phi_{* i}$ describes the evolutions of block-population of ${\mathfrak{B}}(\Hc_i)$ that involve an intermediate transition through $\BR$.
	\end{enumerate}
	Hence, it represents the dynamics of the $i$-th block, collecting all the ``paths" that return to it. In view of Proposition \ref{prop:spectral_radius}, the power series in \eqref{eq:hsc_full} converges to $(\Ic- \bar \Phi_i)^{-1}$, and we can give the following definition.
	
	\begin{definition}[\bf Hybrid stochastic complement] The Hybrid Stochastic Complement (HSC) of $\Phi_{ii}$ is given by: 
		\begin{equation}\label{eq:stoch_comp}
			\Sc_{ii}=\Phi_{ii}+ \Phi_{i*}  (\Ic-\bar \Phi_i)^{-1} \Phi_{*i}.
		\end{equation}
	\end{definition}
	Notice that this expression of the HSC directly matches that of the classical stochastic complement \eqref{eq:csc}. Below we establish some key properties of 
	HSCs, generalising those of their classical counteparts.

	\begin{theorem}[\bf Stationary state of the HSC]\label{thm:hsc_prop}
		Let $\Phi$ be an irreducible quantum channel on a Hilbert space $\Hcp$ and let $\tilde{\mu}=\sum_i \Jc_i(\tilde{\mu}_i)$ be the stationary state of the associated hybrid channel $\Phi_{\oplus}$. Then the stochastic complement ${\Sc_{ii}}$ is an irreducible quantum channel for all $ i\in \{1,\dots, q\}$  and  has a unique stationary state $\tilde \sigma_i$ given by
		$$\tilde{\sigma}_i=\frac{\tilde{\mu}_{i}}{\tr[{\tilde{\mu}_i}]}.$$
	\end{theorem}
	
	\noindent  {Theorem \ref{thm:hsc_prop} includes partial results on CPTP properties of first-recurrence maps from \cite{GrunbaumLardizabalVelazquez2020} and irreducibility and stationary-state structure from \cite{BardetBernardPautrat2017}; a full proof within our framework is proposed in Appendix \ref{app:proofs_hc_hsc}.}
	
	The following result provides a sufficient condition for the primitivity of $\Sc_{ii}$ .

	\begin{proposition}[\bf Primitivity of HSC]\label{prop:hsc_primitivity}Let $\Phi$ be an irreducible quantum channel on a Hilbert space  $\Hcp$, and let $\Phi_{\oplus}$ be the associated hybrid channel. For each $i\in \{1,\dots, q\}$, let $\Sc_{ii}$ denote the stochastic complement of $\Phi_{ii}$ given in~\eqref{eq:stoch_comp}.
		If there exists $n\in \Nb$ such that $\Phi_{ii}^n(\rho)>0 $ for all $ \rho\in \Dc(\Hc_i)$, then $S_{ii}$ is primitive. 	
	\end{proposition}
	\noindent The proof is given in Appendix \ref{app:proofs_hc_hsc}.
	\subsection{Hybrid channels and block-diagonal stationary states}
	In this section, we show that the HSC can be  used to compute stationary states of hybrid channels but also of the original channel, when the  stationary state is block-diagonal. Accordingly, we introduce the following assumption.
	
	\begin{assumption}\label{ass:blk_diag_ss}
		$\Phi$ is irreducible and the stationary state $\tilde {\rho}$ of $\Phi$ is block diagonal with respect to the Hilbert space partition $\Hcp$, (i.e.  $\tilde {\rho}\in \bigoplus_{i=1}^q {\mathfrak{B}}(\Hc_i)$).
	\end{assumption}
	
	The following proposition is preliminary to the main results.
	
	\begin{proposition}[\bf Channels with block diagonal stationary states]\label{prop:block_diag_ss} Under Assumption \ref{ass:blk_diag_ss}, the stationary state of $\Phi$ coincides with the unique stationary state of the associated hybrid channel $\Phi_{\oplus}$.
	\end{proposition}
	The proof can be found in Appendix \ref{app:proofs_hc_hsc}. In agreement with the classical theory, Theorem~\ref{thm:hsc_prop} establishes the existence of a relation between the stationary states of HSCs and the stationary state of the original channel. In particular, for each $i\in \{1,\dots,q\}$ the stationary state of $S_{ii}$, namely $\tilde {\sigma}_{i}$, and the block-population of the stationary state of the hybrid channel $\Phi_{\oplus}$ in the subspace ${\mathfrak{B}}(\Hc_i)$, namely $\tilde {\mu}_{i}$, differ only by a scalar multiple.
	In the following, we show that stochastic complements can be practically used to decouple the associated hybrid channel into separate channels defined on smaller subspaces and, under the block-diagonal stationary state assumption, compute the stationary state of $\Phi$.
	
	\begin{theorem}[\bf Hybrid coupling theorem]\label{thm:coup_thm}
		Under Assumption \ref{ass:blk_diag_ss} the stationary state of $\Phi$ is given by 
		\begin{align}\label{eq:block_reduced_stat_state}
			\tilde{\rho}=\sum_{i=1}^q\xi_i\Jc_i(
			\tilde{\sigma}_i)=\bigoplus_{i=1}^q\xi_i
			\tilde{\sigma}_i.
		\end{align}
		where $\tilde{\sigma}_{i}$ is the stationary state of the stochastic complement $\Sc_{ii},$ $\xi_i\in(0,1),$ and $\sum_i\xi_i=1,$ such that $\xi=\begin{pmatrix}
			\xi_{1}, & \dots &, \xi_{q}
		\end{pmatrix}^\top$ is the unique stationary distribution for the irreducible Markov chain with transition matrix $C$ with entries
		\begin{align}\label{eq:coup_matrix}
			c_{ij}= \tr(\Phi_{ij}(\tilde{\sigma}_{j}))
		\end{align}
		The matrix C is referred as the \textit{hybrid coupling matrix} and $\xi_i$ are called \textit{coupling factors}.
	\end{theorem}
	The proof is provided in Appendix \ref{app:proofs_hc_hsc}.
    
	Under Assumption \ref{ass:blk_diag_ss} stochastic complementation allows us to compute the stationary state of  a quantum channel,  addressing separately each block as outlined in Algorithm \ref{alg:ss_comp}. This can be convenient numerically for large systems, as it can be easily parallelized (or even run sequentially).
	
	\begin{algorithm}[htbp]
		\caption{Stationary state via hybrid stochastic complementation}
		\label{alg:ss_comp}
		\begin{algorithmic}[1]
			\Require A quantum channel $\Phi$ on 
			$\mathcal H=\bigoplus_{i=1}^{q}\mathcal H_i$
			satisfying Assumption~1.
			\Ensure The unique stationary state $\tilde{\rho}$ of $\Phi$.
			
			\For{$i=1,\ldots,q$}
			\State Compute the HSC $\Sc_{ii}$ according to equation \eqref{eq:stoch_comp};
            
            \State Compute its stationary state
			$\tilde{\sigma}_i\in\mathcal D(\mathcal H_i)$
			algebraically by solving
			$
			\mathcal S_{ii}(\tilde{\sigma}_i)
			=\tilde{\sigma}_i.
			$ or by employing power iteration methods: 
            {when $\Sc_{ii}$ is primitive} $$\tilde \sigma_i \approx \Sc_{ii}^t(\sigma_0), \qquad t\gg 0$$ for an arbitrary $\sigma_0\in \mathfrak{D}(\mathcal{H}_i)$;
			\EndFor
			
			\State Given  $\tilde \sigma_i, \ i\in\{1,\dots q\}$, construct the coupling matrix $C$ according to equation \eqref{eq:coup_matrix};
			
			\State Compute its stationary distribution $\xi= (\xi_{i}, \dots, \xi_{q})^\top$ of $C$  algebraically, or {(when C is primitive)} by power iterations:
			$$\xi=C\xi, \qquad \textrm{or}\qquad  \xi\approx C^t \xi_0, \qquad t\gg 0$$
			for an initial guess probability distribution vector $\xi_0\in\mathbb{R}^q$;
			
			\State \Return
			$\displaystyle
			\tilde{\rho}
			=\sum_{i=1}^{q}\xi_i\mathcal J_i(\tilde{\sigma}_i)
			=\bigoplus_{i=1}^{q}\xi_i\tilde{\sigma}_i$.
		\end{algorithmic}
	\end{algorithm}
	
	\section{Nearly completely reducible systems and metastability}\label{sec:ncr}
	
	This section extends the Simon–Ando theory of nearly completely reducible systems \cite{HS-AA:61, PC-RA:77} to quantum systems and, building on the approach of \cite{mayer-sc}, illustrates how the stochastic complement can be used to investigate their dynamics.
	
	We next introduce two definitions for reducibility of a quantum channel with respect to a given Hilbert-space decomposition, based on whether or not the block-coherences are annihilated by the application of the quantum channel. Note that this difference does not exists in the classical case because block-coherences are not present in classical systems.
	
	\subsection{Definitions and properties}
	The following definition makes use of the concept of invariant subspace for a quantum channel introduced in Section \ref{sec:prem_not}.
	\begin{definition}[\bf {$\xi$}-Nearly completely/strongly reducible channels]
		A quantum channel {$\Phi$} is said to be {\em completely reducible} for a Hilbert-space partition $\Hcp$ if every $\Hc_i$ is invariant for {$\Phi$}.   
		If additionally $\Phi(X)=0$ for all $ X\in {\mathfrak{B}}(\Hc)\ominus \big(\bigoplus_i {\mathfrak{B}}(\Hc_i)\big)$ then {$\Phi$} is {\em strongly reducible.}

		A quantum channel $\Phi$ is said to be {\em { $\xi$}-nearly completely reducible} if $\Phi$ is irreducible and can be written as 
		\begin{equation}\label{eq:ncr_eq}
			\Phi=\Phi_R+ \Ec
		\end{equation}
		where $\Phi_R$ is a completely reducible quantum  channel and   
		$\Ec:{\mathfrak{B}}(\Hc)\to {\mathfrak{B}}(\Hc)$ is a linear map such that $\norm{\Ec}_1\leq \xi$. Similarly if such $\Phi_R$ is strongly reducible then $\Phi$ is said to be {\em { $\xi$-}nearly strongly reducible.} 
	\end{definition}
	{In the following we will simply call these channels nearly completely or strongly reducible, dropping the explicit dependence on $\xi$.}
	Since for completely reducible systems the dynamics of the block population is invariant and decoupled from the block coherences, 
	for {\em nearly} completely reducible systems at every transition only a {\em small} portion of the  block-population in ${\mathfrak{B}}(\Hc_i)$ leaks to other subspaces ${\mathfrak{B}}(\Hc_j)$ with $ j\neq i$, and all interactions between block-population and block-coherences must also remain weak.
	We note that strong reducibility is a stricter condition than complete reducibility, and every nearly strongly reducible channel is also nearly completely reducible. In nearly strongly reducible systems, only a ``small portion'' of the block coherences may survive after each transition. This is illustrated in the following example.
	
	\begin{example}[\bf Nearly completely vs. nearly strongly reducible systems]
		{\em We consider a quantum system described on a Hilbert space $\Hc\simeq \mathbb{C}^2$ and the channels, $$\Phi_R^{(c)}(X)=X, \qquad \Phi_R^{(s)}(X)=\sum_{i=1}^2
		\ketbra{i}{i}X\ketbra{i}{i}  .$$ 
		Note that $\Phi_R^{(c)}(\ketbra{i}{i})=\Phi_R^{(s)}(\ketbra{i}{i})=\ketbra{i}{i}$\,, therefore the Hilbert space admits the decomposition $$\mathcal{H}=\bigoplus_{i=1}^2\Hc_i, \qquad \qquad \Hc_i= \mathbb{C}\ket{i},$$
		where every $\Hc_i$ is invariant for the two considered channels.
		
		The identity channel $\Phi_R^{(c)}$ is completely reducible but not strongly reducible, as   $\Phi_R^{(c)}(\ketbra{i}{j})=\ketbra{i}{j}\neq 0 \ \forall i\neq j\in\{1, 2\}$
        On the other hand,  $\Phi_R^{(s)}$ is both completely reducible and strongly reducible, in fact one has $\Phi_R^{(s)}(X)=0$ for all $X\in {\mathfrak{B}}(\Hc)\ominus(\bigoplus_i{\mathfrak{B}}(\Hc_i))$.
		
		Next, consider the channels
		$$\Phi^{(c)}(X)=(1-\gamma)\Phi_R^{(c)}(X)+\frac{\gamma}{2} \tr(X)I_{2} , \qquad \Phi^{(s)}(X)=(1-\gamma)\Phi_R^{(s)}(X)+\frac{\gamma}{2} \tr(X)I_{2},$$
		where $0<\gamma<1$ is a small real positive number. 
		Both channels are irreducible. Since $\Phi^{(c)}$ is a perturbation of a completely reducible channel it is nearly completely reducible, similarly $\Phi^{(s)}$ is a perturbation of a strongly reducible channel therefore it is both nearly completely reducible and nearly strongly reducible. }
	\end{example}

	The following proposition gives an explicit characterization of channels that are strongly or completely reducible in terms of the decomposition introduced in Definition \ref{def:original_channel_decomposition}.
	
	{\begin{proposition}[\bf Operator-sum representations and decomposition of reducible channels]\label{prop:cs_dec}
			Let $\Phi$ be a quantum channel on $\Hcp$. The following are equivalent:
			
			\begin{enumerate}[label=\textup{(\roman*)}]
				\item {$\Phi$} is a completely reducible quantum channel with respect to the partition;
				\item Every operator-sum representation (OSR) $\{K_\alpha\}$ of {$\Phi$} consists of block-diagonal Kraus operators, namely
				\[
				K_\alpha=\sum_{i=1}^q \Pi_i K_\alpha \Pi_i,
				\]
				where $\Pi_i$ denotes the orthogonal projector onto $\mathcal H_i$;
				
				\item {$\Phi$} decomposes as
				\[
				{\Phi}=\Phi_{\rm cr}+\Phi_{\rm oo},
				\]
				where
				\[
				\Phi_{\rm cr}=\sum_{i=1}^q \mathcal J_i\,\Phi_{ii}\,\mathcal J_i^\dagger,
				\qquad
				\Phi_{ii}:{\mathfrak{B}}(\mathcal H_i)\rightarrow\Bf(\mathcal H_i),\quad
				\Phi_{ii}:=\mathcal J_i^\dagger \Phi\,\mathcal J_i,
				\]
				and
				\[
				\Phi_{\rm oo}
				=\sum_{i\neq j}\mathcal P_{ij}\,\Phi\,\mathcal P_{ij},
				\qquad
				\mathcal P_{ij}(X)=\Pi_i X\Pi_j.
				\]
			\end{enumerate}
			Additionally, if {$\Phi$} is completely reducible, the following are equivalent 
			\begin{enumerate}[label=\textup{(\roman*)},resume]
				\item {$\Phi$} is strongly reducible with respect to the partition of $\Hc$;
				\item {$\Phi$} admits an OSR $\{K_\alpha\}$ in which every Kraus operator has support on a single diagonal block, i.e., for every $K_\alpha$ there exists an index $s\in\{1,\dots,q\}$ such that
				\[K_\alpha=\Pi_s K_\alpha \Pi_s ;\]
				
				\item In the decomposition $(iii)$ above
				$\Phi_{\rm oo}=0$.
			\end{enumerate}
	\end{proposition}}
	
	The proof can be found in Appendix \ref{app:proofs_metastability}.

	\smallskip 
	Notice that $\Phi_{\rm cr}$ leaves $\bigoplus_i\!{\mathfrak{B}}(\Hc_i)$ invariant and represents the dynamics on the block-populations, while $\Phi_{\rm oo}$ leaves ${\mathfrak{B}}(\Hc)\ominus\big(\bigoplus_i{\mathfrak{B}}(\Hc_i)\big)$ invariant, capturing the dynamics within the off-diagonal block coherences. 
	The previous proposition highlights how for completely reducible channels each block-populations evolve independently from the others and block-coherences evolve independently from block-population. If the channel is strongly reducible, all block-coherences are mapped to 0.  
	
	{The following proposition characterizes the relation between complete reducibility and strong reducibility of the original and associated hybrid channel.}
    
		\begin{proposition}[Complete reducibility and associated hybrid channel]
			\label{prop:hybrid-complete-reducibility}
			Let $\Phi$ be a quantum channel on a partitioned space
			$\mathcal H=\bigoplus_{i=1}^{q}\mathcal H_i$, and let $\Phi_{\oplus}$ be its
			associated hybrid channel as in Definition~\ref{def:original_channel_decomposition}.
			Then $\Phi$ is completely reducible 
			if and only if $\Phi_{\oplus}$ is completely reducible with respect to the
			same partition. In this case, $\Phi_{\oplus}$ is also strongly reducible.
		\end{proposition}
        The proof of Proposition \ref{prop:hybrid-complete-reducibility} can be found in Appendix \ref{app:proofs_metastability}.

	{
		Note that $\Phi_{\oplus}$ strongly reducible does not imply $\Phi$ is strongly reducible as well. Moreover, we note that the hybrid channel $\Phi_{\oplus}$ associated with a {\em nearly} completely or nearly strongly reducible channel $\Phi=\Phi_R+ \Ec$ does not coincide with $\Phi_R$; indeed, the latter does not allow for transition of block-population between different subspaces whereas $\Phi_{\oplus}$ does.}
	
	\subsection{Deviations from complete and strong reducibility}
	
	In the following, we will often rely on the map aggregating HSCs which we call with a slight abuse of terminology {\em aggregate HSC}. 
	\begin{align}\label{eq:sc_map}
		\Sc:{\mathfrak{B}}(\Hc)&\quad \to \quad {\mathfrak{B}}(\Hc)\\
		X&\quad \to \quad \sum_{i=1}^q\Jc_i \Sc_{ii}\Jc_i^\dag(X)=\bigoplus_{i=1}^q \Sc_{ii}(X_{ii}).\nonumber
	\end{align}
	where $X_{ii}=\Jc_i^\dag(X)\in{\mathfrak{B}}(\Hc_i)$ is the $i$-th diagonal block of $X\in{\mathfrak{B}}(\Hc)$ according to the Hilbert space partition $\Hc=\bigoplus_i\Hc_i$. It is simple to prove that $\Sc$ is a quantum channel. 
	
	Moreover, to streamline the presentation and avoid repeating similar results we introduce the maps 
	$$\WS=\Ic \qquad {\rm and} \qquad  \WC=\sum_i\Pc_i.
    $$
	When applied to the quantum state the first map trivially returns the state itself, whereas the second map returns block-population.
	\begin{theorem}[\bf Deviation from reducibility]\label{thm:qs_dev_comp_red}
		Let $\Phi$ be an irreducible quantum channel on a Hilbert space $\Hcp$, let $\Phi_{\oplus}$ be the associated hybrid channel and $\Phi_{\Delta}$ the reminder map as in Definition \ref{def:original_channel_decomposition}. For each $i\in \{1,\dots, q\}$, let $\Sc_{ii}$ denote the HSC of $\Phi_{ii}$ given in~\eqref{eq:stoch_comp}. Then 
		$$\norm{S_{ii}-\Phi_{ii}}_1=\norm{\Phi_{*i}}_1 . 
		$$
		Let $\Sc$ be the aggregate HSC defined in \eqref{eq:sc_map}, then 
		$$\norm{\Phi_{\oplus}-\Sc}_1= 2 \underset{i\in\{1,\dots q\}\\ }{{\rm max}} \norm{\Phi_{*i}}_1=\delta_h$$
		and
		$$\norm{\Wc_x \Phi-\Sc}_1\leq \norm{\Phi_{\oplus}-\Sc}_1 +\norm{\Wc_x\Phi_{\Delta}}_1 \leq \delta_h +\norm{\Wc_x\Phi_{\Delta}}_1, \qquad x\in \{c,s\}.$$    
	\end{theorem}
	The proof can be found in Appendix \ref{app:proofs_metastability}.
	\begin{definition}[\bf Deviation from complete/strong reducibility]\label{def:qs_deviation}
		With the notation introduced in Theorem \ref{thm:qs_dev_comp_red} we define 
		
		\begin{enumerate}[label=\roman*)]
			\item  
			deviation of $\Phi_{\oplus}$ from complete reducibility:  $\delta_h= 2 \underset{i\in\{1,\dots q\}\\ }{{\rm max}} \norm{\Phi_{*i}}_1$;
			\item 
			deviation of $\Phi$ from strong reducibility: $\delta_s= \delta_h+\norm{\WS \Phi_{\Delta}}_1=\delta_h+\norm{ \Phi_{\Delta}}_1 $;
			\item deviation of $\Phi$ from complete reducibility: 
			$\delta_c= \delta_h+\norm{\WC \Phi_{\Delta}}_1$.
		\end{enumerate}
	\end{definition}
	Exploiting Proposition \ref{prop:cs_dec} it is possible to note that the deviations form complete reducibility are equal to zero if and only if the corresponding channel is completely reducible, and similarly for strong reducibility. The following proposition describes the relationship between deviation for complete reducibility of the associated hybrid channel and deviation from complete reducibility of the original channel.
	
	\begin{proposition}[Comparison of deviations from reducibility]\label{prop:rel_deviations}
		Let $\Phi$ be an irreducible quantum channel on
		$\mathcal H=\bigoplus_{i=1}^q\mathcal H_i$, and let $r$ be its
		Kraus rank. Then
		\[
		\delta_c\leq\delta_s
		\qquad  {\rm
			and}\qquad 
		\delta_c\leq
		\delta_h+r q^2(q-1)\sqrt{\frac{\delta_h}{2}}.
		\]
	\end{proposition}
	
	The proof of Proposition \ref{prop:rel_deviations} can be found in Appendix \ref{app:proofs_metastability}.
	
	The inequality $\delta_c\leq\delta_s$ is consistent with the fact
	that every nearly strongly reducible channel is also nearly
	completely reducible. Moreover, for a fixed Hilbert-space
	decomposition, Proposition \ref{prop:rel_deviations} implies that $\delta_h\to0$ entails
	$\delta_c\to0$. Thus, near complete reducibility of the associated
	hybrid channel implies near complete reducibility of the original
	channel. The same conclusion does not hold for strong
	reducibility.

	\begin{remark}[\bf Spectrum Splitting Effect]\label{rem:ss_eff}
		Let $\Phi$ be an irreducible quantum channel on
		$\mathcal H=\bigoplus_{i=1}^{q}\mathcal H_i$, and let $\Phi_{\oplus}$ be its
		associated hybrid channel. By Theorem \ref{thm:hsc_prop}, each HSC $\mathcal S_{ii}$
		is an irreducible quantum channel and therefore has a simple
		eigenvalue at $1$. Consequently, the aggregate HSC $\mathcal S$  has eigenvalues in 1 with algebraic and geometric multiplicities $q$.
		
		Assume now that for small $\delta_h$ $\Phi_{\oplus}$ has exactly $q$ eigenvalues close to $1$,
		while the remaining eigenvalues have modulus bounded
		away from $1$. Then, by Theorem \ref{thm:qs_dev_comp_red}, $\mathcal S$ is a
		small perturbation of $\Phi_{\oplus}$. Hence, by continuity of the
		eigenvalues, for sufficiently small $\delta_h$, the non-unit
		eigenvalues of $\mathcal S$, and therefore of each
		$\mathcal S_{ii}$, also have modulus bounded away from $1$.
		Similarly, for small $\delta_s$, $\mathcal S$ is a small
		perturbation of $\Phi$, and the same conclusion holds under
		the corresponding spectral assumption on $\Phi$.
	\end{remark}
	
	In the case outlined in the previous remark, assume $\Phi$ has a block-diagonal stationary state and that
	$\Phi$ and each $\mathcal S_{ii}$ are primitive. The asymptotic
	convergence rate of the power sequence
	$\{\mathcal S_{ii}^{t}\}_t$ is governed by the largest modulus
	of its non-unit eigenvalues, which, under the assumptions of
	the previous remark, is bounded away from $1$. As noted after
	Theorem \ref{thm:coup_thm}, power iterations of the stochastic complements can
	be used to compute the local stationary states needed to
	reconstruct the stationary state of $\Phi$. These
	iterations can therefore converge rapidly, although the
	overall computational cost also depends on constructing the
	complements and solving the coupling problem.
	\subsubsection{Characterization of dynamical regimes} 
	\label{sec:dyn_regimes}
	Under some conditions the dynamics of nearly completely or nearly strongly reducible systems can be divided in three different regimes: short-run, middle-run and long-run.
	\subsubsection*{Short-run dynamics}
	
	Given an initial state
	$\rho_0\in\mathcal D(\mathcal H)$, we describe the evolution of the system through repeated application of an irreducible quantum channel $\Phi$ and let
	\begin{equation}\label{eq:sr_dyn}
		\rho_t=\Phi^t(\rho_0)
	\end{equation}
	denote the state after $t$ iterations. We next derive conditions that allow us to characterize when the system shows metastability that is the full state, or its block-population, remains close over a time interval to certain states before converging to theid asymptotic values. The metastable state reached depends on the initial state considered for the evolution and on the stationary states of the HSCs.
	
	\begin{definition}[\bf HSC metastable state]    
		Given an initial state $\rho_0$ for the dynamics \eqref{eq:sr_dyn}, and the stationary states $\sigma_i$ of HSCs $\Sc_{ii}$, we define the HSC metastable state to be
		\begin{equation}
			\sigma_\infty=\sum_{i=1}^q p_i \rho_{\sigma_i}=\bigoplus_{i=1}^q p_i\sigma_i,
		\end{equation}
		where $\rho_{\sigma_i}=\Jc_i(\sigma_i)$ and $p_i={\rm tr}(\Jc_i^\dag (\rho_0))={\rm tr}(\mu_{0,i})$. 
	\end{definition}
	Note that $\rho_{\sigma_i}\in\Dc(\Hc)$ and $p_i\geq 0, \sum_i{p_i}=1$ therefore the HSC metastable state $\sigma_\infty$ is a convex combination of $q$ states in $\Dc(\Hc)$ defined by the stochastic complements.
	Moreover, if each HSC $\Sc_{ii}$ is primitive, let $\Sc$ be the aggregate HSC, then the limit
	\begin{equation}\label{eq:s_infty}
		\Sc_\infty := \lim_{t\to\infty}\Sc^t
	\end{equation}
	exists and satisfies
	\[
	\Sc_\infty(X)
	= \sum_{i=1}^q
	{\rm tr}\!\left[\Jc_i^\dagger(X)\right]\Jc_i(\sigma_i).
	\]
	In particular, $\sigma_\infty=\Sc_\infty(\rho_0)$.
	
	The following theorem characterizes the short-run dynamics of the system and the emergence of the described metastable state. 
	\begin{proposition}[Short-run evolution and stabilization]
		\label{prop:qc_short_run}
		Let $\Phi$ be an irreducible quantum channel on a Hilbert space $\Hcp$ with ${\rm dim}(\Hc)=D$. Let the HSC $\Sc_{ii}$ be primitive for all $i$. 
		Let $\Sc$ be the aggregate HSC defined in \eqref{eq:sc_map}, and assume it is diagonalizable i.e., there exist an invertible linear map \(\Zc\) and a linear map $\Lambda$ such that given an Hilbert–Schmidt orthonormal operator basis $\{F_i\}_{i=1}^{D^2}$
		$$\Sc=\Zc\Lambda\Zc^{-1}, \qquad \Lambda(F_i)=\lambda_iF_i.$$
		Here the $\{\lambda_i\}_{i=1}^{D^2}$ are the eigenvalues of $\Sc$, which we assume, without loss of generality, to be ordered so that
		$|\lambda_1|\geq |\lambda_2|\geq\cdots\geq|\lambda_{D^2}|.$ Consider  $\Sc^\infty,\sigma_\infty$ as in \eqref{eq:s_infty}
		then
		\begin{equation}
			\label{eq:bound}
			\|\Wc_x(\rho_t)-\sigma_\infty\|_1
			\le
			\|\Wc_x\Phi^t-\Sc^\infty\|_1
			\le
			t \delta_{x}
			+\sqrt{D}\kappa |\lambda_{q+1}|^t, \qquad x\in \{c,s\}
		\end{equation}
		where
		$\kappa=\|\Zc\|_2\,\|\Zc^{-1}\|_2.$
	\end{proposition}
	
	The proof of Proposition \ref{prop:qc_short_run} can be found in Appendix \ref{app:proofs_metastability}. \begin{remark}[\bf Relaxing assumptions]
		
		Similarly to the classical case\cite{mayer-sc}, the primitivity assumption in Proposition \ref{prop:qc_short_run} is instrumental to a simpler, direct result; it can be relaxed by considering time averages for both hybrid channel and the stochastic complement in the statement and the proof. 
		If $\mathcal{S}$ is not diagonalizable, one can use stadard bounds for Jordan blocks to obtain a modified version of the bound \eqref{eq:bound} which contains an additional polynomial factor in $t$ in the second term.
	\end{remark}
	
	The above proposition motivates the following definitions.
	
	\begin{definition}[Short-run stabilization interval]\label{def:stabilization_intervals}
		For each $\varepsilon>0$, {$x\in \{c,s\}$ the $x$-short-run stabilization interval} $\mathcal{I}_{x}(\varepsilon)$ is defined as {
			$$\Ic_{x}(\varepsilon) \ =\ \bigl\{t \, \big| \, \|\Wc_x\Phi^t-S^\infty \|_1<\varepsilon\bigr\}.$$}
	\end{definition}
	Proposition \ref{prop:qc_short_run} implies that the {$x$-short-run stabilization interval contains the estimated $x$-short-run stabilization} interval
	{$$E_{x}(\varepsilon)=\{t \ \big| \ t \delta_{x} +\sqrt{D}\kappa |\lambda_{q+1}|^t<\varepsilon \}.$$}
    In view of the above theorem and definitions, we can conclude that if \(\mathcal I_x(\varepsilon)\) is a non-empty, proper subinterval of the considered time domain, with positive length, for \(x\in\{s,c\}\) and sufficiently small \(\varepsilon\), the system exhibits a metastable time regime.

Metastability of the system can be recognized analytically through the estimated interval $E_{x}(\varepsilon)$.
	In particular:
	\begin{enumerate}[label=(\roman*)]
		\item {\em If $\Ic_c(\epsilon)$ is non-empty for small $\epsilon$:} in the short-run the block population $\mu_t=\WC(\rho_t)$ approach the metastable state $\sigma_\infty$ long before converging to the block-components of the stationary state $\tilde \mu=\WC(\tilde \rho)$ and remains close to it for the time-interval $\Ic_c(\epsilon)$.
		\item {\em If $\Ic_s(\epsilon)$ is non-empty for small $\epsilon$:} in the short-run the block-coherences of the system decay, the whole state of the system $\rho_t$ approach the metastable state $\sigma_\infty$ and remain close to it for the time interval $\Ic_s(\epsilon)$.
	\end{enumerate}
	
	\begin{remark}[\bf Reducibility and metastability]
		We note that near complete/strong reducibility is necessary to establish metastability through the estimated short-run stabilization interval \(E_x(\varepsilon)\) $x\in\{s,c\}$. Indeed, if \(\delta_x\) is large compared with the prescribed tolerance \(\varepsilon\), the term \(t\delta_x\) prevents the bound in \eqref{eq:bound} from falling below \(\varepsilon\), and \(E_x(\varepsilon)\) is empty.  
	\end{remark}
	
	\subsubsection*{Middle-run dynamics}
	The middle-run dynamics of the system, i.e. the behavior of $\rho_t$ for $t> {\rm sup}(\Ic_x(\epsilon)), x\in \{c,s\}$ is described by the following theorem.
	{
\begin{proposition}[\bf Middle-run evolution]\label{prop:qc_middle_run}
		
		Let $\Phi$ be an irreducible quantum channel on $\Hcp$ and let $\Phi_{\oplus}$ be the associated hybrid channel. Let the HSCs $\Sc_{ii}$ be primitive and have  stationary state $\tilde \sigma_i$ for all $i$. 
        Given $\epsilon>0$ $x\in \{c,s\}$ such that $\Ic_{x}(\epsilon)\neq \emptyset$ and $t>{\rm sup}\ \Ic_{x}(\epsilon)$, there exist a set of scalars $\{\beta_{t,i}\}_{i=1}^q$ such that for $\upsilon_t=\sum_i\Jc_i(\beta_{t,i} \tilde \sigma_{i})$
		$$\norm{\Wc_x(\rho_t)-\upsilon_t}_1<\epsilon.$$
        In particular for any fixed $h\in \Ic_x{(\epsilon)},$ the coeficients satisfying the above inequality can be chosen as $\beta_{t,i}^{(h)}=\tr[\Jc_i^\dag(\Phi^{t-h}(\rho_0))]$ and they evolve approximately accordingly to the equation
        $$\norm{\beta_{t+1}^{(h)}-C\beta_{t}^{(h)}}_1\leq \delta_c,$$
        for all $t\geq h$ where C is the coupling matrix in Theorem \ref{thm:coup_thm}.
	\end{proposition}
    }
	
	The proof is provided in Appendix \ref{app:proofs_metastability}. The interpretation of the above proposition is as follows: suppose there exist $\epsilon$ small such that the c-short-run stabilization interval $\Ic_c(\epsilon)$ is non-empty. In the middle-run ($t>{\rm sup}\Ic_c(\epsilon)$) the components $\mu_{t,i}=\Jc_i^\dag(\rho_t)$ of the block populations $\mu_t=\Wc(\rho_t)$ start equilibrating. Each block $\mu_{t,i}$ remains approximately proportional to the stationary state of the corresponding stochastic complement (as it was during the short-run stabilization), the only thing that change significantly is the trace of each block {that approximately evolves accordingly to a classical Markov chain.}
	
	Suppose there exist $\epsilon$ small such that the s-short-run stabilization interval $\Ic_s(\epsilon)$ is non-empty. In the middle-run ($t>{\rm sup}\Ic_s(\epsilon)$) the whole system state $\rho_t$  remains approximately block diagonal, with only the block traces undergoing significant change and approximately evolving according to a classical Markov chain.

   The coefficients \(\beta_{t,i}^{(h)}\) in Proposition~\ref{prop:qc_middle_run}
may be replaced by the traces \(p_{t,i}=\operatorname{tr}[\mu_{t,i}]\) of
the block populations at time \(t\).  The approximation error with the
coefficients \(p_{t,i}\) is however at most \(2\epsilon\) for $t > {\rm sup} (\Ic_x(\epsilon))$. If \(\rho_t\) converges to a block-diagonal stationary state, the approximation error then tends to zero
and is less than \(\epsilon\) for all sufficiently large \(t\). The block traces also satisfy the same approximate evolution law under
the coupling matrix \(C\), namely \(\|p_{t+1}-Cp_t\|_1\le\delta_c\)
for every \(t\ge 0\).

	\subsubsection*{Long-run dynamics}
	
	The long-run dynamics of the system, i.e. the behavior of $\rho_t$ for $t\to\infty$ is described by the following proposition.
	
	\begin{proposition}[\bf Long-run evolution and stabilization]\label{prop:qc_long_run} Consider the setup of Proposition \ref{prop:qc_middle_run}, and in addition assume that $\Phi$ is primitive and its stationary state $\tilde \rho$ is block-diagonal with respect to the Hilbert space partition. Then
		$$\lim_{t \rightarrow\infty}\rho_t=\tilde \rho=\sum_i \xi_i \Jc_i (\tilde \sigma_i), \quad {\rm and} \quad \lim_{t\to\infty}\beta^{(h)}_{t,i}=\xi_i,$$
        for every \(i=1,\ldots,q\) and every fixed \(h\in \Ic_x(\epsilon)\), where $\{\xi_i\}_{i=1}^q$ are the coupling factors introduced in Theorem \ref{thm:coup_thm}.
	\end{proposition}
	The proof of Proposition \ref{prop:qc_long_run} can be found in Appendix \ref{app:proofs_metastability}.
	
	Finally, in the long run the system converges to its stationary state. If $\delta_{x}, x\in\{c,s\}$ is sufficiently small and $\Ic_{x}(\epsilon)\neq \emptyset$ for small $\epsilon$, each block population $\mu_{i,t}=\Jc_i^\dag(\rho_t)$ for $t>{\rm inf} \Ic_{x}(\epsilon)$ is proportional to the stationary state of the corresponding stochastic complement, therefore it has already converged except for a scalar weighting the block. In the long-run these scalars converge to the coupling factors introduced in Theorem \ref{thm:coup_thm}.
	
	\begin{remark}
		If the stationary state of the system is not block-diagonal, but the s-short-run stabilization interval $\mathcal{I}_s(\epsilon)$ is non-empty , Proposition \ref{prop:qc_middle_run} implies that the stationary state of the system is approximately block-diagonal.    
	\end{remark}

	\section{Examples: Approximate symmetries for quantum systems}\label{sec:approxsymm}
	We illustrate how metastability in open quantum systems can be induced by the presence of {\em approximate} symmetries. It is well established that strong symmetries induce an Hilbert space decomposition into invariant subspaces, preventing population and coherence transfer between them \cite{grigoletto2025exact}. When these symmetries are perturbatively broken, however, transitions between sectors~become~slow,~leading~to~long-lived~metastable~behavior. 
    
    The theory developed in the previous sections provides a quantitative framework for analyzing this situation. In particular, stochastic complements characterize the local equilibrium within each symmetry subspace and the slow asymptotic relaxation. After formally introducing approximate symmetries,  we illustrate these ideas with two two-qubit examples by introducing two quantum channels: The first is nearly completely reducible but not nearly strongly reducible, whereas the second is nearly strongly reducible. This allows us to highlight the different dynamical regimes predicted by the theory.

	\subsection{Approximate strong symmetries and nearly complete reducibility}
	We recall the definition of strong symmetry for discrete time Markovian systems.
	\begin{definition}[\bf Strong Symmetry]
		Let $\Phi_S$ be a quantum channel on a Hilbert space $\Hc$. A strong symmetry is a unitary operator U that commutes with every Kraus operator $\{K_i\}$ in an arbitrary operator-sum representation of $\Phi$, i.e.,
		\begin{align}\label{eq:sym_cond}
			[K_i, U] = 0 \quad \forall i,
		\end{align}
		which also implies $\Uc \Phi=\Phi \Uc$ where $\Uc(\cdot)=U \cdot U^\dag$.
	\end{definition}
	
	To our aim, the relation between symmetries and invariant subspaces plays a key role. For any symmetry operator $U$ the corresponding eigen-decomposition provide a decomposition of the Hilbert space $\Hc$ as $\Hc=\bigoplus_{i=1}^q \Hc_i$ where $\Hc_i$ are the eigenspaces of $U$.
    It follows that each subspace 
    $\Hc_i$ is invariant for $\Phi$ and  the channel $\Phi_S$ is completely reducible, but in general it needs not be strongly reducible.
	
	We then consider quantum channels $\Phi$ with {\em approximate symmetries}, i.e. channels that do not admit (nontrivial) exact strong symmetries but admit a decomposition
	$$\Phi=\Phi_S+ \Ec,$$
	where $\Phi_\Sc$ is a quantum channel with associated non-trivial strong symmetry while $\Ec$ is a symmetry-breaking linear map on ${\mathfrak{B}}(\Hc)$ such that $ \norm{\Ec}_1\ll 1$. 
	As discussed above $\Phi_S$ is completely reducible, therefore $\Phi$ is a nearly completely reducible quantum channel which can be studied with the theoretical framework introduced in the previous sections 
	\footnote{
		The discussion provided in this section extends naturally to the case in which the system admits a group of symmetries, we restrict here to a single symmetry for simplicity.}.
	
	\subsection{Example 1: Nearly completely reducible two-qubit system}\label{sec:nr_channel}
	As simple example we consider a bipartite quantum system composed of two qubits (which we label with $a$ and $b$). The associated Hilbert space is $\Hc=\Hc_{a}\otimes \Hc_{b}$, where $\Hc_a,\Hc_b \simeq \Cb^2$, $\Hc\simeq \Cb^{4}.$ We denote by $\sigma_\alpha^j$ the Pauli operator $\sigma_\alpha, \alpha\in \{0,x,y,z\}$ acting on qubit $j\in \{a,b\}$, namely $\sigma_\alpha^a=\sigma_\alpha\otimes \one, \sigma_\alpha^b=\one \otimes\sigma_\alpha.$
	
	We assume the system evolves according to the discrete-time time-homogeneous and markovian quantum dynamics described by the channel
	\begin{equation}\label{eq:ncr_channel}
		\Phi=\Phi_S+ \Ec,
	\end{equation}
	where $\Phi_S$ is a symmetric channel and \(\Ec\) is a symmetry-breaking perturbation. The symmetric component $ \Phi_S$ is the channel described by the set of Kraus operators 
	$$\Kc_S=\{K_{S,i}\}_i=\{\sqrt{\alpha_1} \one ,\ \sqrt{\alpha_2} \sigma_x^a\sigma_x^b, \ \sqrt{\alpha_3} \sigma_z^a,\ \sqrt{\alpha_4} \sigma_z^b\}, $$
	with $\alpha_i \in \Rb_{\geq 0} \ \forall i, \sum_i \alpha_i=1$ (in numeric examples, we will consider $\alpha_1=0.49, \alpha_2=0.25$, $\alpha_3=0.25$ and $\alpha_4=0.01$ ). It is straightforward to verify that
	$U=\sigma_z^a\sigma_z^b$
	is a strong symmetry of $\Phi_S$. This symmetry induces a decomposition of $\Hc$ into its eigenspaces,
	$\Hc=\Hc_1\oplus\Hc_2$, where $\Hc_1={\rm span} \Big\{\ket{00},\ket{11} \Big\}$ and $ \Hc_2={\rm span} \Big\{\ket{10},\ket{01}\Big\}$.
	Throughout both examples, matrix representations of operators in ${\mathfrak{B}}(\Hc)$
	are taken with respect to the ordered basis
	$\{\ket{00},\ket{11},\ket{10},\ket{01}\}$.
	Matrix representations of operators in ${\mathfrak{B}}(\Hc_1)$ and ${\mathfrak{B}}(\Hc_2)$
	are taken with respect to the ordered bases $\{\ket{00},\ket{11}\}$
	and $\{\ket{10},\ket{01}\}$, respectively.
	The symmetry-breaking perturbation is chosen as 
	$$\Ec(X)=-\xi X+\xi \sigma_x^a X \sigma_x^a,$$
	where $\norm{\Ec}_1=2\xi$ and $0<\xi\leq\alpha_1$ controls the perturbation strength while ensuring that $\Phi$ is a quantum channel. The norm is computed explicitly in Appendix~\ref{app:symmetry_norms}. Since $\sigma_x^a$ does not commute with $U$, the perturbed channel $\Phi$ does not possess any non-trivial strong symmetry, and $\Ec$ acts as a symmetry-breaking perturbation. 
	For the strictly positive coefficients used here and $0<\xi<\alpha_1$,
	it is possible to check that $\Phi$ is primitive and hence irreducible, with unique stationary state
	$\tilde\rho=I_4/4$. 
    
{    The associated hybrid channel $\Phi_\oplus$ for the considered example has components
	\begin{equation}
		\Phi_{11}=\Phi_{22}=\mathcal T_\xi,
		\qquad \Phi_{12}=\Phi_{21}=\xi\mathcal I,
		\label{eq:symmetry-stationary-blocks}
	\end{equation}
	where
	\begin{equation}
		\mathcal T_\xi(X)=(\alpha_1-\xi)X
		+\alpha_2\sigma_xX\sigma_x
		+(\alpha_3+\alpha_4)\sigma_zX\sigma_z.
		\label{eq:symmetry-stationary-local-map}
	\end{equation}
	Here the $\sigma_\beta$ are the usual two-dimensional Pauli matrices.
	The two stochastic complements therefore coincide:
	\begin{equation}
		\mathcal S_{ii}=\mathcal T_\xi+
		\xi^2(\mathcal I_2-\mathcal T_\xi)^{-1},
		\qquad i=1,2.
		\label{eq:symmetry-stationary-complements}
	\end{equation}
	For the chosen parameters of the model each HSC is primitive and unital, with unique stationary
	state $\tilde\sigma_i=I_2/2$. Refined expression of the HSCs and  all the details on the computation are given in Appendix~\ref{app:symmetry-hsc}.}

	\paragraph*{Metastability.}
	Following Definition~\ref{def:qs_deviation}, the deviations from
	complete reducibility of the associated hybrid channel and of the
	original channel are
	\begin{equation}
		\delta_h=\delta_c=2\xi.
		\label{eq:symmetry-example-one-deviations}
	\end{equation}
	The coherence remainder $\Phi_{\Delta}=\Phi-\Phi_{\oplus}$ has induced trace norm
	\begin{equation}
		\|\Phi_{\Delta}\|_1=1-2\min\{\alpha_3,\alpha_4\},
		\label{eq:symmetry-example-one-coherence-norm}
	\end{equation}
	independently of $\xi$ for $0\leq\xi\leq\alpha_1$.
	Therefore, the deviation from strong reducibility is
	\begin{equation}
		\delta_s=\delta_h+\|\Phi_{\Delta}\|_1
		=2\xi+1-2\min\{\alpha_3,\alpha_4\}.
		\label{eq:symmetry-example-one-strong-deviation}
	\end{equation}
	Detailed computation of the deviations is provided in Appendix~\ref{app:symmetry_norms}. For the coefficients
	used here, $\|\Phi_{\Delta}\|_1=0.98$ and $\delta_s=0.98+2\xi$.
	Thus $\delta_s$ remains close to one even as $\xi\to0$, indicating
	that $\Phi$ is nearly completely reducible but not nearly strongly
	reducible.

	\medskip
	
	We next investigate whether the {\em spectrum splitting} outlined in Remark \ref{rem:ss_eff} is verified in the considered example. 
	Let $\lambda_*(\Ec)$ be the eigenvalue of largest absolute value different from 1 of the channel $\Ec$. According to the spectrum splitting effect when $\delta_h$ is small $\Phi_{\oplus}$ may be characterized by $\lambda_*(\Phi_{\oplus})$ close to one. 
	In contrast, the HSCs may exhibit a substantially smaller $\lambda_*$. In particular we consider $\lambda_*(\Sc)$ with $\Sc$ the aggregated HSC defined in \eqref{eq:sc_map}. $\lambda_*(\Sc)$ coincides with the largest $\lambda_*$ among the stochastic complements.  The spectrum splitting effect for $\Phi_{\oplus}$ and $\Sc$ is illustrated in Table \ref{tab:spectral_gaps} where we also outline $\lambda_*(\Phi)$ for comparison. Even for relatively large parameter $\xi=0.5\delta_h$, the stochastic complement map $\mathcal S$ defined in \eqref{eq:stoch_comp} retains a rather small $\lambda_*(\Sc)$ compared to $\lambda_*(\Phi), \lambda_*(\Phi_{\oplus})$.
	
	\medskip

	\begin{table}[]
		\centering
		\begin{tabular}{|c|c|c|c|}
			\hline
			$\xi$ & $10^{-1}$ & $10^{-2}$ & $10^{-3}$ \\
			
			\hline
			$\Phi$ &  0.98  &0.98 & 0.99\\
			$\Phi_{\oplus}$ & 0.8 \ & 0.98 \ & 0.99 \ \\
			$\Sc$ & 0.42 &0.49 & 0.5\\ \hline
		\end{tabular}
		\caption{Spectrum splitting effect: Eigenvalue of largest absolute value different from 1 of the quantum channel $\Phi$, associated hybrid channel $\Phi_{\oplus}$ and the map $\Sc$ aggregating stochastic complements for different values of the perturbation parameter $\xi$.}
		\label{tab:spectral_gaps}
	\end{table}
	
	We conclude by illustrating the dynamical regimes introduced in
	Section \ref{sec:dyn_regimes}. We set $\xi=10^{-3}$ and generate $N=100$
	initial density matrices using the \texttt{RandomDensityMatrix} routine
	from the QETLAB toolbox for MATLAB. For each initial state
	$\rho_{0,i}$, we consider the evolution
	$\rho_{t,i}=\Phi^t(\rho_{0,i})$, the corresponding block-population 
	$\mu_{t,i}=\mathcal{W}_c(\rho_{t,i})$, and the corresponding
	HSC metastable state
	$\sigma_{\infty,i}=S_\infty(\rho_{0,i})$.
	
	To distinguish the possible relaxation toward the metastable state of the block population and the full systems state, decay
	of block coherences, and convergence to stationarity, we introduce
	\begin{equation}
		\begin{aligned}
			\Delta_{\mathrm{pop},i}(t)
			&= \|\mu_{t,i}-\sigma_{\infty,i}\|_1,\quad 
			\Delta_{\mathrm{full},i}(t)
			= \|\rho_{t,i}-\sigma_{\infty,i}\|_1,\\ 
			\Delta_{\mathrm{coh},i}(t)
			&= \|\rho_{t,i}-\mu_{t,i}\|_1,\quad \ \
			\Delta_{\mathrm{stat},i}(t)
			= \|\rho_{t,i}-\tilde{\rho}\|_1,
		\end{aligned}
		\label{eq:symmetry-distances}
	\end{equation}
	where $\tilde{\rho}=\one/4$ is the stationary state of $\Phi$.
	We denote their empirical averages by
	\begin{equation}
		\bar{\Delta}_a(t)
		= \frac{1}{N}\sum_{i=1}^{N}\Delta_{a,i}(t),
		\qquad
		a\in\{\mathrm{pop},\mathrm{full},\mathrm{stat},\mathrm{coh}\},
		\label{eq:symmetry-average-distances}
	\end{equation}
	where we recall $N$ is the number of considered initial state samples. 
	Proposition~\ref{prop:qc_short_run} provides the state-independent estimates for the distance of block-populations from the metastable state
	\begin{equation}
		\begin{aligned}
			\Delta_{\mathrm{pop},i}(t)
			&\leq B_{\mathcal{I}_c}(t)
			:= \|\mathcal{W}_c\Phi^t-S_\infty\|_{1}
			\leq B_{\mathrm{pop}}(t),\\
		\end{aligned}
		\label{eq:symmetry-numerical-bounds}
	\end{equation}
	where
	\begin{equation}
		\begin{aligned}
			B_{\mathrm{pop}}(t)
			&= t\delta_c+\sqrt{D}\,\kappa|\lambda_{q+1}|^t,\qquad 
		\end{aligned}
		\label{eq:symmetry-analytical-bounds}
	\end{equation}
	These inequalities also hold for the empirical averages.
	The induced-norm quantities determine the c-stabilization
	interval of Definition~\ref{def:stabilization_intervals}:
	\begin{equation}
		\mathcal{I}_c(\epsilon)
		= \{t:B_{\mathcal{I}_c}(t)<\epsilon\}.
		\label{eq:symmetry-stabilization-intervals}
	\end{equation}
	
\begin{figure}
		\centering
		\includegraphics[width=\textwidth]{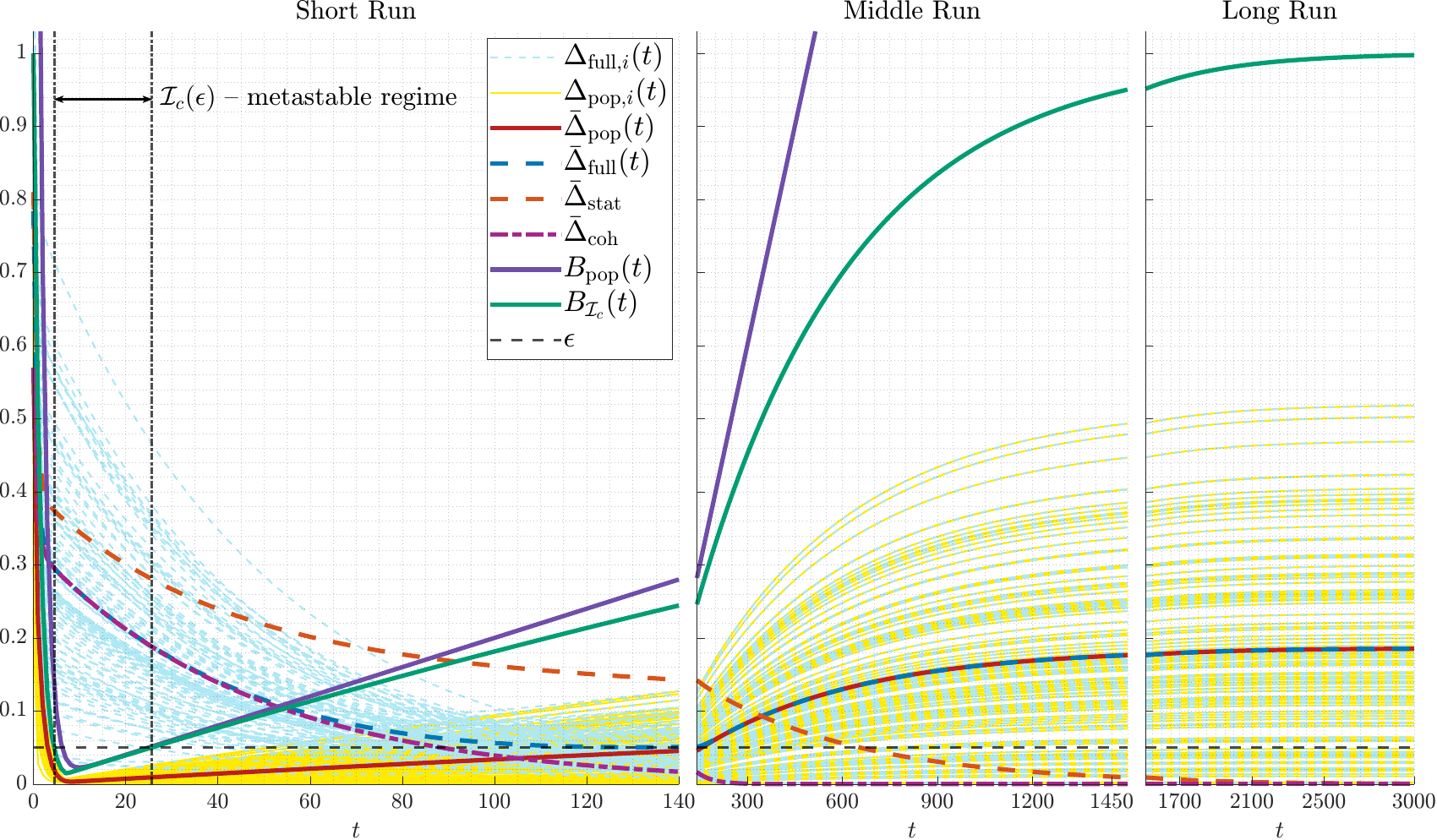}
		
		\caption{
			Dynamics of the nearly completely reducible channel of
			Section \ref{sec:nr_channel}, with $\xi=10^{-3}$, over three time ranges.
			Faint curves show $\Delta_{\mathrm{pop},i}(t)$ and
			$\Delta_{\mathrm{full},i}(t)$ for 100 random initial states;
			thicker curves show their empirical averages, together with
			$\bar{\Delta}_{\mathrm{stat}}(t)$ and
			$\bar{\Delta}_{\mathrm{coh}}(t)$.
			The purple curve is the analytical bound
			$B_{\mathrm{pop}}(t)$, and the green curve is
			$B_{\mathcal{I}_c}(t)
			=\|\mathcal{W}_c\Phi^t-S_\infty\|_{1}$.
			The horizontal dashed line indicates the threshold $\epsilon=0.05$ used for the stabilization interval (the metastable time regime) $\mathcal{I}_c(\epsilon)$;
			the vertical markers delimit the 
			$\mathcal{I}_c(\epsilon)$.
			All panels use linear scales; the time scales (horizontal axis units) of the panels on the right are compressed in order to better appreciate the different dynamical regime. With uniform time scales, the transition between the panes would be smooth.
		}
		
		\label{fig:dyn_reg}
	\end{figure}

	Figure~\ref{fig:dyn_reg}, shows the dynamics
	over three time regimes:
	\begin{enumerate}[label=(\roman*)]
		\item {\em Short-run:} the block-populations rapidly approach their corresponding metastable states. The quantity $B_{\Ic_c}(t)$ falls below the threshold $\epsilon=0.05$, therefore the stabilization interval $\mathcal{I}_c(\epsilon)$ is non-empty indicating a metastable regime in which the proximity to the metastable state of block-population is guaranteed for every initial state. Note that $B_{\Ic_c}(t)$ correctly bound single trajectories ${\Delta}_{\mathrm{pop},i}(t)$ while $B_{pop}(t)$ correctly bounds $B_{\Ic_c}(t)$. The average full-state distance from the metastable state $\bar \Delta_{full}(t)$ do not fall below the threshold since block-coherences persist after the initial block-population relaxation toward the metastable state.
		This is reflected by the separation between $\bar{\Delta}_{\mathrm{full}}(t)$ and $\bar{\Delta}_{\mathrm{pop}}(t)$, together with the slower decay of $\bar{\Delta}_{\mathrm{coh}}(t)$. { The numerical results reflect the theoretical results of  Proposition \ref{prop:qc_short_run}.}
		
		\item {\em Middle-run:} The full-state and block-population distances from the metastable state $\bar{\Delta}_{\mathrm{full}}(t)$ and $\bar{\Delta}_{\mathrm{pop}}(t)$ become nearly indistinguishable after decay of coherences $\bar{\Delta}_{\mathrm{coh}}(t)$. $\bar{\Delta}_{\mathrm{full}}(t)$ and $\bar{\Delta}_{\mathrm{pop}}(t)$ increase indicating the full state  and block-population are moving away from the metastable state. { As indicated by Proposition \ref{prop:qc_middle_run} in this regime each block population remain approximately proportional to the stationary state of the corresponding stochastic complement and to the corresponding component of the HSC metastable state. Only the scalars that weight the blocks change significantly.}
		
		\item {\em Long-run}: In the long run, $\bar{\Delta}_{\mathrm{stat}}(t)$ tends to zero, indicating convergence to stationarity { reflecting the theoretical result of Proposition \ref{prop:qc_long_run}.}
	\end{enumerate}
	{ We remark that all the plotted curves in Figure \ref{fig:dyn_reg} are continuous. The apparent discontinuities between the different panels of the figure are due to a different choice of time scale on the horizontal axis.}
	\medskip
	
	\subsection{Example 2: Nearly strongly reducible two qubit system}\label{sec:ns_channel}
	In this section we illustrate the behavior of a nearly strongly reducible system. As in the previous section we consider a two-qubit system with evolution described by the channel 
	\begin{equation}\label{eq:sr_channel}
		\Phi=\Phi_S+\Ec
	\end{equation}
	where $\Phi_S$ is a channel with strong symmetry 
	$U=\sigma_z^a\sigma_z^b$ and $\Ec$ is a symmetry-breaking perturbation. Unlike the previous example, however, the symmetric component $\Phi_S$ is chosen to be \emph{strongly reducible}. To this end, we consider $\Phi_S$ to be described by Kraus operators \[{\cal K}_S=\{\Pi_j K_{S,i}\Pi_j\}_{j,i},\]
	where $\{K_{S,i}\}$ are the Kraus operators introduced in Section~\ref{sec:nr_channel} and
	$\Pi_i$ denotes the orthogonal projector onto 
	$\mathcal H_i$. By construction, the channel acts independently on each
	symmetry sector and annihilates all block coherences. Since Kraus operators have support on a single diagonal block,
	Proposition~\ref{prop:cs_dec} implies that $\Phi_S$ is strongly reducible.
	Similarly to the  previous example, we introduce the symmetry-breaking perturbation
	
	\[
	\Ec(X)
	=
	-\xi\sum_i\Pi_iX\Pi_i
	+
	\xi\,\sigma_x^aX\sigma_x^a,
	\]
	where $\norm{\Ec}_1=2\xi$ and $0<\xi\leq\alpha_1$ controls the
	perturbation strength while ensuring that $\Phi$ is a quantum channel.
	The explicit norm calculations are given in Appendix~\ref{app:symmetry_norms}.
	As in the previous example 
    if $0<\xi<\alpha_1$, then $\Phi$ is primitive and hence irreducible, with unique stationary state $\tilde\rho=I_4/4$. {The
associated hybrid channel and stochastic complements are the same as in the previous example.}
	
	\paragraph*{Metastability.}
	As in the preceding example,
	\begin{equation}
		\delta_h=\delta_c=2\xi.
	\end{equation}
	The crucial difference is that the symmetric component annihilates all
	block coherences, so
	\begin{equation}
		\Phi_{\Delta}(X)=\xi\,\sigma_x^a
		\bigl(X-\mathcal W_c(X)\bigr)\sigma_x^a,
		\qquad \|\Phi_{\Delta}\|_1=\xi,
		\qquad \delta_s=3\xi.
		\label{eq:symmetry-example-two-deviations}
	\end{equation}
	The limit $\delta_s\to0$ as $\xi\to0$ reflects the near strong reducibility
	of the channel, and hence also its near complete reducibility.

\begin{figure}
		\centering
		\includegraphics[width=0.9\textwidth]{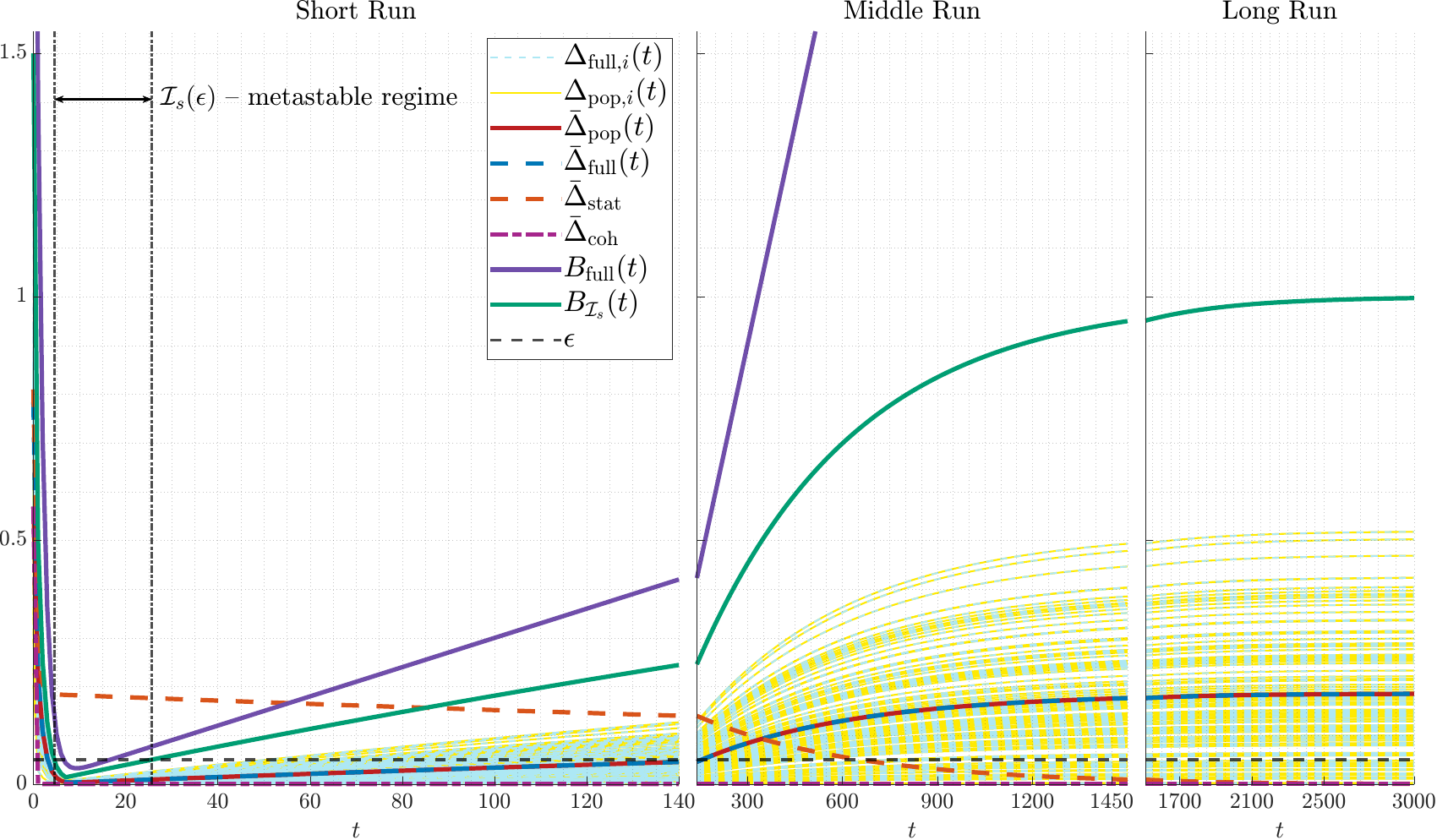}
		\caption{
			Dynamics of the nearly strongly reducible channel of
			Section~\ref{sec:ns_channel}, with $\xi=10^{-3}$, over the three time regimes.
			Faint curves show $\Delta_{\mathrm{pop},i}(t)$ and
			$\Delta_{\mathrm{full},i}(t)$ for 100 random initial states;
			thicker curves show their empirical averages, together with
			$\bar{\Delta}_{\mathrm{stat}}(t)$ and
			$\bar{\Delta}_{\mathrm{coh}}(t)$.
			The purple curve is the analytical bound
			$B_{\mathrm{full}}(t)$, and the green curve is
			$B_{\mathcal{I}_s}(t)$.
			The horizontal dashed line indicates the threshold $\epsilon=0.05$ used for the stabilization interval (the metastable time regime) $\mathcal{I}_s(\epsilon)$;
			the vertical markers delimit 
			$\mathcal{I}_s(\epsilon)$.
			All panels use linear scales, with different horizontal
			scales to display the different dynamical regime.
			Rapid coherence suppression makes $\Delta_{pop}(t)$ and $\Delta_{full}(t)$ nearly overlap.
		}
		\label{fig:nearly-strongly-reducible}
	\end{figure}

	We repeat the numerical experiment of the preceding section
	for the nearly strongly reducible channel, setting
	$\xi=10^{-3}$ and generating 100 random initial density matrices.
	Figure~\ref{fig:nearly-strongly-reducible} shows the individual
	distances and empirical averages defined in
	\eqref{eq:symmetry-distances} and
	\eqref{eq:symmetry-average-distances} over three time ranges,
	with the same threshold $\epsilon=0.05$. Moreover 
	Proposition~\ref{prop:qc_short_run} provides the state-independent estimates for the distance of the full state from the metastable state
	\begin{equation}\label{eq:B_full}
	    \Delta_{\mathrm{full},i}(t)
	\leq B_{\mathcal{I}_s}(t)
	:= \|\Phi^t-S_\infty\|_{1}
	\leq B_{\mathrm{full}}(t),
	\end{equation}
	where
    \begin{equation*}
        B_{\mathrm{full}}(t)
	= t\delta_s+\sqrt{D}\,\kappa|\lambda_{q+1}|^t.
    \end{equation*}

	We make the following observations.
	\begin{enumerate}[label=(\roman*)]
		\item {\em Short-run}: Unlike the preceding example, block-coherences are strongly suppressed on the shortest time scale. The symmetric component of the channel annihilate block-coherences, while the perturbation component acts on them as
		\begin{equation}
			\rho_{O,t+1}
			= \xi\,\sigma_x^a\rho_{O,t}\sigma_x^a.
			\label{eq:strong-symmetry-coherence-evolution}
		\end{equation}
		Since unitary conjugation preserves the trace norm,
		\begin{equation}
			\Delta_{\mathrm{coh},i}(t)
			= \xi^t\Delta_{\mathrm{coh},i}(0).
			\label{eq:strong-symmetry-coherence-decay}
		\end{equation}
		For $\xi=10^{-3}$, this contribution becomes negligible after
		very few iterations. Accordingly,
		$\bar{\Delta}_{\mathrm{full}}(t)$ and
		$\bar{\Delta}_{\mathrm{pop}}(t)$ are nearly indistinguishable
		throughout most of the displayed evolution. The quantity $B_{\mathcal{I}_s}(t)$ is smaller than the threshold $\epsilon$  for a certain time interval establishing a
		non-empty interval $\mathcal{I}_s(\epsilon)$ and therefore a metastable time regime {\em for the full state of the system}.
		Throughout this interval, the {\em full state} remains approximately close to the metastable state. The analytical bound $B_{\mathrm{full}}(t)$
		also falls below the threshold and therefore provides a
		sufficient, generally more conservative estimate of this
		interval and a good indicator of metastability of the system. { The numerical results reflect the theoretical results of  Proposition \ref{prop:qc_short_run}.}
		
		\item {\em Middle and long-run:} the full state of the system remains approximately block-diagonal and moves away from the metastable state. {As indicated by Proposition \ref{prop:qc_middle_run} in this regime each block-diagonal component of the state remain approximately proportional to the stationary state of the corresponding stochastic complement and to the corresponding component of the HSC metastable state. Only the scalars that weight the blocks change significantly. In the long run the state converge to stationary as indicated by $\bar{\Delta}_{\mathrm{stat}}(t)\to 0$ reflecting the result of Proposition \ref{prop:qc_long_run}.}
	\end{enumerate}
	{We remark that all the plotted curves in Figure \ref{fig:nearly-strongly-reducible} are continuous. The apparent discontinuities between the different panels of the figure are due to a different choice of time scale on the horizontal axis.}
	\vspace{20px}
	
		\section{Examples: Approximate quantum nondemolition measurements}\label{sec:QND}
        \subsection{General setup} 
		QND measurements extract information about an observable
		without inducing transitions between its eigenspaces. Consider repeated projective measurements of an observable $Q=\sum_{j=1}^q q_j\Pi_j$,
		with distinct eigenvalues $q_j$, separated by identical intervals of evolution.
		The projections define the subspaces $\Hc_j=\operatorname{range}\Pi_j$.
		Let $\Phi_R$ denote the quantum channel for one interval of evolution
		followed by a measurement, averaged over outcomes.
		The QND condition is
		\cite[Sec.~2.1]{hscIntroBauerBenoistBernard2013}
		\begin{equation}
			\tr\bigl[\Pi_j\Phi_R(\rho)\bigr]=\tr(\Pi_j\rho),
			\qquad j=1,\ldots,q,
		\end{equation}
		for every density operator $\rho$. Thus each subspace is invariant and
		$\Phi_R$ is completely reducible, while dynamics within these subspaces is allowed.
		
		We model {\em approximate} QND dynamics by an irreducible quantum channel
		\begin{equation}
			\Phi=\Phi_R+\Ec,
			\qquad \norm{\Ec}_1\ll1,
		\end{equation}
		where the linear perturbation $\Ec$ breaks the invariance of these subspaces.
		The resulting channel is nearly completely reducible, providing a setting
		for the metastability analysis developed above.
		
		\subsection{Example: A two-level atom and a spin}
		As a simple example, we consider a bipartite quantum system composed of a
		two-level atom, labelled $A$, and a spin-$1/2$ degree of freedom, labelled $B$.
		The associated Hilbert space is $\mathcal H=\mathcal H_A\otimes\mathcal H_B$,
		where $\mathcal H_A,\mathcal H_B\simeq\mathbb C^2$. We perform repeated
		projective measurements of the atom in the orthonormal basis
		$\{|0\rangle_A,|1\rangle_A\}$, with projections
		$\Pi_j=\ketbra{j-1}{j-1}_A\otimes I_B$, $j\in\{1,2\}$.
		Each cycle consists of evolution over an interval $\ti>0$, followed
		by a measurement of the atomic level.
		The corresponding nonselective measurement map is
		$\Wc_c(X)=\sum_{j=1}^2\Pi_jX\Pi_j$.
		
		During this interval, the atom evolves under the Hamiltonian
		\begin{equation*}
			H_{A,\varepsilon}=\frac{\omega}{2}\sigma_z+\varepsilon\Omega\sigma_x,
			\qquad U_{\ti,\varepsilon}=e^{-i\ti H_{A,\varepsilon}},
		\end{equation*}
		where $\sigma_x$ and $\sigma_z$ are the Pauli matrices acting on $\mathcal H_A$
		in the basis $\{|0\rangle_A,|1\rangle_A\}$.
		Here $\omega$ is the level splitting, $\Omega$ sets the transverse coupling
		scale, and $\varepsilon\geq0$ controls its strength. Independently, the spin relaxes towards the thermal state
		associated with the Hamiltonian $H_B=E|1\rangle\langle1|_B$, with $E>0$:
		\begin{equation}
			\tau_\beta=\frac{e^{-\beta H_B}}{\tr(e^{-\beta H_B})}
			=\frac{1}{1+e^{-\beta E}}
			\begin{pmatrix}1&0\\0&e^{-\beta E}\end{pmatrix},
			\qquad 0<\beta<\infty,
			\label{eq:qnd-thermal-state}
		\end{equation}
		We describe this relaxation at rate $\gamma>0$ by the thermalization
		channel
		\begin{equation}
			\Lambda_{\beta,\ti}(Y)=e^{-\gamma\ti}Y
			+(1-e^{-\gamma\ti})\tr(Y)\tau_\beta,
			\qquad Y\in\Bf(\mathcal H_B).
			\label{eq:qnd-thermal-channel}
		\end{equation}
	The thermalization channel is generated by the Lindblad operator $\mathcal L_\beta(Y)=\gamma\bigl(\tr(Y)\tau_\beta-Y\bigr)$ and has the
		unique stationary state $\tau_\beta$. The channel  describing one evolution-measurement cycle is 
		\begin{equation*}
			\Phi_{\ti,\varepsilon}
			=\Wc_c \bigl(\operatorname{Ad}_{U_{\ti,\varepsilon}}\otimes\Lambda_{\beta,\ti}\bigr),
			\qquad \operatorname{Ad}_U(Y)=UYU^\dagger.
		\end{equation*}
		The system therefore evolves according to a discretized time-homogeneous Markovian quantum dynamics.
		The measurement projections induce the decomposition
		\begin{equation*}
			\mathcal H=\mathcal H_1\oplus\mathcal H_2,
			\qquad \mathcal H_j=\operatorname{span}\{|j-1\rangle_A\}\otimes\mathcal H_B,\quad j=1,2.
		\end{equation*}
		With respect to this decomposition of the Hilbert space, the channel can be decomposed in turn as
		\begin{equation}
			\Phi_{\ti,\varepsilon}=\Phi_R+\Ec,
			\label{eq:qnd-decomposition}
		\end{equation}
		where the unperturbed channel $\Phi_R$ obtained by setting $\varepsilon=0$  is completely reducible and described a perfect non demolition dynamics:
		\begin{equation*}
			\Phi_R(X):=\Phi_{\ti,0}(X)
			=\sum_{j=1}^2 \ketbra{j-1}{j-1}_A\otimes\Lambda_{\beta,\ti}(X_{jj}),
			\ \  X_{jj}:=(\langle j-1|_A\otimes I_B)X(|j-1\rangle_A\otimes I_B).
		\end{equation*}
		Moreover, $\Phi_R$ annihilates block-coherences so it is also strongly reducible. Within
		each sector, the spin evolves according to the primitive channel
		$\Lambda_{\beta,\ti}$.
		
		The perturbation $\Ec$ is induced by the transverse Hamiltonian term:
		\begin{equation}
			\Ec:=\Phi_{\ti,\varepsilon}-\Phi_{\ti,0}
			=\Wc_c \Bigl[
			\bigl(\operatorname{Ad}_{U_{\ti,\varepsilon}}-\operatorname{Ad}_{U_{\ti,0}}\bigr)
			\otimes\Lambda_{\beta,\ti}\Bigr].
		\label{eq:qnd-perturbation-map}
		\end{equation}
		The perturbation norm is
		\begin{equation}
			\norm{\Ec}_1
			=\frac{2|\varepsilon\Omega|}{\nu}|\sin(\nu\ti)|,
			\qquad \nu=\sqrt{(\omega/2)^2+\varepsilon^2\Omega^2},
			\label{eq:qnd-perturbation-norm}
		\end{equation}
		The computation is
		given in Appendix~\ref{app:qnd-perturbation-norm}.
		
		The transverse Hamiltonian term couples
		the measured levels and can therefore change their populations between
		successive measurements, breaking the quantum nondemolition property.
		When the transition probability satisfies
		$0<|\langle1|U_{\ti,\varepsilon}|0\rangle|^2<1$, it can be checked that
		the perturbed channel $\Phi_{\ti,\varepsilon}$ is primitive and hence
		irreducible, with unique stationary state
		$\widetilde\rho=(I_A/2)\otimes\tau_\beta$. For sufficiently small
		$\norm{\Ec}_1$, this gives a nearly completely reducible channel to
		which the theoretical framework of the previous sections applies.
		
		We first illustrate the construction of the hybrid stochastic complements
		and the application of the hybrid
		coupling theorem (Theorem~\ref{thm:coup_thm}), following the procedure in Algorithm~\ref{alg:ss_comp}.
		This recovers the known stationary state from the stationary states of
		the smaller channels and the coupling matrix. We then study the
		metastable behavior of the system and its dependence on the measurement
		interval $\ti$.
		
		\paragraph*{Computing stationary states.}
		To compute the hybrid stochastic complements, define
		\begin{equation}
			s:=|\langle1|U_{\ti,\varepsilon}|0\rangle|^2
			=\frac{\varepsilon^2\Omega^2}{\nu^2}\sin^2(\nu\ti),
			\label{eq:qnd-transition-probability}
		\end{equation}
		the transition probability from $|0\rangle_A$ to $|1\rangle_A$ in one
		evolution--measurement cycle. The reverse transition has the same
		probability $s$, while remaining at the same atomic level has probability
		$1-s$. We consider $0<s<1$ throughout. Since
		every output is block diagonal, the associated hybrid channel is
		$\Phi_{\oplus}=\Wc_c \Phi_{\ti,\varepsilon} \Wc_c
		=\Phi_{\ti,\varepsilon} \Wc_c$. The components of the hybrid channel are
		\begin{equation*}
			\Phi_{11}=\Phi_{22}=(1-s)\Lambda_{\beta,\ti},
			\qquad \Phi_{12}=\Phi_{21}=s\Lambda_{\beta,\ti}.
		\end{equation*}
		The stochastic complements therefore coincide and are given by
		\begin{equation}
			\Sc_{jj}=(1-s)\Lambda_{\beta,\ti}
			+s^2\Lambda_{\beta,\ti}\bigl(\Ic-(1-s)\Lambda_{\beta,\ti}\bigr)^{-1}
			\Lambda_{\beta,\ti},
			\qquad j=1,2.
			\label{eq:qnd-stochastic-complements}
		\end{equation}
	 More explicitly, setting
		$\eta=e^{-\gamma\ti}$, these are the channels:
		\begin{equation}
			\Sc_{jj}(Y)=r_{\mathrm{HSC}}Y
			+(1-r_{\mathrm{HSC}})\tr(Y)\tau_\beta,
			\qquad
			r_{\mathrm{HSC}}=(1-s)\eta+
			\frac{s^2\eta^2}{1-(1-s)\eta}.
			\label{eq:qnd-hsc-contraction}
		\end{equation}
		Here $r_{\mathrm{HSC}}$ is the second-largest eigenvalue in modulus of
		each $\Sc_{jj}$. These calculations are given
		in Appendix~\ref{app:qnd-thermal-complements}. 
		
		The hypotheses of the hybrid coupling theorem
		(Theorem~\ref{thm:coup_thm}) are satisfied. Following the procedure in
		Algorithm~\ref{alg:ss_comp}, it is possible to recover the stationary state of the full channel from the HSCs.
		Since $0<r_{\mathrm{HSC}}<1$ and
		$\tau_\beta>0$, each complement is primitive and has the unique
		stationary state $\widetilde\sigma_j=\tau_\beta$.

		 we construct the coupling matrix
		\begin{equation}
			C=[c_{ij}]_{i,j\in\{1,2\}},
			\qquad c_{ij}=\tr\bigl[\Phi_{ij}(\widetilde\sigma_j)\bigr],
			\qquad C=\begin{pmatrix}1-s&s\\s&1-s\end{pmatrix}.
			\label{eq:qnd-coupling-matrix}
		\end{equation}
		Solving $C\xi=\xi$ with $\xi_1+\xi_2=1$ gives the coupling factors
		$\xi_1=\xi_2=1/2$. The final step of the algorithm reconstructs the unique
		stationary state of the full channel as
		\begin{equation}
			\widetilde\rho
			=\sum_{j=1}^2\xi_j\ketbra{j-1}{j-1}_A\otimes\widetilde\sigma_j
			=\frac{I_A}{2}\otimes\tau_\beta.
			\label{eq:qnd-stationary-state}
		\end{equation}
		
		\paragraph*{Metastability.}
		We next focus on the metastable behavior of the system. To this end, we
		first quantify the deviations from reducibility. The deviation of the
		hybrid channel from complete reducibility is
		\begin{equation*}
			\delta_h=2s.
		\end{equation*}
		For the original channel, the remainder is
		$\Phi_{\Delta}=\Phi_{\ti,\varepsilon}-\Phi_{\oplus}$. It satisfies $\Wc_c \Phi_{\Delta}=\Phi_{\Delta}$,
		so the deviations from complete and strong reducibility coincide:
		\begin{equation}
			\delta_c=\delta_s
			=\delta_h+\norm{\Phi_{\Delta}}_1
			=2s+2\sqrt{s(1-s)}.
			\label{eq:qnd-reducibility-deviations}
		\end{equation}
		The computations are given in Appendix~\ref{app:qnd-reducibility-deviations}.
		For fixed $\ti$, $\omega$, and $\Omega$, the transition probability $s$
		and all three deviations tend to zero as $\varepsilon$ tends to zero.
		Thus, for sufficiently small nonzero $s$,
		the channel is both nearly completely reducible and nearly strongly
		reducible.
		
		Since the channel is nearly strongly reducible, we expect the three
		dynamical regimes described in Section~\ref{sec:dyn_regimes} when the
		short-run stabilization interval is nonempty. Their qualitative behavior
		is similar to that of the example in Section~\ref{sec:ns_channel}, so we
		omit detailed simulations of all three regimes.
		
		Here we focus on a distinctive feature of the present system: the dependence of the metastable regime on the measurement
		interval $\ti$, or, equivalently, on the measurement frequency
		$1/\ti$. In the
		frequent-measurement regime $\nu\ti\ll1$, the transition probability
		satisfies $s\simeq\varepsilon^2\Omega^2\ti^2$. 
		Thus, decreasing $\ti$ suppresses population transfer while leaving the
		spin relaxation rate $\gamma$ unchanged, extending the separation between
		local equilibration and global relaxation.
		
		We therefore focus the simulations on the short-run dynamics of the system. We fix $\varepsilon=0.01$, $\omega=2.4$,
		$\Omega=2$, $\gamma=10$, $\beta=1$, and $E=1$, and compare the measurement
		intervals
		$\ti\in\{0.001,0.005,0.01\}$. For each value of $\ti$, we evolve the
		same $N=100$ random initial density matrices $\rho_{0,i}$.
		
		We use the full-state quantities from \eqref{eq:symmetry-average-distances},\eqref{eq:B_full}: the empirical mean distance from the
		HSC metastable state $\bar\Delta_{\mathrm{full}}$, the induced-norm bound
		$B_{\mathcal I_s}$, and the analytical estimate $B_{\mathrm{full}}$. We now
		write their arguments as $(t,\ti)$ to highlight the dependence on the measurement interval $\ti$. The threshold $\epsilon=0.05$ defines the
		stabilization interval $\mathcal I_s(\epsilon,\ti)$. 
		
		\begin{figure}[tbp]
			\centering
			\includegraphics[width=\linewidth]{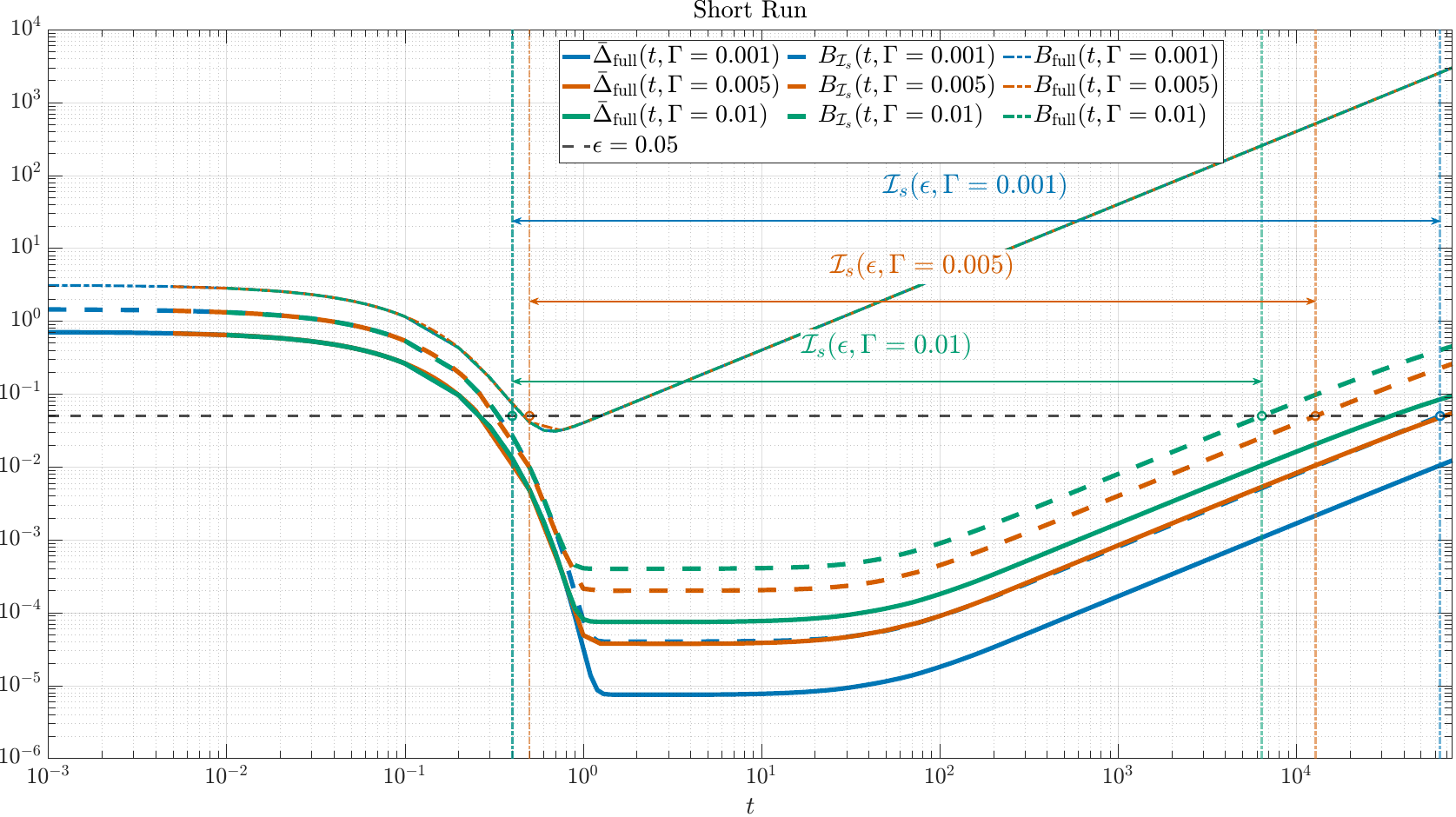}
		
				\caption{Short-run QND dynamics for $\varepsilon=0.01$, $\omega=2.4$,
				$\Omega=2$, and $\gamma=10$. Colors distinguish the measurement intervals:
				blue, $\ti=0.001$; orange, $\ti=0.005$; green, $\ti=0.01$.
				Solid curves show $\bar\Delta_{\mathrm{full}}$, dashed curves are numerical
				estimates of $B_{\mathcal I_s}$, and dash-dotted curves show the analytical
				bound $B_{\mathrm{full}}$. The horizontal line marks $\epsilon=0.05$;
				arrows and vertical markers delimit 
				$\mathcal I_s(\epsilon,\ti)$ on the sampled grid. The horizontal axis
				shows $t=n\ti$, where $n$ is the number of applications of the channel.
				Both axes are logarithmic. The initial point $t=0$ is omitted on the
				logarithmic axis, and each curve starts at $t=\ti$.}
			\label{fig:qnd-short-run}
		\end{figure}
		
		Figure~\ref{fig:qnd-short-run} shows that increasing the measurement frequency extends the metastable interval. 
		The right endpoints of the marked intervals are approximately
		$6.41\times10^4$, $1.28\times10^4$, and $6.41\times10^3$ for
		$\ti=0.001$, $0.005$, and $0.01$, respectively. Thus, reducing the
		measurement interval from $0.01$ to $0.001$ extends the metastable regime
		by approximately a factor of ten. 
		This behavior is consistent with the frequent-measurement estimate
		discussed above. All three choices satisfy $\nu\ti\ll1$, for which the
		transition probability in one evolution--measurement cycle is
		$s\simeq\varepsilon^2\Omega^2\ti^2$. Increasing the measurement
		frequency therefore suppresses transitions between the two subspaces while leaving
		the local relaxation rate unchanged. The resulting separation of time
		scales allows the full state to remain close to its HSC metastable state
		for a longer interval. The analytical bounds $B_{\mathrm{full}}$ remain conservative and the corresponding curves almost overlap.

	\section{Conclusions and outlook}\label{sec:conclusions}
	
	We have shown as hybrid stochastic complementation allows for a
	systematic approach to the {\em stationary-state analysis and metastability}
	of discrete-time quantum dynamics with respect to a given decomposition of the
	Hilbert space into orthogonal sectors. We associate the given quantum
	channel to a hybrid classical--quantum dynamics on the corresponding
	block-diagonal operator algebra, and focus on the stochastic complements, the
	completely positive and trace-preserving maps that combine the direct
	evolution within a chosen sector with all possible excursions through
	other sectors that eventually return to it. Under irreducibility
	assumptions, their stationary states, together with a classical
	coupling matrix, determine the stationary state of the associated
	hybrid channel. When the original channel admits a block-diagonal
	stationary state, this also yields a reconstruction of the stationary
	state of the full dynamics.
	
	The same construction provides a quantitative description of the
	separation between local equilibration and global relaxation in nearly
	reducible dynamics. In particular, the distinction between \emph{nearly
		completely reducible} and \emph{nearly strongly reducible} channels makes
	explicit the role of inter-sector quantum coherences: weak population
	transfer can yield metastability of the block-diagonal components,
	whereas proximity of the entire state to a block-diagonal metastable
	manifold additionally requires sufficiently rapid coherence
	suppression. Under suitable mixing and diagonalizability assumptions,
	our estimates compare the growth of the error due to inter-sector
	leakage with the decay of the intra-sector relaxation term, thereby
	providing sufficient conditions for a nonempty metastable time window.
	The approximate-symmetry examples illustrate how these two mechanisms
	can be separated. More generally, the approach complements the spectral
	theory of quantum metastability
	\cite{macieszczak2016towards,macieszczak2021classical,macieszczak2021operational}:
	rather than inferring long-lived phases solely from the slow spectrum,
	it starts from a physically motivated sector decomposition and
	constructs exact quantum return channels for the associated hybrid
	dynamics.
	
	Several directions for further investigation emerge from these
	results.
	A natural and interesting extension of this framework would cover {\em continuous quantum dynamical semigroups}: a continuous-time theory would allow direct comparison with adiabatic elimination and master equation methods based on projections \cite{alicki2007semigroups,NS:58,ZR:60}. The most prominent application of a continuous framework would be to examine the emergence of thermal
	equilibrium through two successive relaxation mechanisms in generators compatible with the Davies's weak-coupling limit \cite{hscIntroDavies1974,hscIntroSpohnLebowitz1978}. 
	A natural starting point would be to consider a Hamiltonian $H=\sum_i E_i\Pi_i$ and an open-system dynamics that mixes states rapidly within its energy eigenspaces, or within suitably coarse-grained energy shells, while inducing much slower transitions between them. 
	The HSCs, or its continuous equivalent, could
	describe a fast local equilibration process, possibly to microcanonical ensambles, whereas the hybrid coupling matrix
	would determine the sector weights in the global stationary state, which would be of Gibbsian. 
	
	Beyond stationary-state
	reconstruction and time-scale separation, it would be valuable to establish whether quantum
	detailed balance induces a corresponding reversibility property of the
	hybrid coupling matrix, and under which scaling limits the slow sector
	populations admit a closed kinetic description. 
	Overall, this would connect HSCs more directly to thermal transition rates, entropy
	production, and the approach to Gibbs equilibrium.
	
	Another potential development is motivated by quantum information (QI) applications: the present stabilization bounds use primitive stochastic complements,
	so that each sector relaxes to a unique stationary state, but in
	scenarios of interest for QI  the nominal dynamics may
	preserve a whole algebra of states within a sector, as occurs for
	noiseless subsystems and encoded logical degrees of freedom. The corresponding complements need not be
	primitive and may have multidimensional fixed-point spaces. Extending
	the theory to this setting would replace individual local stationary
	states by invariant operator spaces and could yield quantitative
	bounds on the preservation of encoded quantum information. 
	A limitation of the present hybrid construction is that it directly
	considers block-diagonal dynamics, while long-lived inter-sector
	coherences can carry additional metastable degrees of freedom. A more
	general theory should therefore relate stochastic complementation to
	reductions on noncommutative invariant algebras, including
	noiseless-subsystem and information-preserving structures \cite{BlumeKohoutEtAl2010}. 
	Another challenge is to identify suitable sectors from the dynamics
	itself, rather than assuming a partition in advance. Determining how to identify
	approximately invariant algebras, weakly broken symmetries \cite{grigoletto-inprep}, and slow
	spectral subspaces of quantum channels \cite{adiabaticalgebraic} could provide 
	systematic ways for selecting the needed decomposition and for comparing the
	resulting metastable manifolds with those obtained spectrally.
	
	Together, these developments could connect the present structural
	approach to metastability with thermalization, protected quantum
	memories, and the stochastic dynamics of continuously monitored open
	systems.
	
	\vspace{4mm}
	
	 \section*{Acknowledgments}
	M.G. acknowledges funding from the  Engineering and Physical Sciences Research Council, grant no. EP/T022140/1. F.T. acknowledges funding from  European Union’s Horizon Europe research and innovation programme under the project “Quantum Secure Networks Partnership” (QSNP, grant agreement No 101114043) and by the Italian Ministry of University and Research under the PRIN project ``Extracting essential information and dynamics from complex networks”, grant No.\,2022MBC2EZ.
    M.P. thanks Maël Bompais, Tommaso Grigoletto and Miguel Angel Casanova Medina for helpful discussions. 

    \section*{Author contributions}
    M.P. is the first author and led the conceptualization and writing of the manuscript, and developed the code for the examples. M.G. contributed to the conceptualization and helped improve the presentation of the results. F.T. contributed to the conceptualisation and writing of the manuscript.
    Generative AI was used for adversarial review of initial manuscript versions, writing part of the code used to analyze the examples and generate the figures, and improving the literature review.

	\clearpage

	\appendix
\section{Classical Stochastic Complementation}\label{sec:csc}
	In order to keep the presentation self-contained, we briefly recall some key results from the theory of stochastic complementation for classical Markov chains \cite{mayer-sc}. 
    We denote classical probability distributions by \(\Pb\) and represent discrete distributions as column vectors.  This convention yields similar expressions for classical transitions and  quantum maps in the Schr\"odinger picture, which are central to this paper.
	
	\subsection{Definition, properties and uncoupling of Markov chains}
	
	We consider a homogeneous and irreducible Markov chain $\{X_t \, :\, t\in \mathbb{N} \}$ with state space $\Nc=\{1,\dots, n\}$. We denote by $\pi_t$ the distribution vector of $X_t$, whose $i$-th entry is $\pi_{t,i}=\Pr[X_t=i]$.
	The transition matrix for the chain is a stochastic matrix $P=[p_{ij}]$ with $p_{ij}=\Pr[X_{t+1}=i | X_{t}=j]$.  We assume the state space is partitioned in $q \geq 2$ disjoint subclasses $\Qc_i$ for $i\in \{1,\dots, q\}$, i.e. $\Qc_i\cap\Qc_j=\emptyset$ for  $i\neq j$ and $\bigcup_{i} \Qc_i=\Nc$. This partition induces in turn a block-partition of the  transition probability matrix and unique stationary distribution $\tilde \pi$, (i.e. the distribution satisfying $\tilde \pi=P \tilde \pi$) of the chain:
	
	\begin{equation}\label{eq:transition_matrix}
		P =
		\begin{pmatrix}
			P_{11} & P_{12} & \dots & P_{1q}\\
			P_{21} & P_{22}&\dots & P_{2q}\\
			\vdots & \vdots & \ddots & \vdots\\
			P_{q1} & P_{q2}&\dots & P_{qq}\\
		\end{pmatrix},
		\qquad \tilde \pi=\begin{pmatrix}
			\tilde \pi_1\\ \tilde \pi_2\\\vdots \\ \tilde \pi_q
		\end{pmatrix}.
	\end{equation}

	Here, the block $P_{ij}, \ i,j\in\{1,\dots ,q\}$ denotes the submatrix of $P$ whose rows are indexed by $\Qc_i$ and columns by $\Qc_j$, and represents the probability of a transition from a state in $\Qc_j$ to a state in $\Qc_i$. Similarly $\tilde \pi_i$ denotes the subvector of $\tilde \pi$ whose entries are indexed by $\Qc_i$.
	
	In the following, we make use of the following maps: 
	\begin{itemize}
		\item $P_i$ is the principal sub-matrix of $P,$ obtained by deleting the $i$-th row and $i$-th column in the block decomposition of $P$. It describes the probability of transitioning from a state in $\Nc\setminus\Qc_i$ to $\Nc\setminus\Qc_i$.
		\item $P_{*i}=\begin{pmatrix}
			P_{1i}^\top & \dots & P^\top_{i-1,i}& P^\top_{i+1,i}& \dots & P^\top_{q,i} 
		\end{pmatrix}^\top$,  describes the probability of transitioning from a state in $\Qc_i$ to its complement.
		\item $P_{i*}=\begin{pmatrix} P_{i1} &\dots& P_{i,i-1} &P_{i,i+1} &\dots & P_{i,q} \end{pmatrix}$, describes the probability of transitioning to $\Qc_i$ from a state in the complement.
	\end{itemize}
	To introduce the concept of stochastic complement,consider the following expression: 
	\begin{equation}\label{eq:sc_ext}
		S_{ii}=  P_{ii}+ P_{i *}\sum_{k=0}^\infty P^k_{i} \, P_{* i}.
	\end{equation}
	From a probabilistic point of view, the  above is the transition probability matrix of a new process, called a {\it reduced chain}, that is obtained by observing the {original chain} only when it is in a state belonging to one of the subclasses. In particular, it describes transitions within a subset $\mathcal{Q}_i$ that may involve excursions through the complementary subset $\Nc\setminus\mathcal{Q}_i$ before returning. 
	The first term in the R.H.S of the above expression is associated with a direct transition from $\Qc_i$ to $\Qc_i$, while the second term accounts for all possible transitions from a state in $\Qc_i$ to another state in $\Qc_i$ that involve an intermediate transition through states in $\Nc\setminus\Qc_i$.
	
	The well-definiteness of \eqref{eq:sc_ext} is guaranteed by the fact that $\one-P_{i}$ is a non-singular $M$-matrix, which implies that the power series in the expression converge to  $(\one-P_{i})^{-1}$ \cite{mayer-sc}. In view of this and according to \cite{mayer-sc} we give the following definition.

	\begin{definition}[{\bf Classical Stochastic complement}] The stochastic complement of $P_{ii}$ is the matrix
		\begin{align}\label{eq:csc}
			S_{ii}=P_{ii}+ P_{i *}(\one-P_{i})^{-1} P_{* i}.
		\end{align}
	\end{definition}

	In the following theorem we collect the key properties of the classical stochastic complements in \cite[Theorems 2.1,\,2.2,\,2.3]{mayer-sc}. 
	
	\begin{theorem}[\bf Properties of the stochastic complement]\label{thm:sc_prop}
		If the transition matrix P in \eqref{eq:transition_matrix} is irreducible, then $S_{i i}$ is an irreducible stochastic matrix for all $i\in\{1,\dots , q\}$. 
		Moreover, the unique stationary distribution $ s_{i}$ of $\Sc_{ii}$ satisfies 
		$$ s_i=\frac{\tilde \pi_{i}}{\sum_j \tilde \pi_{i,j}},$$
		where $\tilde \pi_{i,j}$ is the $j$-th component of $\tilde \pi_i.$
	\end{theorem}
	
	Another property of interest is primitivity. The following result provides a sufficient condition \cite[Theorem 5.1 and 5.2]{mayer-sc}.
	\begin{theorem}[\bf Primitivity of the stochastic complement]
		If $P_{ii}$ is primitive or has at least a non-zero diagonal entry, then also the corresponding stochastic complement $S_{ii}$ must be primitive. 
	\end{theorem}
	Theorem~\ref{thm:sc_prop}, establish the existence of a relation between stochastic complements, corresponding to reduced Markov chains,  and the stationary distribution of the original Markov chain. In particular, for each set $\Qc_i$, the vectors $\tilde s_{i}$ (which we recall is the stationary state of $S_{ii}$) and $\tilde \pi_{i}$ differ only by a scalar multiple. The following theorem shows that these scalars can be determined independently of the stationary distribution. Stochastic complements can therefore be used to decouple a Markov chain into smaller chains and compute the stationary distribution of the original chain from the reduced ones.
	\begin{theorem}[\bf Coupling theorem]\label{thm:coup_thm_cl}
		If the transition matrix P in \eqref{eq:transition_matrix} is  irreducible, its stationary distribution $\tilde \pi$ is given by 
		$$\tilde \pi=\begin{pmatrix}
			\xi_{1}  s^\top_{1} & \dots &  \xi_{q}  s^\top_{q}
		\end{pmatrix}.$$
		where $s_{i}$ is the stationary distribution for the stochastic complement $S_{ii}$ and $\xi=(
		\xi_{1} \ \dots \ \xi_{q}
		)^\top$ is the unique stationary distribution for the Markov chain with transition probability matrix $C$ with entries
		$$c_{ij}=\ones^\top P_{ij} s_j.$$
	\end{theorem}
	
	The matrix $C$ is called the \textit{coupling matrix} and the $\xi_i$ are called \textit{the coupling factors}.

	\subsection{Nearly completely reducible systems and metastability}
	
	We next introduce nearly completely reducible Markov chains and, by exploiting stochastic complementation, characterize their convergence behavior.

	\begin{definition}[\bf Nearly completely reducible Markov Chain]
		A Markov Chain and the associated transition matrix are said to be completely reducible if there exists a decomposition of the state space in subclasses $\{\Qc_i\}_{i=1}^q$ that induces the following block-partition of the transition matrix:
		$$P^* = \begin{pmatrix}
			P^*_{11} & 0 & \dots & 0\\
			0 & P^*_{22}&\dots & 0\\
			\vdots & \vdots & \ddots & \vdots\\
			0 & 0&\dots & P^*_{qq}\\
		\end{pmatrix}.$$
		A Markov chain with transition matrix $P$ is said to be nearly completely reducible if 
		$$P=P^*+ \Delta$$
		where $P^*$ is the transition matrix of a completely reducible Markov chain, and
		$\Delta=[\Delta_{ij}]$ is such that $\norm{\Delta}_1=\max_j \sum_i |\Delta_{ij}|<\epsilon$ and $\epsilon$ is a small real positive number. 
	\end{definition}

	In other words, the subclasses $\mathcal{Q}_i$ in the above definition are only weakly coupled together: the probability of transitioning from a state in one subclass to a state in another is small. 
	\begin{definition}[\bf Deviation from complete reducibility] Let P be the irreducible transition probability matrix in \eqref{eq:transition_matrix} and let $S={\rm diag(S_{11},\dots,S_{qq})}$, where the $S_{ii}$ are the stochastic complements. The number 
		
		$$\delta=\norm{P-S}_1=2 \max_i \| P_{*i}\|_1$$
		is called deviation from complete reducibility.
	\end{definition}
	Note when $\delta=0$ $P$ is completely reducible.
	
	The stochastic complement shows interesting spectral properties when $\delta$ tends to 0  \cite[Theorem 6.2]{mayer-sc}; in this case if $P$ has exactly $q-1$ non-unit eigenvalues close to 1 and the others have modulus bounded away from it, each $S_{ii}$ has no eigenvalues near its unit eigenvalue (it has a large spectral gap). When the states in a nearly completely reducible chain are naturally ordered,
	and when the transition matrix $P$ is partitioned in the natural way according to the closely coupled subclasses, then the spectrum splitting effect almost always occurs \cite{mayer-sc}. Therefore, in the above scenario {(whenever $P$ and $S_{ii}$ are primitive)} the convergence to stationarity of the full chain is slow while stochastic complements approach stationarity on a shorter time scale. 
	
	The following theorem, based on the concept of stochastic complement, establishes an upper bound on the short-run behavior of the original Markov chain, thereby enabling the characterization of possible metastable behaviors. Remarkably, the metastable probability distributions for the original Markov chain are {block-wise} proportional to $s=\begin{pmatrix}
		s_1^\top,\dots , s_q^\top
	\end{pmatrix}^\top$, where $s_i$ is the stationary distribution of $S_{ii}$.
	
	\begin{theorem}[\bf Short-run dynamics bound]
		\label{TH:short_run_classical} Let the transition matrix P in \eqref{eq:transition_matrix} be irreducible. Let
		$S=\operatorname{diag}(S_{11},\dots,S_{qq})$ where the  $S_{ii}$ are the stochastic complements. Assume that each $S_{ii}$ is primitive and that $S$ is
		diagonalizable, i.e., $S=ZDZ^{-1}$ with $Z$ invertible and $D$
		diagonal. Let the eigenvalues of $S$ be such that $|\lambda_1|\geq |\lambda_2| \geq \dots \geq |\lambda_n|$.
		Let $S_\infty=\lim_{t\to\infty}S^t$. For any initial probability
		column vector $\pi_0$, define $\pi_t=P^t\pi_0$ and
		$s_\infty=S_\infty\pi_0$. Then, for every integer $t\ge1$,
		\begin{equation}\label{eq:classical_bound}
			\|\pi_t-s_\infty\|_1
			\le \|P^t-S_\infty\|_1
			\le t\delta+\kappa|\lambda_{q+1}|^t,
		\end{equation}
		where $\delta$ is the deviation from complete reducibility and
		$\kappa=\|Z^{-1}\|_1\|Z\|_1$.
	\end{theorem}
	As observed in \cite{mayer-sc}, the primitivity and diagonalizability assumptions can be relaxed to obtain a more general version of the bound above. Here, we retain these assumptions to simplify the exposition and follow the presentation of the results in \cite{mayer-sc}.
	
	The above bound can be used to characterize the metastable time regime of the chain, i.e. a time regime where the probability distribution remain approximately stationary before converging to the true stationary distribution.  
	
	\begin{definition}[\bf Short-run stabilization interval]
		For each $\varepsilon>0$, there is a short-run stabilization interval $\mathcal{I}(\varepsilon)$ which is defined to be the set 
		$$\Ic(\varepsilon) \ =\ \bigl\{t \, :\, \|P^t-S^\infty \|_1<\varepsilon\bigr\}.$$
	\end{definition}
	Using the upper bound in Theorem \ref{TH:short_run_classical} we find that the short-run stabilization interval contains the  interval  
	$E(\varepsilon)=\{t \,:\, \ t \delta +\kappa |\lambda_{q+1}|^t<\varepsilon \}$, which can be used to estimate analytically $\Ic(\epsilon)$.
	If $\Ic(\epsilon)$ is non-empty for $\epsilon$ sufficiently small, the chain has a metastable behavior. $\Ic(\epsilon)$ is the metastable time-regime.
	
	The overall convergence of the original nearly completely reducible chain to its stationary distribution can be characterized as follows \cite[Theorem 10.1]{mayer-sc}.

	Consider a nearly completely reducible original chain with primitive transition matrix $P$ and primitive stochastic complements $S_{ii}$.
	We assume that \(\mathcal I(\varepsilon)\) is nonempty for some small tolerance \(\varepsilon>0\). For an initial probability distribution $\pi_0$, the distribution $\pi_t=P^t \pi_0$ evolves as follows:
	\begin{itemize}
		\item {\em Short run evolution and stabilization ($t\leq\sup\,\Ic(\varepsilon)$)}: $\pi_t$ evolve such that subclasses tend toward their own approximate local equilibrium, which is proportional to the stationary distribution of the associated stochastic complement. When $t\in \Ic(\varepsilon)$ $\pi_t \approx\begin{pmatrix}
			\alpha_1 s_1^\top & \dots \alpha_q s_q^\top\end{pmatrix}$, and the chain is in a metastable time regime.
		
		\item {\em Middle run evolution ($t>\sup \,\Ic(\varepsilon)$):}  $\pi_t$ moves away from the approximate local equilibrium, but subclasses maintain a relative stability, in particular there exists $\beta_i(t)\in [0,1] $ with $\sum_{i=1}^q \beta_i(t)=1$ such that $\pi_t\approx(\beta_1(t)s_1^\top , \dots,$ $\beta_q(t) s_q^\top)^\top$.
		
		\item {\em Long run evolution and stabilization ($t\to\infty$):} $\pi_t$ converge to the stationary distribution of the chain, which is $\tilde \pi=\begin{pmatrix}
			\xi_1 s_1^\top, \dots, \xi_q s_q^\top
		\end{pmatrix}^\top$ (Theorem \ref{thm:coup_thm_cl}), in particular, each $\beta_i(t)$ coverage to the corresponding coupling factor $\xi_i$.
	\end{itemize}
	The estimated short-run stabilization interval $E_c(\epsilon)$ can be used to analytically determine the emergence of the metastable behavior outlined above.

	\begin{remark}[Reducibility and metastability]
		We note that near complete reducibility is necessary to establish metastability through the estimated short-run stabilization interval \(E(\varepsilon)\). Indeed, if \(\delta\) is large compared to the prescribed tolerance \(\varepsilon\), the term \(t\delta\) prevents the bound in \eqref{eq:classical_bound} from falling below \(\varepsilon\), and \(E(\varepsilon)\) is empty.  
	\end{remark}

	\section{Proofs of the Properties of the Hybrid Channel and Stochastic Complement}\label{app:proofs_hc_hsc}
	\begin{proof}[Proof of Proposition \ref{prop:hc_props} (Irreducibility of associated hybrid channel)]
		We first prove $\Phi_{\oplus}$ is a CPTP map, i.e. a quantum channel.
		To prove it we exploit the fact that the composition of CPTP maps is CPTP. Indeed $$\Phi_{\oplus} := \sum_{i,j=1}^q \Jc_i \Phi_{ij} \Jc_j^\dag =\sum_{i,j=1}^q \Pc_i \Phi \mathcal{P}_j=(\sum_{i=1}^q \Pc_i) \Phi (\sum_{j=1}^q \Pc_j),$$ where $\mathcal{P}_i$ are CP and their sum is CPTP.
		
		We now prove irreduciblity of $\Phi_{\oplus}$. Given a Kraus representation of $\Phi(\cdot)=\sum_i K_i \cdot K_i^\dag,$ $\Phi_{\oplus}$ admits Kraus representation
		$$ \Phi_{\oplus}(\cdot)=
		\sum_{i,j=1}^q \Pc_i \Phi \mathcal{P}_j(\cdot)=\sum_{i,j=1}^q\sum_s\Pi_iK_s\Pi_j\cdot  \Pi_j K_s^\dag \Pi_i,$$
		where we exploited the fact that $\Pc_i(\cdot)=\Pi_i\cdot\Pi_i$. 
		We label $E_{i,s,j}=\Pi_iK_s\Pi_j.$
		
		Suppose, by contradiction, that $\Phi_{\oplus}$ is reducible, then there exists a nontrivial subspace $\Zc\subset \Hc$ which is invariant for $\Phi_{\oplus}$. Let $\ket{\Phi_{\oplus}}\in \Zc$, then it holds $E_{i,s,j}\ket{\Phi_{\oplus}}\in \Zc, \forall s$ and for all $i,j\in\{1,\dots q\}$. However, 
		$$\sum_{i,j=1}^q E_{i,s,j}\ket{\Phi_{\oplus}}=(\sum_{i=1}^q \ \Pi_i) K_s (\sum_{j=1}^q \Pi_j) \ket{\Phi_{\oplus}}=K_s\ket{\Phi_{\oplus}},$$
        for all $s$.
		This shows that $K_s \ket{\Phi_{\oplus}} \in \mathcal{Z}$, and hence $\mathcal{Z}$ must be invariant for $\Phi$ contradicting irreducibility of the quantum channel.
	\end{proof}
	
	Before proving Proposition \ref{prop:spectral_radius}, we introduce the following lemma \cite[Theorem 6.4, 6.5]{wolf-qc}.
	\begin{lemma}\label{lem:sr_pos_maps}
		Let $\Ec$ be a CP map on $\mathfrak{B}(\mathcal{H})$ and $r(\Ec)$ its spectral radius i.e. $ r(\Ec) = \sup \{ |\lambda|, \lambda \in {\rm{sp}}(\Ec) \}$. Then there exists a non-zero eigenoperator $X \ge 0$ such that
		\[
		\Ec(X) = r(\Ec) X.
		\]
		Moreover, $\Ec$ is irreducible if and only if $r(\Ec)$ is a non degenerate eigenvalue and there exist operators $X,Y>0$ such that $\Ec(X)=r(\Ec)X, \ \Ec^\dag(Y)=r(\Ec)Y$. 
	\end{lemma}

	\begin{proof}[Proof of Proposition \ref{prop:spectral_radius} (Spectral properties of complementation maps)]
		
		Without loss of generality, we prove the claim for $\Bar{\Phi}_i$. The proof for $\Phi_{ii}$ follows the same lines.
		We first prove that $\Bar{\Phi}_i$ has all eigenvalues inside the unit disk, we successively prove all eigenvalues are strictly inside the unit disk.
		
		In the following of the proof we consider the bipartition of the Hilbert space $\Hc=\Hc_i\oplus\bar{\Hc_i}$ as in equation \eqref{eq:hs_bip}. Block-diagonal operators in ${\mathfrak{B}}(\Hc)$ will be decomposed according to the bipartition as $X=X_{ii} \oplus \bar{X}_{ii}$ with $X_{ii}\in {\mathfrak{B}}(\Hc_i)$ and $\bar X_{ii}\in\bigoplus_{j\neq i} {\mathfrak{B}}(\Hc_j)$.
		
		By Proposition \ref{prop:russo_dye} in  Appendix \ref{app:norms} and the definition of superoperator norm, we have
		\begin{equation}\label{eq:rd}
			\norm{\Phi_{\oplus}^\dag(X)}_\infty \leq \norm{\Phi_{\oplus}^\dag(I)}_\infty \norm{X}_\infty=\norm{X}_\infty \qquad \forall X.
		\end{equation} 
		Now let $X$ be such that $X_{ii} = 0$. Using \eqref{eq:blk_hch_i}, we obtain
		\begin{align}\label{eq:r_block}
			\norm{\Phi_{\oplus}^\dag(X)}_\infty\!&=\!\norm{\Phi_{*i}^\dag(\bar{X}_{ii})\oplus \Bar{\Phi}^\dag_i(\bar{X}_{ii})}_\infty
			\!\!=\!{{\rm max}} \Bigl\{ \norm{\Phi_{*i}^\dag(\bar{X}_{ii})}_\infty\!,\!\norm{\Bar{\Phi}_i^\dag(\bar{X}_{ii})}_\infty \Bigr\}.
		\end{align} 
		Combining \eqref{eq:rd} and \eqref{eq:r_block} it follows $\norm{\Bar{\Phi}_i^\dag(\bar{X}_{ii})}_\infty\leq \norm{\bar{X}_{ii}}_\infty \ \forall \bar{X}_{ii}$. In particular if $\bar{X}_{ii}$ is an eigenoperator of $\Bar{\Phi}^\dag_i$ with eigenvalue $\lambda$, then         
		$$\norm{\Bar{\Phi}_i^\dag(\bar{X}_{ii})}_\infty=|\lambda| \norm{\bar{X}_{ii}}_\infty \leq \norm{\bar{X}_{ii}}_\infty,$$ 
		which implies $|\lambda|\leq 1$.  Hence all eigenvalues of $\Bar{\Phi}_i^\dag$, and therefore of $\Bar{\Phi}_i$, lie inside the unit disk.
		We now show all the eigenvalues of $\Bar{\Phi}_i$ are strictly inside the unit disk. Suppose, by contradiction, that this is not the case. Then, by Lemma \ref{lem:sr_pos_maps},  there exists a nonzero operator
		$\bar{X}_{ii}\! \in\! \bigoplus_{j \ne i} \mathfrak{B}(\mathcal{H}_j),  \bar{X}_{ii} \ge 0,$
		s.t.  $\Bar{\Phi}_i(\bar{X}_{ii})\!=\!\bar{X}_{ii}$.
		
		Then, given  $X=X_{ii}\oplus \bar{X}_{ii}$  with $X_{ii}=0$,  using again \eqref{eq:blk_hch_i}, we obtain     
		$$\Phi_{\oplus}(X)=\Phi_{i*}(\bar{X}_{ii})\oplus\Bar{\Phi}_i(\bar{X}_{ii})=\Phi_{i*}(\bar{X}_{ii})\oplus \bar{X}_{ii}.$$
		
		Since $\Phi_{\oplus}$ is trace-preserving,
		$$ \tr{(\Phi_{\oplus}( X))}=\tr{(\Phi_{i*}(\bar{X}_{ii}))}+\tr{(\bar{X}_{ii})}=\tr{(\bar{X}_{ii})},$$
		Hence $\operatorname{tr}(\Phi_{i*}(\bar{X}_{ii})) = 0$. As $\Phi_{i*}$ is CP and $\bar{X}_{ii}\geq0$ the above equation implies $\Phi_{i*}(\bar{X}_{ii})=0$. 
		
		Therefore 
		$$\Phi_{\oplus}(X)= X=0\oplus \bar{X}_{ii,}$$
		so $X$ is an eigenoperator of $\Phi_{\oplus}$ with eigenvalue 1. However, $X$ does not have full support on $\Hc$, contradicting the irreducibility of  $\Phi_{\oplus}$.
	\end{proof}
	
	\begin{proof}[Proof of Theorem \ref{thm:hsc_prop} (Properties of the HSC)]
		We focus on $\Sc_{ii}$ for fixed $i$, the proof follows the same lines $\forall$ $i\in\{1,\dots,q\}$. We consider the bipartition of the Hilbert space $\Hc=\Hc_i\oplus\bar{\Hc_i}$ as in equation \eqref{eq:hs_bip}. In the following of the proof,   block-diagonal operators in ${\mathfrak{B}}(\Hc)$ will be decomposed according to the bipartition as $X=X_{ii} \oplus \bar{X}_{ii}$ with $X_{ii}\in {\mathfrak{B}}(\Hc_i)$ and $\bar X_{ii}\in\bigoplus_{j\neq i} {\mathfrak{B}}(\Hc_j)$. In particular  
		since $\Phi$ is irreducible, by Proposition \ref{prop:hc_props}, $\Phi_{\oplus}$ is irreducible as well and the unique stationary state of $\Phi_{\oplus}$ will be decomposed according to the bipartition as $\tilde{\mu}=\tilde \mu_i\oplus \tilde \mu_*$.

		We first prove $\mathcal{S}_{ii}$ is a quantum channel. 
		The complete positivity of $\mathcal{S}_{ii}$ follows directly from its definition \(\eqref{eq:hsc_full}\). Indeed, it is given by the sum and composition of completely positive maps, and both of these operations preserve complete positivity.  To prove that $\Sc_{ii}$ is also TP, we use the fact that a completely positive map is trace-preserving if and only if its adjoint (with respect to the Hilbert--Schmidt inner product) is unital.
		Since $\Phi_{\oplus}^\dag$ is unital, i.e. $\Phi_{\oplus}^\dag(I)=I$, from equation \ref{eq:blk_hch_i} we have
		\begin{align*}
			\Phi_{\oplus}^\dag(I)=(\Phi_{ii}^\dag (I_{ii})+\Phi_{*i}^\dag (\bar{I}_{ii}))\oplus(\Phi_{i*}^\dag(I_{ii})+\Bar{\Phi}_i^\dag (\bar{I}_{ii}))=I_{ii}\oplus\bar{I}_{ii}		,
		\end{align*}
		where $I_{ii},\bar{I}_{ii}$ are the identity operators on $\Hc_i$ and $\bar{\Hc}_i$ respectively.
		It follows from the second block equivalence that 
		\begin{equation}
			(\Ic-{\bar\Phi_i}^\dag)^{-1} \Phi_{i*}^\dag(I_{ii})=\bar{I}_{ii},
		\end{equation}
		where the invertibility of ($\Ic-\bar \Phi_{i}^\dag$) is guaranteed by  Proposition  \ref{prop:spectral_radius}.
		This implies
		\begin{equation*}
			\Sc^\dag_{ii}(I_{ii})=\Phi_{ii}^\dag(I_{ii})+ \Phi_{*i}^{\dag} (\Ic-\bar \Phi_i^\dag)^{-1} \Phi_{i*}^\dag(I_{ii})=\Phi_{ii}^\dag(I_{ii})+ \Phi_{*i}^\dag(\bar{I}_{ii})=I_{ii}.
		\end{equation*}
		Therefore $\Sc_{ii}$ is completely positive and trace preserving, i.e. a quantum channel.
		
		We we now prove $ \tilde \sigma_i$ is a stationary state for $\Sc_{ii}$ and that  $\Sc_{ii}$ is irreducible. To this aim we consider the following invertible maps\footnote{Invertibility is granted by the fact that these linear operators admit a block-triangular matrix representation, whose diagonal blocks are identity maps.} from $\bigoplus_{j} {\mathfrak{B}}(\Hc_j)$ to itself
		\begin{align*}
			\Ec_1(X)&:= (X_{ii} + \Phi_{i*}  (\Ic-\bar \Phi_i)^{-1}(\bar X_{ii}))\oplus\bar X_{ii}\\
			\Ec_2(X)&:=  X_{ii} \oplus ((\Ic-\bar \Phi_i)^{-1} \Phi_{*i}(X_{ii})+\bar{X}_{ii}).
		\end{align*}
		
		From \eqref{eq:blk_hch_i}, for any $X={X}_{ii}\oplus \bar{X}_{ii}$ we have
		\begin{align*}
			(\Ic-\Phi_{\oplus})(X)=
			( X_{ii}-\Phi_{ii}( X_{ii})-\Phi_{i*}( \bar X_{ii}))\oplus
			(\bar X_{ii}-\Phi_{*i}(X_{ii})-\bar\Phi_i( \bar X_{ii} )).
		\end{align*}
		Let $B=\bar X_{ii}- \Phi_{*i}(X_{ii})-\bar \Phi_{i}(\bar X_{ii})$, then
		\begin{align*}
			&\Ec_1(\Ic-\Phi_{\oplus})(X)=\\
			&={X_{ii}\! -\!\Phi_{ii}(X_{ii})\! -\!\Phi_{i*}(\bar X_{ii})}+\Phi_{i*}   (\Ic \! -\!\bar \Phi_{i})^{\! -\!1}({\bar X_{ii}\! -\!\Phi_{*i}(X_{ii})\! -\!\bar \Phi_{i}(\bar X_{ii})}) \oplus B \\
			&=X_{ii}- (\Phi_{ii}(X_{ii})+ \Phi_{i*} (\Ic -\bar \Phi_{i})^{-1}\Phi_{*i}(X_{ii}))+ \Phi_{i*}  (\Ic-\bar \Phi_i)^{-1}(\bar X_{ii}-\!\bar \Phi_{i}(\bar X_{ii}))\oplus B\\
			&=(X_{ii}-\Sc_{ii}(X_{ii})+\Phi_{i*}(\bar X_{ii})-\Phi_{i*}(\bar X_{ii}))\oplus (\bar X_{ii}- \Phi_{*i}(X_{ii})-\bar \Phi_{i}(\bar X_{ii}))\\
			&=(X_{ii}-\Sc_{ii}(X_{ii}))\oplus (\bar X_{ii}- \Phi_{*i}(X_{ii})-\bar \Phi_{i}(\bar X_{ii})).
		\end{align*}
		Since $\tilde{\mu}$ is a stationary state for $\Phi_{\oplus}$, $(\Ic-\Phi_{\oplus})(\tilde\mu)=0$. Therefore substituting $\tilde \mu$ in the previous equation we obtain $\tilde\mu_i-\Sc_{ii}(\tilde\mu_i)=0$, which implies $\tilde \mu_i$ is a fixed point for $\Sc_{ii}$. Therefore, the state obtained by normalizing $\tilde \mu_i$, i.e. $\tilde \sigma_i=\tilde \mu_i/{\rm tr}(\tilde \mu_i)$ is a stationary state for $\Sc_{ii}$. Since $ \Phi_{\oplus}$ is irreducible Lemma \ref{lem:sr_pos_maps} implies $\tilde \mu_i$ and therefore $\tilde \sigma_i$ are of full-rank.
		
		To prove irreduciblity of $\Sc_{ii}$ we let $A=(X_{ii}-\Sc_{ii}(X_{ii}))$, then
		\begin{align*}
			&\Ec_1(\Ic-\Phi_{\oplus})\Ec_2(X)=\\
			&=A\oplus(  (\Ic-\bar \Phi_i)^{-1} \Phi_{*i}(X_{ii})+\bar{X}_{ii}- \Phi_{*i}(X_{ii})-\bar \Phi_{i}(
			(\Ic-\bar \Phi_i)^{-1} \Phi_{*i}(X_{ii}))+\bar \Phi_{i}(\bar{X}_{ii}))\\
			&=A\oplus ((\Ic -\bar \Phi_i)(\Ic-\bar \Phi_i)^{-1} \Phi_{*i}(X_{ii})-\Phi_{*i}+\bar X_{ii} -\bar \Phi_i(\bar X_{ii}))\\
			&=(X_{ii}-\Sc_{ii}(X_{ii}))\oplus (\bar X_{ii} -\bar \Phi_i(\bar X_{ii}))
		\end{align*}
		We let $\Phi_{R}$ be the restriction of $\Phi_{\oplus}$ to $
        \bigoplus_{j} {\mathfrak{B}}(\Hc_j)$, since $\Phi_{\oplus}$ is irreducible and $\tilde \mu\in
        \bigoplus_{j} {\mathfrak{B}}(\Hc_j)$, then $\Phi_R$ has fixed-point space \(\operatorname{span}\{\widetilde\mu\}\). Then from the previous equation
		\begin{align*}
			\dim(\ker(\Ic-\Phi_{R}))\!=\!\dim(\ker({\Ec_1} (\Ic-\Phi_{\oplus}){\Ec_{2}}))\!=\!\dim(\ker(\Ic-{\Sc_{ii}}))+\dim(\ker(\Ic-{\bar \Phi_i}))=1.
		\end{align*}
		By Proposition \ref{prop:spectral_radius} $\dim(\ker(\Ic-{\bar \Phi_i}))=0$, hence  the previous equation implies $\Sc_{ii}$ is a CPTP map with a unique stationary state $\tilde \sigma_i$ which is of full rank. By Lemma \ref{lem:sr_pos_maps}, $\Sc_{ii}$ is irreducible.
	\end{proof}
	\begin{proof}[Proof of Proposition \ref{prop:hsc_primitivity} (Primitivity of HSC)]\label{proof:S_primitivity} Recall the definition of stochastic complement in equation \eqref{eq:stoch_comp}, since $\Phi_{ii}$  and $\Phi_{i*} (\Ic-\bar \Phi_i)^{-1} \Phi_{*i}$ are CP maps, then for every $n \in \mathbb{N}$
		$$\Sc_{ii}^n=(\Phi_{ii})^n+\Ec,$$
		where $\Ec$ is a CP map. Since $\exists n$ s.t. $  (\Phi_{ii})^n(\rho_i)>0$, and  $\Ec(\rho_i)\geq 0 \  \forall$ $\rho_i \in \Dc(\Hc_i)$, then $\Sc_{ii}^n(\rho_i)>0 \ \forall$ $\rho_i \in \Dc(\Hc_i)$. Therefore $\Sc_{ii}$ is primitive.
	\end{proof}
	
	\begin{proof}[Proof of Proposition \ref{prop:block_diag_ss}]
		Let $\widetilde\rho$ be the stationary state of $\Phi$.
		Under Assumption~1, it is block diagonal, so
		\[
		\sum_j \Pc_j(\widetilde\rho)=\widetilde\rho.
		\]
		Using $\Phi_{\oplus}=\sum_{i,j}\Pc_i\Phi\Pc_j$ and
		$\Phi(\widetilde\rho)=\widetilde\rho$, we obtain
		\[
		\begin{aligned}
			\Phi_{\oplus}(\widetilde\rho)
			&=\sum_i \Pc_i\Phi
			\left(\sum_j\Pc_j(\widetilde\rho)\right)
			=\sum_i\Pc_i\Phi(\widetilde\rho)
			=\sum_i\Pc_i(\widetilde\rho)
			=\widetilde\rho.
		\end{aligned}
		\]
		Thus $\widetilde\rho$ is a stationary state of $\Phi_{\oplus}$.
		By Proposition~1, $\Phi_{\oplus}$ is irreducible and therefore has
		a unique stationary state, which consequently coincides
		with $\widetilde\rho$.
	\end{proof}
	
	\begin{proof}[Proof of Theorem \ref{thm:coup_thm} (Hybrid coupling theorem)]
		We first prove that $C$ is a transition probability matrix. Each $\Phi_{ji}$ is a CP map, and by Theorem \ref{thm:hsc_prop} $\tilde \sigma_i\in \Dc(\Hc_i)$ is a full rank state. 
		Together, these facts ensure that $\tr(\Phi_{ji}(\tilde \sigma_i))\geq 0 \ \forall i,j$. Furthermore, since $\Phi_{\oplus}$ is trace-preserving (TP), we have
		$$\tr(\Phi_{\oplus}(\Jc_i
		(\tilde\sigma_i)))=\tr(\sum_j \Jc_j\Phi_{ji} \Jc_i^\dag \Jc_i( \tilde\sigma_i))=\sum_j\tr(\Phi_{ji}(\tilde\sigma_i))=1.$$  
		Hence, each column of $C$ is a probability distribution, i.e. $C$ is a transition probability matrix.
		
		Let $\tilde \mu$ be the stationary state of $\Phi_{\oplus}$, which  by Proposition \ref{prop:block_diag_ss} coincides with that of $\Phi$, we now show $\xi=(\xi_1 \dots \xi_q)^\top,$ with $\xi_i=\tr(\tilde \mu_i)$ and $\tilde \mu_i=\Jc^\dag_i(\tilde \mu)$ is a stationary probability distribution for $C$. 
		
		Fist note that, by exploiting the decomposition of $\Phi$ in Definition \ref{def:original_channel_decomposition}, 
		$$\tilde\mu_j =\Jc_j^\dag  (\tilde \mu)= \Jc_j^\dag\Phi_{\oplus}(\tilde \mu) =\Jc_j^\dag \sum_{i,s} \Jc_s \Phi_{si}\Jc_i^\dag (\tilde \mu)=\sum_i \Phi_{ji}(\tilde \mu_i).$$
		Then, by the definition of $C$, the $j$-th entry of $C\xi$ is given by
		\begin{align}
			[C \xi]_j&=\sum_i c_{ji}\xi_i=\sum_i \tr(\Phi_{ji}(\xi_i \tilde \sigma_i))=\tr(\sum_i\Phi_{ji}(\tilde \mu_i))=\tr(\tilde \mu_j)=\xi_j,
		\end{align}
		therefore $C\xi=\xi$, i.e. $\xi$ is a stationary distribution for the Markov Chain with transition probability matrix $C$.
		
		We now prove that the irreducibility of $\Phi_{\oplus}$ implies irreducibility of $C$. 
		By contradiction suppose $\Phi_{\oplus}$ is irreducible and $C$ is not. 
		$C$ is reducible if and only if $\exists \ \mathcal{K} \subseteq\{1,\dots q\}$ such that $c_{ji}=0 \ \forall i\in \mathcal{K}, j\in \mathcal{\overline{K}}=\{1,\dots, q\} \setminus \Kc$. 
		Since the stochastic complements $\Sc_{ii} \ i\in\{1,\dots,q\}$ are irreducible (Theorem \ref{thm:sc_prop}), by Lemma \ref{lem:sr_pos_maps} their stationary states $\tilde \sigma_{i}$ are positive definite, moreover the $\Phi_{ij}$ $ i,j\in\{1,\dots,q\}$ are CP. Then $c_{ji}=\tr(\Phi_{ji}(\tilde \sigma_i))=0$ $\forall i\in \mathcal{K}, j\in \mathcal{\overline{K}}$ , implies $\Phi_{ji}=0$  $\forall i\in \mathcal{K}, j\in \mathcal{\overline{K}}$. Let $\Pi_\Kc$ be the projector on $\bigoplus_{i\in \Kc}\Hc_i$, then $\Phi_{\oplus}(\Pi_\Kc{\mathfrak{B}}(\Hc) \Pi_\Kc)\subseteq \Pi_\Kc {\mathfrak{B}}(\Hc) \Pi_\Kc$ which implies $\Phi_{\oplus}$ is not irreducible, leading to the contradiction.
	\end{proof}
	
	\section{Proofs of results on nearly completely reducible/strongly reducible\\ systems and metastability}\label{app:proofs_metastability}
	\begin{proof}[Proof of Proposition\! \ref{prop:cs_dec}\! (Decomposition of completely/strongly reducible channels)]
		
		We first prove  \textup{(i)} $\Leftrightarrow$ \textup{(ii)} . Let $\{K_\alpha\}$ be an OSR for the quantum
		channel {$\Phi$}. Fix a bipartition
		$\mathcal H=\mathcal H_i\oplus\overline{\mathcal H}_i,
		\overline{\mathcal H}_i=\bigoplus_{j\neq i}\mathcal H_j.$ $\Hc_i$ is invariant for
		{$\Phi$} if and only if
		\cite[Proposition 1]{IC-FT:15} 
		\[
		\Pi_iK_\alpha\Pi_i
		+\Pi_iK_\alpha\overline{\Pi}_i
		+\overline{\Pi}_iK_\alpha\overline{\Pi}_i
		=K_\alpha,\qquad
		\]
		for every $\alpha$, where $\Pi_i$ and $\overline{\Pi}_i$ are the
		orthogonal projectors onto $\mathcal H_i$ and
		$\overline{\mathcal H}_i$, respectively. Since for a completely reducible channel invariance holds for all $\Hc_i$ the above condition yields. 
		\[
		K_\alpha=\sum_{i=1}^q\Pi_iK_\alpha\Pi_i,
		\qquad \forall\alpha.
		\]
		We now prove \textup{(i)} $\Rightarrow$ \textup{(iii)}. Let
		$X=X_D+X_O\in\Bf(\mathcal H),$
		where
		\[
		X_D=\sum_i\Pi_iX\Pi_i
		\in\bigoplus_i\Bf(\mathcal H_i),
		\qquad
		X_O=\sum_{i\neq j}\Pi_iX\Pi_j
		\in
		\Bf(\mathcal H)\ominus
		\bigoplus_i \Bf(\mathcal H_i).
		\]
		Using the block-diagonal form of the Kraus operators from \textup{(i)}, we obtain
		\[
		\begin{aligned}
			{\Phi}(X_D)
			=
			\sum_{\alpha,i}
			\Pi_iK_\alpha\Pi_i
			X_D
			\Pi_iK_\alpha^\dagger\Pi_i
			=
			\sum_i
			\mathcal P_i{\Phi}\mathcal P_i(X_D)=
			\sum_i
			\mathcal J_i\Phi_{ii}\mathcal J_i^\dagger(X_D),
		\end{aligned}
		\]
		where $\Phi_{ii}:=\mathcal J_i^\dagger{\Phi}\mathcal J_i.$
		Similarly,
		\[
		\begin{aligned}
			{\Phi}(X_O)
			&=
			\sum_{\alpha}
			\sum_{i\neq j}
			\Pi_iK_\alpha\Pi_i
			X_O
			\Pi_jK_\alpha^\dagger\Pi_j
			&=
			\sum_{i\neq j}
			\mathcal P_{ij}{\Phi}\mathcal P_{ij}(X_O).
		\end{aligned}
		\]
		Hence
		\[
		{\Phi}=\Phi_{\rm cr}+\Phi_{\rm oo},
		\]
		with
		\[
		\Phi_{\rm cr}=
		\sum_i\mathcal J_i\Phi_{ii}\mathcal J_i^\dagger,
		\qquad
		\Phi_{\rm oo}=
		\sum_{i\neq j}
		\mathcal P_{ij}{\Phi}\mathcal P_{ij}.
		\]
		
		Finally we prove \textup{(iii)} $\Rightarrow$ \textup{(i)} note that whenever \textup{(iii)} holds 
		\begin{align*}
			{\Phi}\bigl(\Pi_i{\mathfrak{B}}(\mathcal H)\Pi_i\bigr)&=\Phi_{\rm cr}(\Pi_i{\mathfrak{B}}(\mathcal H)\Pi_i)
			=\sum_j \Jc_j \Phi_{jj} \Jc_j^\dag(\Pi_i{\mathfrak{B}}(\Hc)\Pi_i)\\&=\Jc_i\Phi_{ii}\Jc_i^\dag(\Pi_i{\mathfrak{B}}(\Hc)\Pi_i)\subseteq \Pi_i{\mathfrak{B}}(\Hc)\Pi_i \ \forall i.
		\end{align*}
		Therefore {$\Phi$} is completely reducible, i.e. \textup{(i)} holds.
		
		We now prove 
		\textup{(iv)} $\Rightarrow$ \textup{(vi)}.
		If {$\Phi$} is strongly reducible, then it is, in particular,
		completely reducible. Hence it admits the decomposition in
		\textup{(iii)}. Moreover, by the definition of strong reducibility,
		\[
		{\Phi}(X_O)=0
		\qquad \forall \ 
		X_O\in
		\Bf(\mathcal H)\ominus
		\bigoplus_{i=1}^q\Bf(\mathcal H_i).
		\]
		On this subspace, the decomposition in \textup{(iii)} gives
		\[
		{\Phi}(X_O)=\Phi_{\rm oo}(X_O).
		\]
		Therefore $\Phi_{\rm oo}(X_O)=0$
		for every off-diagonal operator $X_O$, and hence $
		\Phi_{\rm oo}=0.$

		\textup{(vi)} $\Rightarrow$ \textup{(v)}.
		Assume that $\Phi_{\rm oo}=0$. Then
		\[
		{\Phi}
		=
		\Phi_{\rm cr}
		=
		\sum_{i=1}^q
		\mathcal J_i\Phi_{ii}\mathcal J_i^\dagger.
		\]
		For each $i$, choose an OSR $\{L_{i,\beta}\},$  for the CP map 
		$\Phi_{ii}$, where $L_{i,\beta}\in\Bf(\mathcal H_i) \ \forall \beta$ and 
		define the operators 
		\[
		K_{i,\beta}:=\Jc_i(L_{i,\beta})\in {\mathfrak{B}}(\Hc).
		\]
		Each of these operators is supported on a single diagonal block, i.e. 
		$
		K_{i,\beta}
		=
		\Pi_iK_{i,\beta}\Pi_i$ and $
		\Pi_jK_{i,\beta}\Pi_j=0$ for $j\neq i$.
		Moreover,
		\[
		\begin{aligned}
			\sum_{i,\beta}
			K_{i,\beta}XK_{i,\beta}^\dagger
			&=
			\sum_i
			\mathcal J_i
			\left(
			\sum_\beta
			L_{i,\beta}\,
			\mathcal J_i^\dagger(X)\,
			L_{i,\beta}^\dagger
			\right)
			&=
			\sum_i
			\mathcal J_i\Phi_{ii}\mathcal J_i^\dagger(X)
			&=
			{\Phi}(X).
		\end{aligned}
		\]
		Thus $\{K_{i,\beta}\}_{i,\beta}$ is an OSR of {$\Phi$} in which each
		Kraus operator has support on a single diagonal block.

		\textup{(v)} $\Rightarrow$ \textup{(iv)} Let $\{K_\alpha\}$ be an OSR for {$\Phi$} satisfying \textup{(v)}, then the Kraus operators are block diagonal with respect to the Hilbert space partition, \textup{(ii)} holds and {$\Phi$} is completely reducible. Now let $X_{ij}=\Pi_iX\Pi_j, i\neq j,$ by \textup{(v)} for every $\alpha$ there exists $s\in \{1,\dots q\}$ such that,
		\[
		K_\alpha X_{ij}K_\alpha^\dagger
		=
		\Pi_sK_\alpha\Pi_s
		\Pi_iX\Pi_j
		\Pi_sK_\alpha^\dagger\Pi_s
		=0.
		\] 
		Hence
		\[
		{\Phi}(X_{ij})=0
		\qquad
		\text{for every }i\neq j.
		\]
		By linearity, {$\Phi$} annihilates the entire off-diagonal operator
		subspace. Thus {$\Phi$} is strongly reducible.
	\end{proof}

    \begin{proof}[Proof of Proposition \ref{prop:hybrid-complete-reducibility}\! (Complete reducibility and associated hybrid channel)]
			Let $\Pi_i$ denote the orthogonal projector onto $\mathcal H_i$,
			and let $\{K_\alpha\}_\alpha$ be an arbitrary operator-sum
			representation of $\Phi$. By Definition~\ref{def:original_channel_decomposition},
			\[
			\Phi_{\oplus}(X)
			=\sum_{i,j=1}^{q}\Pi_j\Phi(\Pi_i X\Pi_i)\Pi_j
			=\sum_{\alpha,i,j}L_{\alpha,ji}X L_{\alpha,ji}^{\dagger},
			\qquad
			L_{\alpha,ji}:=\Pi_jK_\alpha\Pi_i.
			\]
			Thus $\{L_{\alpha,ji}\}_{\alpha,i,j}$ is an operator-sum
			representation of $\Phi_{\oplus}$.
			Suppose first that $\Phi$ is completely reducible with respect
			to the given partition. By Proposition~\ref{prop:cs_dec}, each $K_\alpha$ is
			block diagonal. Hence $L_{\alpha,ji}=0$ whenever $j\ne i$, and
			\[
			\Phi_{\oplus}(X)=\sum_{\alpha,i}
			(\Pi_iK_\alpha\Pi_i)X(\Pi_iK_\alpha\Pi_i)^\dagger.
			\]
			It follows that each $\mathcal H_i$ is invariant under $\Phi_{\oplus}$,
			so $\Phi_{\oplus}$ is completely reducible with respect to the same
			partition.
			
			Conversely, suppose that $\Phi_{\oplus}$ is completely reducible with
			respect to this partition. By Proposition~\ref{prop:cs_dec}, every Kraus operator
			in every operator-sum representation of $\Phi_{\oplus}$ is block diagonal, that is
			
			$$L_{\alpha,ji}=\Pi_j K_\alpha \Pi_i=0 \qquad {\rm for} \qquad j\neq i.$$
			Consequently,
			\[
			K_\alpha
			=\sum_{i,j=1}^{q}\Pi_jK_\alpha\Pi_i
			=\sum_{i=1}^{q}\Pi_iK_\alpha\Pi_i
			\qquad\text{for every }\alpha.
			\]
			Thus each $\mathcal H_i$ is invariant under $\Phi$, proving
			complete reducibility with respect to the given partition.
			Finally, the defining expression for $\Phi_{\oplus}$ shows that
			$\Phi_{\oplus}(X)=0$ for every $X\in{\mathfrak{B}}(\Hc)\ominus(\bigoplus_i{\mathfrak{B}}(\Hc_i))$.
			Therefore, whenever $\Phi_{\oplus}$ is completely reducible, it is also
			strongly reducible.
	\end{proof}
    {
	\begin{proof}[Proof of Theorem \ref{thm:qs_dev_comp_red} (Deviation from reducibility)]
    Since $\Phi$ is an irreducible quantum channel, by Theorem 1
the HSC $S_{ii}$ is a quantum channel for all
$i\in\{1,\ldots,q\}$. Moreover, by definition of the HSC,
\[
S_{ii}-\Phi_{ii}
=\Phi_{i*}(I-\bar\Phi_i)^{-1}\Phi_{*i}
=\sum_{k=0}^{\infty}\Phi_{i*}\bar\Phi_i^k\Phi_{*i},
\]
which is CP. Now, consider a density operator
$\rho_i\in\mathcal D(\mathcal H_i)$.
Since $S_{ii}$ is trace preserving, we have
\[
\operatorname{tr}[(S_{ii}-\Phi_{ii})(\rho_i)]
=\operatorname{tr}[\rho_i]
-\operatorname{tr}[\Phi_{ii}(\rho_i)].
\]

Moreover, since $\Phi_\oplus$ admits the 
block decomposition 
in equation (6) and is TP, we have
\[ 
\Phi_\oplus(\rho_i\oplus 0)
=\Phi_{ii}(\rho_i)\oplus\Phi_{*i}(\rho_i), \qquad \operatorname{tr}[\rho_i]
=\operatorname{tr}[\Phi_{ii}(\rho_i)]
+\operatorname{tr}[\Phi_{*i}(\rho_i)].
\]
for every $\rho_i\in \Bf(\Hc_i).$ Combining the above identities, we obtain
\[
\operatorname{tr}[(S_{ii}-\Phi_{ii})(\rho_i)]
=\operatorname{tr}[\Phi_{*i}(\rho_i)].
\]

Finally, since both $S_{ii}-\Phi_{ii}$ and $\Phi_{*i}$
are completely positive maps, their induced trace norms
are attained on density operators (see Proposition \ref{prop:russo_dye} in Appendix \ref{app:norms}). Therefore,
\begin{equation}
\begin{aligned}
\|S_{ii}-\Phi_{ii}\|_1
&=\max_{\rho_i\in\mathcal D(\mathcal H_i)}
\operatorname{tr}[(S_{ii}-\Phi_{ii})(\rho_i)]
=\max_{\rho_i\in\mathcal D(\mathcal H_i)}
\operatorname{tr}[\Phi_{*i}(\rho_i)]
=\|\Phi_{*i}\|_1.
\end{aligned}
\label{eq:HSC_norm_equality}
\end{equation}
This establishes the first equality in the statement of the theorem. 

Both $\Phi_{\oplus}$ and $\mathcal S$ annihilate off-diagonal input blocks  therefore the induced superoperator norm of $\mathcal T:=\Phi_{\oplus}-\mathcal S$ is obtained on on block-diagonal operators.
		
		Let $X\in\mathcal {\mathfrak{B}}(\mathcal H)$ satisfy $\|X\|_\infty=1$ and write
		\[
		\mathcal W_{\mathrm c}(X)=\bigoplus_{i=1}^q X_{ii},
		\qquad
		X_{\bar i}:=\bigoplus_{j\ne i}X_{jj}.
		\]
        Since  $\mathcal W_{\mathrm c}$ is a unital completely positive map 
		 $\|\mathcal W_{\mathrm c}\|_\infty=1$ (see Proposition \ref{prop:russo_dye} in Appendix \ref{app:norms}) and 
		\[
		\|X_{ii}\|_\infty\leq 1,
		\qquad
		\|X_{\bar i}\|_\infty\leq 1
		\quad\text{for every }i.
		\]
		Using the block form of $\Phi_{\oplus}^\dagger$ from \eqref{eq:blk_hch_i} it follows that  the $i$-th diagonal
		block of $\mathcal T^\dagger(X)$ is
		\[
		\bigl(\mathcal T^\dagger(X)\bigr)_{ii}
		= (\Phi_{ii}^\dagger-\mathcal S_{ii}^\dagger)(X_{ii})
		+\Phi_{\ast i}^\dagger(X_{\bar i}).
		\]
		Therefore,
		\begin{align*}
			\left\|\bigl(\mathcal T^\dagger(X)\bigr)_{ii}\right\|_\infty
			&\leq
			\|\Phi_{ii}^\dagger-\mathcal S_{ii}^\dagger\|_\infty
			\|X_{ii}\|_\infty
			+\|\Phi_{\ast i}^\dagger\|_\infty
			\|X_{\bar i}\|_\infty
			\leq
			\|\mathcal S_{ii}-\Phi_{ii}\|_1
			+\|\Phi_{\ast i}\|_1
			=2\|\Phi_{\ast i}\|_1.
		\end{align*}
		Since $\mathcal T^\dagger(X)$ is block diagonal, its operator norm is
		the maximum of the operator norms of its diagonal blocks.  Thus
		\[
		\|\mathcal T^\dagger(X)\|_\infty
		\leq 2\max_{i\in\{1,\ldots,q\}}\|\Phi_{\ast i}\|_1.
		\]
		Taking the supremum over $\|X\|_\infty=1$ and using duality (Proposition \ref{prop:norm_duality} in Appendix \ref{app:norms}) once more
		yields
		\begin{equation}\label{eq:delta_h_upper}
		    \|\Phi_{\oplus}-\mathcal S\|_1
		=\|\mathcal T^\dagger\|_\infty
		\leq 2\max_{i\in\{1,\ldots,q\}}\|\Phi_{\ast i}\|_1
		=:\delta_h.
		\end{equation}
        
    Furthermore, since $\Sc_{ii}-\Phi_{ii}$ and $\Phi_{*i}$
are completely positive, for every density operator
$\rho_i\in\mathcal D(\Hc_i)$ exploiting \eqref{eq:HSC_norm_equality} we have
\begin{align*}
\norm{(\Phi_{\oplus}-\mathcal S)
(\Jc_i(\rho_i))}_1
&=\norm{((\Phi_{ii}-\Sc_{ii})(\rho_i))
\oplus(\Phi_{*i}(\rho_i))}_1\\
&=\tr[(\Sc_{ii}-\Phi_{ii})(\rho_i)]
+\tr[\Phi_{*i}(\rho_i)]
=2\tr[\Phi_{*i}(\rho_i)].
\end{align*}

Taking the maximum
over density operators and then over
$i\in\{1,\ldots,q\}$, we obtain
\begin{align*}
\norm{\Phi_{\oplus}-\mathcal S}_1
&\geq \max_i
\max_{\rho_i\in\mathcal D(\Hc_i)}
\norm{(\Phi_{\oplus}-\mathcal S)
(\Jc_i(\rho_i))}_1
=2\max_i\norm{\Phi_{*i}}_1
=\delta_h.
\end{align*}

Combining this inequality with
\eqref{eq:delta_h_upper}, we conclude that
\[
\norm{\Phi_{\oplus}-\mathcal S}_1
=2\max_{i\in\{1,\ldots,q\}}\norm{\Phi_{*i}}_1
=\delta_h.
\]

Finally, we note that
$\Phi=\Phi_{\oplus}+\Phi_{\Delta}$ and the identity
$\mathcal W_x\Phi_{\oplus}=\Phi_{\oplus}$ imply
\begin{align*}
\norm{\mathcal W_x\Phi-\mathcal S}_1
&\leq
\norm{\mathcal W_x\Phi_{\oplus}-\mathcal S}_1
+\norm{\mathcal W_x\Phi_{\Delta}}_1
=\norm{\Phi_{\oplus}-\mathcal S}_1
+\norm{\mathcal W_x\Phi_{\Delta}}_1
=\delta_h+\norm{\mathcal W_x\Phi_{\Delta}}_1.
\end{align*}

\end{proof}}
	\begin{proof}[Proof of proposition \ref{prop:rel_deviations} (Relationship between deviations from complete reducibility)]
		
		To prove the first bound we exploit the fact that $\WC$ is a CPTP map and its adjoint with respect to the Hilbert-Schmidt inner product is unital.
			By the properties of the superoperator norms in Appendix \ref{app:norms}, we get
			\begin{align*}
				\delta_c=\delta_h+\norm{\WC \Phi_{\Delta}}_1\leq \delta_h+\norm{\WC^\dag(\one)}_\infty \norm{\Phi_{\Delta}}_1=\delta_h+\norm{\Phi_{\Delta}}_1.
			\end{align*}
			 Let us consider an OSR for $\Phi(\cdot)=\sum_{l=1}^r K_{l}\cdot K_{l}^\dag$, where $r$ is the Kraus rank for $\Phi$.
             
             To prove the second bound, we first establish an instrumental result, that is $\norm{K_{ji,l}}_\infty^2\leq \delta_h/2,$ if $i\neq j.$
            Let  $K_{ji,l}=\Pi_j K_l \Pi_i,$  then, by the properties of the superoperator norms in Appendix \ref{app:norms}, for any $i\neq j$ we get
			
				\begin{align*}
					\norm{K_{ji,l}}_\infty^2& =\sup_{\norm{v}=1}\bra{v}K_{ji,l}^\dag K_{ji,l}\ket{v}\leq \sup_{\norm{v}=1}\Bigl| \bra{v}\sum_{j\neq i}\sum_l K_{ji,l}^\dag K_{ji,l}\ket{v}\Bigr|,
				\end{align*}
                where \(\|v\|_2=\sqrt{\langle v\mid v\rangle}\) is the Eucledean norm for vectors in $\Hc$. 
				Next, by the Cauchy-Schwarz inequality, we have
				\begin{align*}
					&\sup_{\norm{v}=1}\Bigl| \bra{v}\sum_{j\neq i}\sum_l K_{ji,l}^\dag K_{ji,l}\ket{v}\Bigr| \leq \sup_{\norm{v}=1} \norm{v} \norm{(\sum_{j\neq i}\sum_l K_{ji,l}^\dag K_{ji,l}) \ket{v}}.
				\end{align*}
				By exploiting the definition of operator norm and the quantum channel decomposition in Definition \ref{def:original_channel_decomposition}, we can then obtain
				\begin{align*}    
					\sup_{\norm{v}=1} \norm{v} \norm{(\sum_{j\neq i}\sum_l K_{ji,l}^\dag K_{ji,l}) \ket{v}}
					&\leq  \sup_{\norm{v}=1} \norm{v} \norm{\sum_{j\neq i} \sum_{l=1} K_{ji,l}^\dag K_{ji,l}}_\infty \norm{v}\\
					&=
					\norm{\sum_{j\neq i} \sum_{l=1} K_{ji,l}^\dag K_{ji,l}}_\infty=\norm{\sum_{j\neq i} \Jc_i \Phi_{ji}^\dag\Jc_j^\dag(I)}_\infty.
				\end{align*}
				Finally, exploiting again the norm properties and using the definition of the map $\Phi_{*i}$ in Section \ref{sec:hsc_intro}, we get 
				\begin{align*}
					&\norm{\sum_{j\neq i} \Jc_i \Phi_{ji}^\dag\Jc_j^\dag(I)}_\infty=\norm{\sum_{j\neq i} \Jc_j \Phi_{ji}\Jc_i^\dag}_1
					=\sup_{\substack{\norm{X}_1=1\\X\in {\mathfrak{B}}(\Hc_i)\oplus0_R }}\norm{\sum_{j\neq i} \Jc_j \Phi_{ji}\Jc_i^\dag(X)}_1=\\& \sup_{\substack{\norm{X}_1=1\\X\in{\mathfrak{B}}(\Hc_i)}}\norm{\bigoplus_{j\neq i}\Phi_{ji}(X)}_1= 
					\norm{\Phi_{*i}}_1\leq\delta_h/2.
				\end{align*}	
			
			Then, in order to obtain the second bound, we can first write:
			\begin{align*}
				\norm{\mathcal W_c\Phi_{\Delta}}_1=\norm{\mathcal (\WC\Phi_{\Delta})^\dag}_\infty
				&=\sup_{\norm{X}_\infty=1}\norm{\sum_{i,j=1}^q\sum_{k\neq j}\sum_{l=1}^r
					K_{ij,l}^\dag X K_{ik,l}}_\infty\\&\leq
				\sum_{i,j=1}^q\sum_{ k\neq j}\sum_{l=1}^r
				\norm{K_{ij,l}}_\infty\norm{K_{ik,l}}_\infty.
			\end{align*}
			Because $ k\neq j$, for each summand at least one between $K_{ij,l}$ and
			$K_{ki,l}$ is such that $i\neq j$ or $k\neq i$ respectively. Moreover there are $q^2(q-1)r$ summands. Combining these two facts with the previously established  bound $\|K_{ij,l}\|_\infty\leq\sqrt{\delta_h/2}$
			whenever $i\neq j$, and the more general bound $\|K_{ij,l}\|_\infty\leq1$ for all $i,j,l$, we obtain the inequality of the statement
			\[\delta_c=\delta_h+
			\|\mathcal W_c\Phi_{\Delta}\|_1
			\leq\delta_h + r q^2(q-1)\sqrt{\frac{\delta_h}{2}}.
			\]

	\end{proof}
	
	\begin{proof}[Proof of Proposition \ref{prop:qc_short_run} (Short-run evolution and stabilization)]
		We first prove by induction that the following identity holds $\forall t>0$
		
		\begin{align}\label{eq:power_identity}
			\Phi^t-\Sc^t=\sum_{j=0}^{t-1} \Sc^{t-1-j} (\Phi-S)\Phi^j.
		\end{align}
		Note that for $t=1$ the equality is trivially satisfied. Now suppose the equality is true for $t$, we prove it holds for $t+1$. We have  
		
		\begin{align*}
			\Phi^{t+1}-\mathcal S^{t+1}
			&= (\Phi^t-\mathcal S^t)\Phi+\mathcal S^t(\Phi-\mathcal S) = \sum_{j=0}^{t-1}\mathcal S^{t-1-j}(\Phi-\mathcal S)\Phi^{j+1}
			+\mathcal S^t(\Phi-\mathcal S) \\&= \sum_{j=0}^{t}\mathcal S^{t-j}(\Phi-\mathcal S)\Phi^j,
		\end{align*}
		which proves the claim.
		
		Since both $\Phi$ and $\Sc$ are CP maps and their adjoints are unital, by exploting the proerties of the norms introduced in Appendix \ref{app:norms} we have
		$$\norm{\Phi^j}_1=\norm{\Phi^{j\dag}(\one)}_\infty =1 \quad {\rm and } \quad \norm{S^j}_1=\norm{S^{j\dag}(\one)}_\infty=1\qquad \forall j\in \mathbb{N}.$$
		Therefore, exploiting the two above facts, the relation $\Wc_x \Sc=\Sc \Wc_x=\Sc$ and \eqref{eq:power_identity} we conclude that 
		\begin{align*}
			&\norm{\Wc_x\Phi^t-\Sc^t}_1=\norm{\Wc_x(\Phi^t-\Sc^t)}_1=\norm{\Wc_x \sum_{j=0}^{t-1} \Sc^{t-1-j} (\Phi-S)\Phi^j}_1\\
			&= \norm{\sum_{j=0}^{t-1} \Sc^{t-1-j}\Wc_x (\Phi-S)\Phi^j}_1 \leq\sum_{j=0}^{t-1}\underbrace{ \norm{\Sc^{t-1-j}}_1}_{=1} \norm{\Wc_x(\Phi-\Sc)}_1\underbrace{\norm{\Phi^j}_1}_{=1}\leq t \delta_{x}.
		\end{align*}
		In view of the diagonalizability assumption
		$$\Sc=\Zc\Lambda\Zc^{-1}, \qquad \Lambda(F_i)=\lambda_iF_i,$$
		where w.l.o.g the $\{\lambda_i\}_{i=1}^{D^2}$ are the eigenvalues of $\Sc$ ordered so that
		$|\lambda_1|\geq |\lambda_2|\geq\cdots\geq|\lambda_{D^2}|.$
		Since the $\Sc_{ii}$ are  primitive $\forall i$, $\Sc$ has exactly $q$ eigenvalues equal to 1, all remaining eigenvalues have modulus strictly smaller than 1. Consequently $\Sc^t$ has eigenvalues $\lambda_1^t=\lambda_2^t=\dots=\lambda_q^t=1$, the remaining eigenvalues are $\{\lambda_i^t\}_{i=q+1}^{D^2}$ such that $|\lambda_{q+1}|^t\geq |\lambda_{q+2}|^t\geq \dots\geq |\lambda_{D^2}|^t$. Moreover, ${S}^{\infty}={\rm lim}_{n\rightarrow \infty}{S}^t$ has eigenvalues $\{\lambda_i\}_{i=1}^q$ equal to 1 and $\{\lambda_i\}_{i=q+1}^{D^2}$ equal to 0.
		In view of the above facts 
		$${\Sc}^{t}-{\Sc}^{\infty}=\Zc \Dc \Zc^{-1}, $$
		where $\Dc$ is a  linear map diagonal in the same some operator basis of before such that 
		$$\Dc(F_i)=0\  {\rm for }\ i\in \{1,\dots q\} \qquad \vee \qquad \Dc(F_i)=\lambda_i^t F_i\  {\rm for }\ i\in \{q+1,\dots D^2\}.$$        
		We then exploit 
		the relation between super operator norms (Lemma \ref{lem:relation_between_superop_norms}) and Proposition \ref{prop:diagonalizable-map} in Appendix \ref{app:norms} to obtain 
		\begin{align*}
			\norm{\Sc^{t}-\Sc^{\infty}}_1&\leq \sqrt{D}\norm{\Sc^{t}-\Sc^{\infty}}_2\leq \sqrt{D} \kappa |\lambda_{q+1}|^t.
		\end{align*}
		where $\kappa=\norm{\Zc}_2\norm{\Zc^{-1}}_2$ 
		and we recall $\lambda_{q+1}$ is the eigenvalue of $\Sc$ different from 1 with largest absolute value.
		We conclude by observing that
		\begin{align*}
			\norm{\Wc_x\Phi^t\!-\!\Sc^{\infty}}_1\!&=\!\norm{\WCS\Phi^t\!-\!\Sc^t+\Sc^t\!-\!\Sc^{\infty}}_1\!\\
			&\leq\!  \norm{\Wc_x \Phi^t\!-\!\Sc^t}_1+\norm{\Sc^t\!-\!\Sc^{\infty}}_1 \leq t \delta_{x} +\sqrt{D}\kappa |\lambda_{q+1}|^t,
		\end{align*}
		and  
		\begin{align*}
			\norm{\Wc_x(\rho_t)-\sigma_\infty}_1&=\norm{\Wc_x\Phi^t(\rho_0)-S^\infty(\rho_0)}_1
			&\leq
			\norm{\Wc_x\Phi^t-S^\infty}_1\leq t \delta_{x} +\sqrt{D} \kappa |\lambda_{q+1}|^t.
		\end{align*}
	\end{proof}
	
	{\begin{proof}[Proof of Proposition \ref{prop:qc_middle_run} (Middle-run evolution)]
    Fix \(h\in \Ic_x(\epsilon)\), for any integer \(t>\sup \Ic_x(\epsilon)\), set \(r=t-h>0\).
    Then
    \[
        \rho_t=\Phi^h(\rho_r).
    \]
    For each $i=1,\ldots,q$, define
    \[
        \mu_{r,i}:=\Jc_i^\dag(\rho_r),
        \qquad
        \beta_{t,i}^{(h)}
        :=\tr[\mu_{r,i}]
        =\tr\!\left[
            \Jc_i^\dag\!\left(\Phi^{t-h}(\rho_0)\right)
        \right].
    \]
    By primitivity of each HSC,
    \[
        \Sc_{ii}^{\infty}(X)
        :=\lim_{m\to\infty}\Sc_{ii}^{m}(X)
        =\tr[X]\tilde{\sigma}_i,
        \qquad X\in\mathfrak{B}(\Hc_i).
    \]
    Consequently, setting
    \begin{align*}
        \upsilon_t
        :=\Sc^{\infty}(\rho_r)
        =\sum_{i=1}^q
            \Jc_i\!\left(\Sc_{ii}^{\infty}(\mu_{r,i})\right)
        =\sum_{i=1}^q\Jc_i\!\left(\beta_{t,i}^{(h)}\tilde{\sigma}_i\right),
    \end{align*}
    we obtain
    \begin{align*}
        \norm{\Wc_x(\rho_t)-\upsilon_t}_1
        =
        \norm{
            \left(\Wc_x\Phi^h-\Sc^{\infty}\right)(\rho_r)
        }_1\leq
        \norm{\Wc_x\Phi^h-\Sc^{\infty}}_1
        \norm{\rho_r}_1
        <\epsilon,
    \end{align*}
    where the last inequality follows from $h\in\Ic_x(\epsilon)$.
    
To prove the estimate for the coefficients, fix \(t\geq h\) and write
\(r=t-h\) as above. For each \(i\), the block population \(\mu_{r,i}=\Jc^\dag(\rho_r)\)
is Hermitian positive semidefinite. Since \(\Phi_{*i}\) is completely positive,
\(\Phi_{*i}(\mu_{r,i})\) is also hermitian positive semidefinite. Their trace
norms therefore equal their traces, and the definition
\(\delta_h=2\max_j\|\Phi_{*j}\|_1\) gives
\[
  \operatorname{tr}[\Phi_{*i}(\mu_{r,i})]
  = \|\Phi_{*i}(\mu_{r,i})\|_1
  \le \|\Phi_{*i}\|_1\,\|\mu_{r,i}\|_1
  \le \frac{\delta_h}{2}\operatorname{tr}[\mu_{r,i}].
\]
By trace preservation of $\Phi_\oplus$
\[
  \operatorname{tr}[\Phi_{ii}(\mu_{r,i})]
  = \operatorname{tr}[\mu_{r,i}]
    - \operatorname{tr}[\Phi_{*i}(\mu_{r,i})]
  \ge \left(1-\frac{\delta_h}{2}\right)\beta_{t,i}^{(h)},
\]
where \(\beta_{t,i}^{(h)}=\operatorname{tr}[\mu_{r,i}]\).
Since the contributions from the other blocks are nonnegative
\[
  \begin{aligned}
  \sum_j \operatorname{tr}[\Phi_{ij}(\mu_{r,j})]
  = \operatorname{tr}[\Phi_{ii}(\mu_{r,i})]
     + \sum_{j\ne i}\operatorname{tr}[\Phi_{ij}(\mu_{r,j})]
  \ge \operatorname{tr}[\Phi_{ii}(\mu_{r,i})]
   \ge \left(1-\frac{\delta_h}{2}\right)\beta_{t,i}^{(h)}.
  \end{aligned}
\]

Recall that the entries of the coupling matrix $C$ defined in Theorem \ref{thm:coup_thm} are $c_{ji}=\tr[\Phi_{ji}(\tilde \sigma_i)]$.
Following the same reasoning as above we get 
\(c_{ii}=\operatorname{tr}[\Phi_{ii}(\tilde\sigma_i)]
\ge 1-\delta_h/2\), since \(\operatorname{tr}[\tilde\sigma_i]=1\) which implies 
\[
  \begin{aligned}
  (C\beta_t^{(h)})_i
  =c_{ii}\beta_{t,i}^{(h)}+\sum_{j\ne i}c_{ij}\beta_{t,j}^{(h)}
  \ge c_{ii}\beta_{t,i}^{(h)}
   \ge (1-\delta_h/2)\beta_{t,i}^{(h)}.
  \end{aligned}
\]

By trace preservation of $\Phi_\oplus$
it follows that
\[
  \sum_i\sum_j \operatorname{tr}[\Phi_{ij}(\mu_{r,j})]
  = \sum_j \operatorname{tr}[\mu_{r,j}] = 1,
  \qquad
  \sum_i (C\beta_t^{(h)})_i
  = \sum_j \beta_{t,j}^{(h)} = 1.
\]
Then
\[
  \begin{aligned}
  \sum_i \left|
    \sum_j \operatorname{tr}[\Phi_{ij}(\mu_{r,j})]
    - (C\beta_t^{(h)})_i
  \right|
  &\le \sum_i \left(
    \sum_j \operatorname{tr}[\Phi_{ij}(\mu_{r,j})]
    + (C\beta_t^{(h)})_i
    - 2(1-\delta_h/2)\beta_{t,i}^{(h)}
  \right) \\
  &= 1+1-2(1-\delta_h/2)
   = \delta_h.
  \end{aligned}
\]

Finally, since \(\rho_{r+1}=\Phi(\rho_r)\) and
\(\Phi=\Phi_\oplus+\Phi_\Delta\), we have
\[
  \beta_{t+1,i}^{(h)}
  = \operatorname{tr}[\mathcal J_i^\dagger(\Phi(\rho_r))]
  = \sum_j \operatorname{tr}[\Phi_{ij}(\mu_{r,j})]
    + \operatorname{tr}[\mathcal J_i^\dagger(\Phi_\Delta(\rho_r))].
\]
This combined with the bounds proved above yields
\[
  \begin{aligned}
  \|\beta_{t+1}^{(h)}-C\beta_t^{(h)}\|_1
  &\le \delta_h
    + \sum_i \left|
      \operatorname{tr}[\mathcal J_i^\dagger(\Phi_\Delta(\rho_r))]
      \right|
  \le \delta_h
    + \sum_i \|\mathcal J_i^\dagger(\Phi_\Delta(\rho_r))\|_1 \\
  &= \delta_h + \|\mathcal W_c\Phi_\Delta(\rho_r)\|_1
  \le \delta_h + \|\mathcal W_c\Phi_\Delta\|_1
   = \delta_c.
  \end{aligned}
\].
	\end{proof}}
	\begin{proof}[Proof of Proposition \ref{prop:qc_long_run} (Long-run evolution and stabilization)]
		
		Theorem \ref{thm:coup_thm} combined with primitivity of $\Phi$ guarantees that there exist $\xi_{i}, i\in \{1,\dots,q\}$ such that 
		$$\lim_{t\to \infty} \rho_t =\tilde \rho=\sum_i \xi_i \Jc_i (\tilde \sigma_i)=\bigoplus_i \xi_{i} \tilde \sigma_{i}.$$

        Fix $h\in\Ic_x(\epsilon)$, by the explicit choice in Proposition \ref{prop:qc_middle_run},
    \[
        \beta_{t,i}^{(h)}
        =\tr\!\left[\Jc_i^\dag(\rho_{t-h})\right].
    \]
    Since $h$ is fixed, 
    \[
        \lim_{t\to\infty}\beta_{t,i}^{(h)}
        =\tr\!\left[
            \Jc_i^\dag\!\left(
                \lim_{t\to\infty}\rho_{t-h}
            \right)
        \right]
        =\tr\!\left[\Jc_i^\dag(\tilde{\rho})\right]
        =\xi_i.
    \]
    This holds for every $i=1,\ldots,q$ and every fixed
    $h\in\Ic_x(\epsilon)$.
	\end{proof}
	
	\section{Useful norms and properties}\label{app:norms}
	In the main text we introduced the following operator norms on \({\mathfrak{B}}(\Hc)\) and the corresponding induced norms on linear maps \(\Ec:{\mathfrak{B}}(\Hc)\to{\mathfrak{B}}(\Hc)\):
	\begin{align} \label{eq:sup_op_norms_def}
		&\norm{A}_\infty=\sqrt{r(A^\dag A )},  \quad &&\norm{\Ec}_\infty= \underset{\norm{A}_\infty=1}{\rm sup} \norm{\Ec(A)}_\infty,\\
		&\norm{A}_1=\tr\Bigl(\sqrt{A^\dag A}\Bigr),  &&\norm{\Ec}_1=\underset{\norm{A}_1=1}{\rm sup}\norm{\Ec(A)}_1,\nonumber\\
		& \norm{A}_2=\sqrt{\tr(A^\dag A)}, 
		&& \norm{\Ec}_2=\underset{{\norm{A}_2=1}}{{\rm sup}}\norm{\Ec(A)}_2.\nonumber
	\end{align}
	Here $r(\cdot)$ denotes the spectral radius.
	
	In this appendix we collect several useful elementary properties of these norms.
	
	\begin{proposition}\label{prop:russo_dye}
		Let $\Ec:{\mathfrak{B}}(\Hc)\to{\mathfrak{B}}(\Hc)$ be a completely positive map. {Then
		\[\|\Ec\|_\infty=\|\Ec(I)\|_\infty, \qquad \|\Ec\|_1=\sup_{A\in\Dc(\Hc)} \norm{\Ec(A)}_1.\]}
	\end{proposition}
	This identity follows from the Russo--Dye theorem; see, for
example, \cite[proof of Proposition~6.1]{wolf-qc}, \cite[Theorem 3.39]{watrous2018theory}.
\begin{proposition}\label{prop:norm_duality}
		Let $\Ec:{\mathfrak{B}}(\Hc)\to{\mathfrak{B}}(\Hc)$ be a linear map. Then
		\[\|\Ec^\dagger\|_\infty=\|\Ec\|_1.\]
	\end{proposition}
The identity follows from the duality of the trace and operator
norms \cite[Eq.~(1.173)]{watrous2018theory}, together with the
definitions \eqref{eq:sup_op_norms_def}.
	
	\begin{proposition}[Submultiplicativity]
		
		\label{prop:submult}
		
		Let $\Ec,\mathcal{P}:{\mathfrak{B}}(\Hc)\to{\mathfrak{B}}(\Hc)$ be linear maps. For every
		$p\in\{1,2,\infty\}$,
		\begin{equation}
			\norm{\Ec\mathcal{P}}_p
			\leq
			\norm{\Ec}_p\norm{\mathcal{P}}_p.
		\end{equation}

		Thus, the superoperator norms induced by the trace, Hilbert--Schmidt, and
		operator norms are submultiplicative under composition.
	\end{proposition}
	This follows directly from the definition of the induced norms. For the trace norm, see e.g. \cite[Proposition~3.41]{watrous2018theory}.
	
	\begin{lemma}[Relation between superoperator norms]\label{lem:relation_between_superop_norms}
		Let $\mathcal{H}$ be a $D$-dimensional Hilbert space, and let
		$\Ec:{\mathfrak{B}}(\mathcal{H})\rightarrow{\mathfrak{B}}(\mathcal{H})$
		be a linear map. Then the following relation between superoperators norms induced by the trace and the Hilbert-Schmidt norm holds
		\[
		\frac{1}{\sqrt D}\,\|\Ec\|_{2}
		\le
		\|\Ec\|_{1}
		\le
		\sqrt D\,\|\Ec\|_{2},
		\]
	\end{lemma}
	The result follows directly from \eqref{eq:sup_op_norms_def} and \cite[Equation 1.168 and 1.170]{watrous2018theory}. 
    
	We next relate the induced Hilbert--Schmidt norm of a superoperator to the
	spectral norm of its matrix representation.

	Let
	$\{F_a\}_{a=1}^{D^2}$ and $\{G_b\}_{b=1}^{D^2}$ be Hilbert--Schmidt
	orthonormal operator bases, so that
	${\rm tr}(F_a^\dag F_{a'})=\delta_{aa'}$ and similarly for the $G_b$.  We can introduce a matrix representation of a linear map $\Ec:{\mathfrak{B}}(\Hc)\to{\mathfrak{B}}(\Hc)$ as 
	\begin{equation}
		(\widehat\Ec)_{ba}={\rm tr}({G_b}^\dag{\Ec(F_a)}).
	\end{equation}
	This vectorization is an isometry from the Hilbert--Schmidt operator
	space to Euclidean space.  Consequently,
	\begin{equation}\label{eq:liouville-isometry}
		\norm{\Phi}_{2}=\norm{\widehat\Phi}_\infty,
	\end{equation}
	where the norm on the right is the usual matrix spectral norm, i.e., the largest
	singular value.
	\begin{proposition}[Norm of diagonalizable maps]\label{prop:diagonalizable-map}
		Let
		$\mathcal E:\mathcal {\mathfrak{B}}(\mathcal H)\to\mathcal {\mathfrak{B}}(\mathcal H)$ be a diagonalizable linear map 
		that is there exist an invertible linear map \(\Vc\) and a linear map $\Dc$ such that given an Hilbert–Schmidt orthonormal operator basis $\{F_i\}_{i=1}^{D^2}$
		$$\Ec=\Vc\Dc\Vc^{-1}, \qquad \Dc(F_i)=\lambda_iF_i.$$
		Here the $\{\lambda_i\}_{i=1}^{D^2}$ are the eigenvalues of $\Ec$.
		Define
		$\kappa
		:=
		\|\mathcal V\|_2\,
		\|\mathcal V^{-1}\|_2.$
		Then for every integer $t\geq 0$,
		$$\|\mathcal E^t\|_2
		\leq
		\kappa\max_j|\lambda_j|^t.$$
	\end{proposition}
	The above result follows from \eqref{eq:liouville-isometry} together with submultiplicativity of the norms.

	\begin{remark}[Normal maps]
		If $\Ec$ is normal with respect to the Hilbert--Schmidt inner product, then
		$\kappa=1$ and $\norm{\Ec^t}_{2}=\max_j|\lambda_j|^t$ for all $t\geq0$.
	\end{remark}
	
	\section{Detailed computations for the approximate symmetries examples}
	\label{app:symmetry_norms}
	In this appendix we provide explicit computation of some quantities used in the examples on approximate symmetries. 
	We denote by $\Phi^{(C)}$ and $\Phi^{(S)}$ the full perturbed channels of Examples~1 and~2, respectively, reserving $\Phi_S$ for the symmetric, unperturbed component used in the main text.
	
	\subsubsection*{Norm of perturbation channels}
	We compute the {\em norm of the perturbation} of the original quantum channel for both the example on near-complete reducibility and near-strong reducibility. We let $U=\sigma_z^a\sigma_z^b$ be the common symmetry for the two original channels.
	We denote the perturbations in the two examples by $\mathcal E_C$
	and $\mathcal E_S$, respectively:
	\[
	\mathcal E_C(X)=\xi(\sigma_x^aX\sigma_x^a-X),\qquad
	\mathcal E_S(X)=\xi\bigl(\sigma_x^aX\sigma_x^a-\mathcal W_c(X)\bigr).
	\]
	 In both examples the considered hilbert space partition is $\Hc=\Hc_1\oplus \Hc_2$ where orthogonal projectors on $\Hc_1$ and $\Hc_2$ are respectively $\Pi_1=(I+U)/2$ and $\Pi_2=(I-U)/2$. Consequently,
	\begin{equation*}
		\begin{aligned}
			\mathcal W_c(X)
			&=\sum_{i=1}^2\Pi_iX\Pi_i
			=\Pi_1X\Pi_1+\Pi_2X\Pi_2\\
			&=\frac{1}{4}\bigl((I+U)X(I+U)+(I-U)X(I-U)\bigr)
			=\frac{X+UXU}{2}.
		\end{aligned}
		\label{eq:symmetry-norm-projections}
	\end{equation*}
    Since $\Wc_c$ is CP and unital the properties of Appendix \ref{app:norms} ensure $\|\mathcal W_c\|_1=1$.
	Let $\mathcal U_{\sigma_x^a}(X)=\sigma_x^aX\sigma_x^a$. Since $\sigma_x^a$ is unitary,
	$\|\mathcal U_{\sigma_x^a}\|_1=1$. Using $\|\mathcal W_c\|_1=1$ and the triangle inequality,
	\begin{align*}
		\|\mathcal E_C\|_1
		&=\xi\|\mathcal U_{\sigma_x^a}-\Ic\|_1
		\leq\xi\bigl(\|\mathcal U_{\sigma_x^a}\|_1+\|\Ic\|_1\bigr)
		=2\xi,\\
		\|\mathcal E_S\|_1
		&=\xi\|\mathcal U_{\sigma_x^a}-\mathcal W_c\|_1
		\leq\xi\bigl(\|\mathcal U_{\sigma_x^a}\|_1+\|\mathcal W_c\|_1\bigr)
		=2\xi.
	\end{align*}
	Note that for $X=\ket{00}\bra{00}$
	\[
	\norm{\mathcal E_Z(X)}_1
	=\xi\norm{\bigl(\ket{10}\bra{10}-\ket{00}\bra{00}\bigr)}_1=2\xi,
	\qquad Z\in\{S,C\},
	\]
	therefore the upper bound is achieved implying 
		${\|\mathcal E_C\|_1
			=\|\mathcal E_S\|_1=2\xi.}$
	
	\subsubsection*{Common associated hybrid channel}
	
	We compare the full perturbed channels $\Phi^{(C)}$ and $\Phi^{(S)}$ for given $\alpha_1,\ldots,\alpha_4$ and $\xi$.
	The unperturbed channel in Example~2 is obtained by composing the
	unperturbed channel in Example~1 with the map $\mathcal W_c$. Therefore
	\[
	\Phi^{(S)}(X)=\Phi^{(C)}\bigl(\mathcal W_c(X)\bigr)
	+\xi \sigma_x^a\bigl(X-\mathcal W_c(X)\bigr)\sigma_x^a.
	\]
	Since $\mathcal W_c^2=\mathcal W_c$, this implies
	$\Phi^{(S)}\mathcal W_c=\Phi^{(C)}\mathcal W_c$.
	Thus the associated hybrid channel is the same in both examples:
	\[
	\Phi_{\oplus}=\mathcal W_c\Phi^{(C)}\mathcal W_c
	=\mathcal W_c\Phi^{(S)}\mathcal W_c,
	\qquad
	\Phi_{ij}=\Jc_i^\dag\Phi^{(C)}\Jc_j
	=\Jc_i^\dag\Phi^{(S)}\Jc_j.
	\]
	Since the two channels have the same associated hybrid channel, as we show next, they have the same stochastic complement.

	\subsubsection*{Stochastic complements}
	\label{app:symmetry-hsc}
	
	Assume $0<\xi\leq\alpha_1$. To compute the components of the hybrid channel and compute explicitly the  stochastic complement we consider the matrix  representation of operators in ${\mathfrak{B}}(\Hc) $ with respect to the basis $(\ket{00},\ket{11},\ket{10},\ket{01})$ for $\Hc=\Hc_1\oplus \Hc_2$, and the martix representation of operators in ${\mathfrak{B}}(\Hc_1),{\mathfrak{B}}(\Hc_2)$   with respect to the basis $(\ket{00},\ket{11})$ for $\mathcal H_1$ and $(\ket{10},\ket{01})$ for $\mathcal H_2$.
	With respect to this basis, the operators appearing
	in the symmetric channel and its perturbation have the block forms
	\begin{align*}
		\sigma_x^a&=\begin{pmatrix}0&I_2\\ I_2&0\end{pmatrix},
		\qquad \sigma_x^a\sigma_x^b&=
		\begin{pmatrix}\sigma_x&0\\0&\sigma_x\end{pmatrix},\qquad
		\sigma_z^a&=\begin{pmatrix}\sigma_z&0\\0&-\sigma_z\end{pmatrix},
		\qquad\sigma_z^b&=\begin{pmatrix}\sigma_z&0\\0&\sigma_z\end{pmatrix}.
	\end{align*}
	Here $I_2\in\mathbb{C}^2$ is the identity matrix, and
	$\sigma_x$ and $\sigma_z$ are the usual Pauli matrices.
	The components of the associated hybrid channel are obtained directly
	from either full channel $\Phi\in\{\Phi^{(C)},\Phi^{(S)}\}$ through the embedding and selection maps. In the considered examples, the embedding matrices are
	$J_1=\begin{pmatrix}I_2 & 0\end{pmatrix}^\dag,
	J_2=\begin{pmatrix}0 & I_2\end{pmatrix}^\dag,$
	so that $\Jc_j(Y)=J_jYJ_j^\dag$ and
	$\Jc_i^\dag(X)=J_i^\dag XJ_i$.  The matrix representation
	of each component of the channel is
	\[
	\Phi_{ij}(Y)=\Jc_i^\dag\Phi^{(S)}\Jc_j=J_i^\dag\Phi(J_jYJ_j^\dag)J_i.
	\]
	Let
	\begin{equation}
		\begin{aligned}
			\mathcal T_\xi(Y)
			&=(\alpha_1-\xi)Y
			+\alpha_2\sigma_xY\sigma_x
			+(\alpha_3+\alpha_4)\sigma_zY\sigma_z,\\
		\end{aligned}
	\end{equation}
	then 
	\begin{align*}
		\Phi_{11}&=\Jc_1^\dag \Phi \Jc_1=\mathcal T_\xi,\ \ 
		&\Phi_{22}&=\Jc_2^\dag \Phi \Jc_2=\mathcal T_\xi,\ \ &
		\Phi_{12}&=\Jc_1^\dag \Phi \Jc_2=\xi\mathcal I,\ 
		&\Phi_{21}&=\Jc_2^\dag \Phi \Jc_1=\xi\mathcal I.
	\end{align*}
	The stochastic complement is then obtained as 
	\begin{equation}\label{eq:symmetry-hsc-resolvent}
		\mathcal S_{ii}
		=\Phi_{ii}+\Phi_{ij}
		(\mathcal I_2-\Phi_{jj})^{-1}\Phi_{ji}\notag=\mathcal T_\xi+
		\xi^2(\mathcal I_2-\mathcal T_\xi)^{-1},
		\qquad j=3-i.
	\end{equation}
	
	$ $ 
	To evaluate the inverse explicitly, not that the map $\mathcal T_\xi$ is diagonal in the Pauli basis, i.e. 
	$\mathcal T_\xi(\sigma_\beta)=\lambda_\beta\sigma_\beta, \beta\in\{0,x,y,z\},$
	where
	$\lambda_0=1-\xi,
		\  \lambda_x=1-\xi-2(\alpha_3+\alpha_4),\notag \ 
		\lambda_y=2\alpha_1-1-\xi,
		\ \lambda_z=1-\xi-2\alpha_2.$
	Thus
	\begin{equation}
	(\mathcal I_2-\mathcal T_\xi)^{-1}(\sigma_\beta)
	=\frac{\sigma_\beta}{1-\lambda_\beta},
	\qquad
		\mathcal S_{ii}(\sigma_\beta)
		=\left(\lambda_\beta+
		\frac{\xi^2}{1-\lambda_\beta}\right)\sigma_\beta.
		\label{eq:symmetry-hsc-pauli-action}
	\end{equation}
	The eigenvalues of the complements are therefore 1 and 
	\begin{equation}
		\eta_\beta=\lambda_\beta+
		\frac{\xi^2}{1-\lambda_\beta}
		\qquad \beta\in\{x,y,z\}.
		\label{eq:symmetry-hsc-eigenvalues}
	\end{equation}
	Since an arbitrary operator in $\Bf(\Hc_i), i=1,2$ can be written as	\begin{equation}
	X\!=\!\frac{1}{2}\!\left[\operatorname{tr}(X)I_2+
	\!\!\!\!\!\sum_{\beta\in\{x,y,z\}}\!\!\!\!\!
	\operatorname{tr}(\sigma_\beta X)\sigma_\beta\right],\ 
		\mathcal S_{ii}(X)=\!\frac{1}{2} \!\left[
		\operatorname{tr}(X)I_2+
		\!\!\!\!\! \sum_{\beta\in\{x,y,z\}}\!\!\!\!\! \eta_\beta
		\operatorname{tr}(\sigma_\beta X)\sigma_\beta\right]\!. 
		\label{eq:symmetry-hsc-explicit}
	\end{equation}
	
	\subsubsection*{Deviations from complete reducibility}
	The two channels have the same associated hybrid channel and
	stochastic complements. In particular, the common off-diagonal
	blocks $\Phi_{12}=\Phi_{21}=\xi\mathcal I$ give
	$\|\Phi_{*i}\|_1=\xi$ for $i=1,2$. Hence, in both examples,
	\begin{equation}
		\delta_h=2\max_{i\in\{1,2\}}\|\Phi_{*i}\|_1=2\xi.
		\label{eq:symmetry-hybrid-deviation}
	\end{equation}
	
	To compute $\delta_c$, define $\Qc=\mathcal I-\mathcal W_c$, which
	retains the off-diagonal blocks with respect to
	$\Hc=\Hc_1\oplus\Hc_2$:
	\[
	\Qc(X)=X-\mathcal W_c(X)
	=\Pi_1X\Pi_2+\Pi_2X\Pi_1
	=\frac{X-UXU}{2}.
	\]
	
	It is possible to notice that, for either $\Phi\in\{\Phi^{(C)},\Phi^{(S)}\}$,
	$\mathcal W_c\Phi=\Phi\mathcal W_c$. Using
	$\Phi_{\oplus}=\mathcal W_c\Phi\mathcal W_c$ and
	$\mathcal W_c\Qc=\Qc\mathcal W_c=0$, we obtain
	\[
	\Phi_{\Delta}=\Phi-\mathcal W_c\Phi\mathcal W_c=\Qc\Phi\Qc,
	\qquad
	\mathcal W_c\Phi_{\Delta}=\mathcal W_c\Qc\Phi\Qc=0.
	\]
	Therefore Definition~\ref{def:qs_deviation} gives, for both channels,
	\[
	\delta_c=\delta_h+\|\mathcal W_c\Phi_{\Delta}\|_1=2\xi.
	\]
	
	\subsubsection*{Deviations from strong reducibility}
	Since $\delta_h=2\xi$, Definition~\ref{def:qs_deviation} gives
	$\delta_s=2\xi+\|\Phi_{\Delta}\|_1$, so it remains to compute the norm
	of the coherence remainder in each example.
	
	\paragraph*{Example 1: Nearly completely reducible two qubit system}
	
	Since $U\mathcal Q(X)U=-\mathcal Q(X)$ and
	$\sigma_z^a=U\sigma_z^b$, we have $
	\sigma_z^a\mathcal Q(X)\sigma_z^a
	=-\sigma_z^b\mathcal Q(X)\sigma_z^b.$
	which yeald
	\begin{equation}
		\begin{aligned}
			\Phi_{\Delta}(X)={\mathcal Q }\Phi{ \mathcal Q}(X)&=(\alpha_1-\xi)\mathcal Q(X)
			+\alpha_2\sigma_x^a\sigma_x^b\mathcal Q(X)\sigma_x^a\sigma_x^b\\
			&+(\alpha_4-\alpha_3)\sigma_z^b\mathcal Q(X)\sigma_z^b
			+\xi \sigma_x^a\mathcal Q(X)\sigma_x^a.
		\end{aligned}
		\label{eq:symmetry-coherence-map}
	\end{equation}
	Since $\alpha_1-\xi\geq0$, the triangle inequality yields
	\begin{align}
		\|\Phi_{\Delta}(X)\|_1
		\leq\bigl(\alpha_1-\xi+\alpha_2
		+|\alpha_4-\alpha_3|+\xi\bigr)\|\mathcal Q(X)\|_1
		\leq\bigl(\alpha_1+\alpha_2
		+|\alpha_4-\alpha_3|\bigr)\|X\|_1.
		\label{eq:symmetry-coherence-upper-bound}
	\end{align}
	where we used the fact that unitary invariance and
	the triangle inequality give $\|\Qc(X)\|_1\leq\|X\|_1$.
	To show that this bound is sharp, define
	$\ket{\pm}=(\ket{0}\pm\ket{1})/\sqrt{2}$ and take the density
	operator $\rho_x=\ket{++}\bra{++}$, which has trace norm one.
	Its off-diagonal component is
	\[
	\mathcal Q(\rho_x)
	=\frac{\ket{++}\bra{++}-\ket{--}\bra{--}}{2}.
	\]
	This operator is invariant under conjugation by $\sigma_x^a$ and by
	$\sigma_x^a\sigma_x^b$, whereas
	\[
	\sigma_z^b\mathcal Q(\rho_x)\sigma_z^b
	=\frac{\ket{+-}\bra{+-}-\ket{-+}\bra{-+}}{2}.
	\]
	These two off-diagonal operators have orthogonal supports and both have
	trace norm one. In view of this Equation~\eqref{eq:symmetry-coherence-map} gives
	\begin{align}
		\norm{\Phi_{\Delta}(\rho_x)}_1&
		=\norm{(\alpha_1+\alpha_2)\mathcal Q(\rho_x)
			+(\alpha_4-\alpha_3)\sigma_z^b\mathcal Q(\rho_x)\sigma_z^b}_1\\ &= \norm{(\alpha_1+\alpha_2)\mathcal Q(\rho_x)}_1
		+\norm{(\alpha_4-\alpha_3)\sigma_z^b\mathcal Q(\rho_x)\sigma_z^b}_1\\&=\alpha_1+\alpha_2+|\alpha_3-\alpha_4|
		=1-2\min\{\alpha_3,\alpha_4\}.
		\label{eq:symmetry-coherence-norm-exact}
	\end{align}
	Therefore,
	\begin{equation}
		{\delta_s=2\xi+1-2\min\{\alpha_3,\alpha_4\}.}
		\label{eq:symmetry-strong-deviation-one}
	\end{equation}
	For $(\alpha_1,\alpha_2,\alpha_3,\alpha_4)=(0.49,0.25,0.25,0.01)$,
	this gives $\|\Phi_{\Delta}\|_1=0.98$ and $\delta_s=0.98+2\xi$.
	
	\paragraph*{Example 2: Nearly strongly reducible two qubit system}
	
	The strongly reducible symmetric component in~\eqref{eq:sr_channel}
	annihilates $\Qc(X)$. Using $\Phi_{\Delta}=\Qc\Phi\Qc$, we therefore obtain
	\[
	\Phi_{\Delta}(X)=\Qc \Phi \Qc(X)= \Qc(\Phi_S+\Ec)\Qc(X)= \Qc\Ec\Qc(X)= \xi \sigma_x^a\mathcal Q(X)\sigma_x^a.
	\]
	Unitary invariance gives
	$\|\Phi_{\Delta}(X)\|_1=\xi\|\mathcal Q(X)\|_1\leq \xi \norm{X}_1$ for every $X$, 
	since $\|\mathcal Q\|_1=1$. The upper bound is attained, for example,
	on the same input $\rho_x$ used in the previous section. Therefore we obtain
	$\|\Phi_{\Delta}\|_1=\xi, \delta_s=3\xi.$
	
	\section{Detailed computations for the approximate QND example}
	\label{app:qnd-example}
	
	We use the notation of the QND example and write
	$\Phi=\Phi_{\ti,\varepsilon}$, $U=U_{\ti,\varepsilon}$, and
	$\eta=e^{-\gamma\ti}$. We assume $\gamma>0$, $\ti>0$, $E>0$,
	and $0<\beta<\infty$, so $0<\eta<1$ and $\tau_\beta>0$.
	The stochastic-complement and stationary-state calculations use $0<s<1$,
	with $s$ defined in~\eqref{eq:qnd-transition-probability}; the norm formulas
	also apply at the endpoints by continuity.
	
	\subsubsection*{Norm of the perturbation}
	\label{app:qnd-perturbation-norm}
	
	We compute the induced trace norm of the perturbation defined
	in~\eqref{eq:qnd-perturbation-map},
	\[
	\begin{aligned}
		\Ec&=\Phi_{\ti,\varepsilon}-\Phi_{\ti,0}
		=\Wc_c\Bigl[
		\bigl(\operatorname{Ad}_{U_{\ti,\varepsilon}}
		-\operatorname{Ad}_{U_{\ti,0}}\bigr)
		\otimes\Lambda_{\beta,\ti}\Bigr].
	\end{aligned}
	\]
	To obtain the action of $\Ec$, we expand $X$ in the product operator basis:
	\[
	 X=\sum_{k,\ell,\mu,\nu=0}^1 x_{k\mu,\ell\nu}
	 \ketbra{k}{\ell}_A\otimes\ketbra{\mu}{\nu}_B,
	 \qquad
	 x_{k\mu,\ell\nu}
	 =(\langle k|_A\otimes\langle\mu|_B)X
	 (|\ell\rangle_A\otimes|\nu\rangle_B).
	\]
	We recall the  partial trace over the first factor is
	\begin{align*}
	 \tr_A(X)
	 &=\sum_{k,\ell,\mu,\nu}x_{k\mu,\ell\nu}
	 \tr\bigl(\ketbra{k}{\ell}_A\bigr)
	 \ketbra{\mu}{\nu}_B.
	\end{align*}
	
	The tensor-product map acts separately on the two factors of each basis
	operator. Applying $\Wc_c(Y)=\sum_{j=0}^1\Pi_{j+1}Y\Pi_{j+1}$, with
	$\Pi_{j+1}=\ketbra{j}{j}_A\otimes I_B$, then gives
	\begin{align*}
	 &\Ec(X)
	 =\Wc_c\Bigl(\bigl[
	 (\operatorname{Ad}_{U_{\ti,\varepsilon}}-\operatorname{Ad}_{U_{\ti,0}})
	 \otimes\Lambda_{\beta,\ti}\bigr](X)\Bigr)\\
	 &=\sum_{k,\ell,\mu,\nu}x_{k\mu,\ell\nu}\,
	 \Wc_c\Bigl(\bigl[
	 U_{\ti,\varepsilon}\ketbra{k}{\ell}_A U_{\ti,\varepsilon}^\dagger-U_{\ti,0}\ketbra{k}{\ell}_A U_{\ti,0}^\dagger
	 \bigr]\otimes\Lambda_{\beta,\ti}(\ketbra{\mu}{\nu}_B)\Bigr)\\
	 &=\sum_{j,k,\ell,\mu,\nu}x_{k\mu,\ell\nu}
	 \bigl[\langle j|U_{\ti,\varepsilon}|k\rangle
	 \langle\ell|U_{\ti,\varepsilon}^\dagger|j\rangle
	 -\langle j|U_{\ti,0}|k\rangle
	 \langle\ell|U_{\ti,0}^\dagger|j\rangle\bigr]
	 \ketbra{j}{j}_A\otimes\Lambda_{\beta,\ti}(\ketbra{\mu}{\nu}_B)\\
	 &
	 =\sum_j\ketbra{j}{j}_A\otimes\Lambda_{\beta,\ti}\Bigl(
	 \sum_{k,\ell,\mu,\nu}x_{k\mu,\ell\nu}
	 \bigl[\langle j|U_{\ti,\varepsilon}|k\rangle
	 \langle\ell|U_{\ti,\varepsilon}^\dagger|j\rangle-\langle j|U_{\ti,0}|k\rangle
	 \langle\ell|U_{\ti,0}^\dagger|j\rangle\bigr]
	 \ketbra{\mu}{\nu}_B\Bigr)\\
	 &=\sum_j\ketbra{j}{j}_A\otimes\Lambda_{\beta,\ti}\Bigl(
	 \sum_{k,\ell,\mu,\nu}x_{k\mu,\ell\nu}
	 \tr\bigl((U_{\ti,\varepsilon}^\dagger\ketbra{j}{j}_A
	 U_{\ti,\varepsilon}
	 -U_{\ti,0}^\dagger\ketbra{j}{j}_A
	 U_{\ti,0})\ketbra{k}{\ell}_A\bigr)
	 \ketbra{\mu}{\nu}_B\Bigr)\\
	 &=\sum_j\ketbra{j}{j}_A\otimes\Lambda_{\beta,\ti}\Bigl(
	 \tr_A\Bigl[\Bigl(\bigl(
	 U_{\ti,\varepsilon}^\dagger\ketbra{j}{j}_A U_{\ti,\varepsilon}-U_{\ti,0}^\dagger\ketbra{j}{j}_A U_{\ti,0}
	\bigr)\otimes I_B\Bigr)X\Bigr]\Bigr).
	\end{align*}
Let
	\begin{equation}
	\begin{aligned}
	 D&=U_{\ti,\varepsilon}^\dagger\ketbra{0}{0}_A U_{\ti,\varepsilon}
	 -U_{\ti,0}^\dagger\ketbra{0}{0}_A U_{\ti,0}
	 =U_{\ti,\varepsilon}^\dagger\ketbra{0}{0}_A U_{\ti,\varepsilon}
	 -\ketbra{0}{0}_A.
	\end{aligned}
	\end{equation}
	Using $\ketbra{1}{1}_A=I_A-\ketbra{0}{0}_A$ and unitarity
	\[
	 U_{\ti,\varepsilon}^\dagger\ketbra{1}{1}_A U_{\ti,\varepsilon}
	 -U_{\ti,0}^\dagger\ketbra{1}{1}_A U_{\ti,0}=-D.
	\]
	Substituting these two expressions into the preceding sum gives
	\begin{equation}
	\begin{aligned}
		\Ec(X)&=\ketbra{0}{0}_A\otimes\Lambda_{\beta,\ti}
		\Bigl(\tr_A\bigl[(D\otimes I_B)X\bigr]\Bigr)
		-\ketbra{1}{1}_A\otimes\Lambda_{\beta,\ti}
		\Bigl(\tr_A\bigl[(D\otimes I_B)X\bigr]\Bigr)\\
		&=\sigma_z\otimes\Lambda_{\beta,\ti}
		\Bigl(\tr_A\bigl[(D\otimes I_B)X\bigr]\Bigr),
	\end{aligned}
		\label{eq:qnd-perturbation-factorization}
	\end{equation}
	where $\sigma_z=\ketbra{0}{0}_A-\ketbra{1}{1}_A$.
	The Hilbert--Schmidt adjoint of
	$X\mapsto\tr_A[(D\otimes I_B)X]$ is $Y\mapsto D\otimes Y$.
	By duality of the trace and operator norms,
	\begin{equation}
		\bigl\|\tr_A[(D\otimes I_B)X]\bigr\|_1
		\leq\|D\|_\infty\|X\|_1.
	\end{equation}
	Since $\|\Lambda_{\beta,\ti}\|_{1}=1$ and $\|\sigma_z\|_1=2$,
	multiplicativity of the trace norm on tensor products
	and~\eqref{eq:qnd-perturbation-factorization} give
	\begin{equation}
		\|\Ec(X)\|_1\leq 2\|D\|_\infty\|X\|_1.
		\label{eq:qnd-perturbation-upper-bound}
	\end{equation}
	
	To evaluate $\|D\|_\infty$, use
	\begin{equation}
		U_{\ti,\varepsilon}=\cos(\nu\ti)I_A
		-i\frac{\sin(\nu\ti)}{\nu}
		\left(\frac{\omega}{2}\sigma_z+\varepsilon\Omega\sigma_x\right).
	\end{equation}
	By the Born rule, the probability of measuring the atom in $|1\rangle_A$
	at the end of a cycle, given that it starts in $|0\rangle_A$, is
	\begin{equation}
		s:=|\langle1|U_{\ti,\varepsilon}|0\rangle|^2
		=\frac{\varepsilon^2\Omega^2}{\nu^2}\sin^2(\nu\ti).
	\end{equation}
	The reverse transition has the same probability $s$, since the two
	off-diagonal entries of $U_{\ti,\varepsilon}$ coincide. Thus $1-s$ is
	the probability of remaining at the same atomic level. In particular,
	unitarity gives
	\[
		|\langle0|U_{\ti,\varepsilon}|0\rangle|^2
		+|\langle1|U_{\ti,\varepsilon}|0\rangle|^2=1.
	\]
	As the difference of two rank-one projections, $D$ is Hermitian and
	traceless. Each projection equals its square and has trace one;
	expanding $D^2$ and using cyclicity of the trace therefore yields
	\begin{equation}
	\begin{aligned}
		\tr(D^2)
		&=2-2\tr\bigl(U_{\ti,\varepsilon}^\dagger
		\ketbra{0}{0}_A U_{\ti,\varepsilon}\ketbra{0}{0}_A\bigr)\\
		&=2\bigl(1-|\langle0|U_{\ti,\varepsilon}|0\rangle|^2\bigr)
		=2s.
	\end{aligned}
	\end{equation}
	Since $D$ acts on the two-dimensional space $\mathcal H_A$, its two
	real eigenvalues sum to zero and can be written as $\lambda$ and
	$-\lambda$. Thus $2\lambda^2=\tr(D^2)=2s$, so the eigenvalues are
	$\pm\sqrt{s}$ and $\|D\|_\infty=\sqrt{s}$.
	
	Finally, let $|d\rangle$ be a normalized eigenvector of $D$ with eigenvalue
	$\sqrt{s}$. The density operator $\ketbra{d}{d}\otimes\tau_\beta$
	has unit trace norm and, since $\Lambda_{\beta,\ti}(\tau_\beta)=\tau_\beta$,
	\begin{equation}
		\Ec\bigl(\ketbra{d}{d}\otimes\tau_\beta\bigr)
		=\sqrt{s}\,\sigma_z\otimes\tau_\beta.
	\end{equation}
	Its trace norm is $2\sqrt{s}$, attaining the bound
	in~\eqref{eq:qnd-perturbation-upper-bound} and proving
	\begin{equation}
	{\displaystyle
			\norm{\Ec}_1=2\sqrt{s}
			=\frac{2|\varepsilon\Omega|}{\nu}|\sin(\nu\ti)|.}
	\end{equation}
	
	\subsubsection*{Associated hybrid channel}
	Since $\Wc_c \Phi=\Phi$, the associated hybrid channel is
	$\Phi_{\oplus}=\Wc_c \Phi \Wc_c=\Phi \Wc_c$.
	The embedding and block-selection maps are
	\begin{align*}
		\Jc_j(Y)&=\ketbra{j-1}{j-1}_A\otimes Y,\\
		\Jc_j^\dag(X)&=(\langle j-1|_A\otimes I_B)X(|j-1\rangle_A\otimes I_B),
		\qquad j\in\{1,2\}.
	\end{align*}
	For an input supported on subspace $j$, the channel acts as
	\begin{align*}
		\Phi\bigl(\Jc_j(Y)\bigr)
		&=\Wc_c\Bigl(U\ketbra{j-1}{j-1}_A U^\dagger
		\otimes\Lambda_{\beta,\ti}(Y)\Bigr)\\
		&=\sum_{i=1}^2\ketbra{i-1}{i-1}_A\otimes
		\langle i-1|U|j-1\rangle\langle j-1|U^\dagger|i-1\rangle
		\Lambda_{\beta,\ti}(Y).
	\end{align*}
	Here $\Wc_c$ retains only the diagonal atomic blocks, and
	$\langle i-1|U|j-1\rangle\langle j-1|U^\dagger|i-1\rangle
	=|\langle i-1|U|j-1\rangle|^2$. Applying the block-selection map
	$\Jc_i^\dag$ gives
	\begin{equation}
		\Phi_{ij}(Y)=\Jc_i^\dag\Phi\Jc_j(Y)
		=|\langle i-1|U|j-1\rangle|^2\Lambda_{\beta,\ti}(Y).
	\end{equation}
	Unitarity and $s=|\langle1|U|0\rangle|^2$ imply
	$\Phi_{jj}=(1-s)\Lambda_{\beta,\ti}$ and
	$\Phi_{3-j,j}=s\Lambda_{\beta,\ti}$.
	
	\subsubsection*{Stochastic complements}
	\label{app:qnd-thermal-complements}
	
	For $j\in\{1,2\}$ and $i=3-j$, substituting the hybrid-channel components gives
	
	\begin{align} \label{eq:app_hsc}
		\Sc_{jj}
		=\Phi_{jj}+\Phi_{ji}(\Ic-\Phi_{ii})^{-1}\Phi_{ij}
		=(1-s)\Lambda_{\beta,\ti}
		+s^2\Lambda_{\beta,\ti}
		\bigl(\Ic-(1-s)\Lambda_{\beta,\ti}\bigr)^{-1}
		\Lambda_{\beta,\ti}.
	\end{align}
	Thus the two stochastic complements coincide.
	
	Let $\mathcal T_\beta(Y)=\tr(Y)\tau_\beta$. Since $\tr(\tau_\beta)=1$,
	we have $\mathcal T_\beta^2=\mathcal T_\beta$, and the channel decomposes as
	\begin{equation}
		\Lambda_{\beta,\ti}
		=\mathcal T_\beta+\eta(\Ic-\mathcal T_\beta),
		\qquad \eta=e^{-\gamma\ti}.
	\end{equation}
	It follows that
	\[
		\Ic-(1-s)\Lambda_{\beta,\ti}
		=s\mathcal T_\beta
		+\bigl[1-(1-s)\eta\bigr](\Ic-\mathcal T_\beta).
	\]
	The maps $\mathcal T_\beta$ and $\Ic-\mathcal T_\beta$ are complementary
	projectors: they are idempotent and satisfy
	$\mathcal T_\beta(\Ic-\mathcal T_\beta)
	=(\Ic-\mathcal T_\beta)\mathcal T_\beta=0$.
	For $s>0$, both coefficients are positive, therefore the previous observation implies 
	\begin{equation}
		\bigl(\Ic-(1-s)\Lambda_{\beta,\ti}\bigr)^{-1}
		=\frac{1}{s}\mathcal T_\beta
		+\frac{1}{1-(1-s)\eta}(\Ic-\mathcal T_\beta).
	\end{equation}
	Substitution into~\eqref{eq:app_hsc} yields
	\begin{equation}
		\Sc_{jj}=\mathcal T_\beta
		+r_{\mathrm{HSC}}(\Ic-\mathcal T_\beta),
	\end{equation}
	with $r_{\mathrm{HSC}}$ given in~\eqref{eq:qnd-hsc-contraction}.
	For $0<s<1$ and $0<\eta<1$, one has $0<r_{\mathrm{HSC}}<1$; indeed,
	\begin{equation}
		1-r_{\mathrm{HSC}}
		=\frac{(1-\eta)\bigl(1-(1-2s)\eta\bigr)}{1-(1-s)\eta}>0.
	\end{equation}
	Thus, for every nonzero $Y\geq0$,
	$\Sc_{jj}(Y)\geq(1-r_{\mathrm{HSC}})\tr(Y)\tau_\beta>0$.
	Each complement is therefore primitive, with stationary state $\tau_\beta$.
	Its action on an operator $Y$ is
	\begin{equation}
		\Sc_{jj}(Y)=r_{\mathrm{HSC}}Y
		+(1-r_{\mathrm{HSC}})\tr(Y)\tau_\beta.
	\end{equation}
	
	To recognize the eigenvalues of the HSCs, 
	since $\tr(\tau_\beta)=1$ and the Pauli operators are traceless,
	\begin{align*}
		\Sc_{jj}(\tau_\beta)&=\tau_\beta,\\
		\Sc_{jj}(\sigma_k^B)&=r_{\mathrm{HSC}}\sigma_k^{B},
		\qquad k\in\{x,y,z\}.
	\end{align*}
	The three Pauli operators span the traceless operators on $\mathcal H_B$.
	Together with $\tau_\beta$, which has nonzero trace, they form an
	eigenbasis of $\Bf(\mathcal H_B)$. Hence each $\Sc_{jj}$ has
	eigenvalue $1$ with multiplicity one and eigenvalue $r_{\mathrm{HSC}}$
	with multiplicity three. Since $0<r_{\mathrm{HSC}}<1$, the latter is
	the second-largest eigenvalue in modulus.
		
	\subsubsection*{Computing stationary states}
	For $0<s<1$, it can be checked that $\Phi$ is primitive, and hence
	irreducible, with unique stationary state
	$\widetilde\rho=(I_A/2)\otimes\tau_\beta$. This state is block diagonal
	with respect to the Hilbert space partition
	$\mathcal H=\mathcal H_1\oplus\mathcal H_2$, so the hypotheses of
	Theorem~\ref{thm:coup_thm} are satisfied.

	The complements have stationary states
	$\widetilde\sigma_j=\tau_\beta$, so the coupling matrix has entries
	\[
	 c_{ij}=\tr\bigl[\Phi_{ij}(\tau_\beta)\bigr]
	 =|\langle i-1|U|j-1\rangle|^2,
	 \qquad C=\begin{pmatrix}1-s&s\\s&1-s\end{pmatrix}.
	\]
	For $s>0$, solving $C\xi=\xi$ with $\xi_1+\xi_2=1$ gives
	$\xi_1=\xi_2=1/2$. Algorithm~\ref{alg:ss_comp} therefore reconstructs
	the stationary state~\eqref{eq:qnd-stationary-state}.
	
	\subsubsection*{Deviations from reducibility}
	\label{app:qnd-reducibility-deviations}
	
	For this partition into two subspaces, the off-diagonal hybrid-channel
	components are $\Phi_{12}=\Phi_{21}=s\Lambda_{\beta,\ti}$. Therefore,
	\begin{equation}
		\delta_h
		=2\max\bigl\{\norm{\Phi_{21}}_1,\norm{\Phi_{12}}_1\bigr\}
		=2s,
	\end{equation}
	where we used $\norm{\Lambda_{\beta,\ti}}_1=1$.
	
	Since $\Phi_{\oplus}=\Phi \Wc_c$, the coherence remainder is
	$\Phi_{\Delta}=\Phi (\Ic-\Wc_c)$. Write
	$X_{ij}=(\langle i-1|_A\otimes I_B)X(|j-1\rangle_A\otimes I_B)$,
	$i,j\in\{1,2\}$, and let $P_j=\ketbra{j-1}{j-1}_A$. Then
	\begin{align*}
	 X&=\ketbra{0}{0}_A\otimes X_{11}
	 +\ketbra{0}{1}_A\otimes X_{12}
	 +\ketbra{1}{0}_A\otimes X_{21}
	 +\ketbra{1}{1}_A\otimes X_{22},\\
	 \Wc_c(X)&=\sum_{j=1}^2(P_j\otimes I_B)X(P_j\otimes I_B)
	 =\ketbra{0}{0}_A\otimes X_{11}
	 +\ketbra{1}{1}_A\otimes X_{22}.
	\end{align*}
	Therefore
	\[
	 (\Ic-\Wc_c)(X)=X-\Wc_c(X)
	 =\ketbra{0}{1}_A\otimes X_{12}
	 +\ketbra{1}{0}_A\otimes X_{21}.
	\]
	Set
	\begin{equation}
	 a=\langle0|U|0\rangle\,\overline{\langle0|U|1\rangle},
	 \qquad |a|=\sqrt{s(1-s)}.
	\end{equation}
	Applying the channel to these two coherence components yields
	\begin{align}
	 \Phi_{\Delta}(X)
	 &=\Phi\bigl((\Ic-\Wc_c)(X)\bigr)\notag\\
	 &=\Phi\bigl(\ketbra{0}{1}_A\otimes X_{12}
	 +\ketbra{1}{0}_A\otimes X_{21}\bigr)\notag\\
	 &=\Wc_c\Bigl(U\ketbra{0}{1}_A U^\dagger
	 \otimes\Lambda_{\beta,\ti}(X_{12})
	 +U\ketbra{1}{0}_A U^\dagger
	 \otimes\Lambda_{\beta,\ti}(X_{21})\Bigr)\notag\\
	 &=\sum_{j=1}^2P_j\otimes\Lambda_{\beta,\ti}\Bigl(
	 \langle j-1|U|0\rangle\langle1|U^\dagger|j-1\rangle\,X_{12}
	 +\langle j-1|U|1\rangle
	 \langle0|U^\dagger|j-1\rangle\,X_{21}\Bigr)\notag\\
	 &=P_1\otimes\Lambda_{\beta,\ti}
	 \bigl(aX_{12}+\overline a X_{21}\bigr)
	 +P_2\otimes\Lambda_{\beta,\ti}
	 \bigl(-aX_{12}-\overline a X_{21}\bigr)\notag\\
	 &=(P_1-P_2)\otimes\Lambda_{\beta,\ti}
	 \bigl(aX_{12}+\overline a X_{21}\bigr),
	 \label{eq:qnd-coherence-remainder}
	\end{align}
	where in the above expression we used the fact that definition of the adjoint gives
	$\langle1|U^\dagger|j-1\rangle
	=\overline{\langle j-1|U|1\rangle}$, while 
	$\langle0|U|0\rangle\,\overline{\langle0|U|1\rangle}
	 +\langle1|U|0\rangle\,\overline{\langle1|U|1\rangle}=0.
	$
	For any operator $X$, trace-norm contractivity of $\Lambda_{\beta,\ti}$ gives
	\begin{equation}
		\|\Phi_{\Delta}(X)\|_1
		\leq 2|a|\bigl(\|X_{12}\|_1+\|X_{21}\|_1\bigr)
		=2|a|\|(\Ic-\Wc_c)(X)\|_1.
	\end{equation}
	Let $V=(P_1-P_2)\otimes I_B$ and observe that since V is unitary
	\begin{equation}
		(\Ic-\Wc_c)(X)=\frac{X-VXV^\dagger}{2},
		\qquad \|(\Ic-\Wc_c)(X)\|_1\leq\|X\|_1.
	\end{equation}
	Thus $\norm{\Phi_{\Delta}}_1\leq2|a|$. $\|\Phi_{\Delta}(X_*)\|_1=2|a|$ is  attained for $X_*=|0\rangle\langle1|_A\otimes\tau_B$, $\|X_*\|_1=1$
	. Consequently,
	\begin{equation}
		\norm{\Phi_{\Delta}}_1=2\sqrt{s(1-s)}.
	\end{equation}
	Finally, since the output of $\Phi_{\Delta}$ is block diagonal,
	$\Wc_c \Phi_{\Delta}=\Phi_{\Delta}$, and therefore
	\begin{equation}
		\delta_c=\delta_s=\delta_h+\norm{\Wc_c \Phi_{\Delta}}_1
		=2s+2\sqrt{s(1-s)}.
	\end{equation}
  
	\bibliographystyle{IEEEtran}
	\bibliography{bibliography}

\end{document}